\documentclass[11pt,a4paper,nofootinbib,showpacs,preprint,utf8]{article}
\usepackage{jheppub}
\pdfoutput=1
\usepackage{amsmath,amssymb}
\usepackage{amsfonts}
\usepackage{epsfig}
\usepackage{graphicx}
\usepackage{bbm}
\usepackage{pifont}
\usepackage[usenames,dvipsnames]{xcolor}
\definecolor{warmblack}{rgb}{0.0, 0.26, 0.26}
\definecolor{mediumtealblue}{rgb}{0.0, 0.33, 0.71}
\usepackage{slashed}
\usepackage[compat=1.1.0]{tikz-feynman}
\usepackage{tikzsymbols}
\usepackage{comment}
\usepackage[normalem]{ulem}

\preprint{MITP-25-049}

\title{Leptogenesis from Dark Matter Coannihilation}

\author[a]{Simran Arora}
\author[b]{, Debasish Borah}
\author[c]{, Arnab Dasgupta}
\author[d,e]{, P. S. Bhupal Dev}
\author[f]{, Devabrat Mahanta}

\affiliation[a]{Department of Physics and Astronomical Science, Central University of Himachal Pradesh, Dharamshala, Himachal Pradesh 176215, India}
\affiliation[b]{Department of Physics, Indian Institute of Technology Guwahati, Assam 781039, India}
\affiliation[c]{Pittsburgh Particle Physics, Astrophysics, and Cosmology Center, Department of Physics and Astronomy, University of Pittsburgh, Pittsburgh, Pennsylvania 15260, USA}
\affiliation[d]{Department of Physics and McDonnell Center for the Space Sciences, Washington University in St. Louis, Missouri 63130, USA}
\affiliation[e]{PRISMA$^+$ Cluster of Excellence \& Mainz Institute for Theoretical Physics, 
Johannes Gutenberg-Universit\"{a}t Mainz, 55099 Mainz, Germany}
\affiliation[f]{Department of Physics, Pragjyotish College, Guwahati, Assam 781009, India}

\emailAdd{009simranarora@gmail.com}
\emailAdd{dborah@iitg.ac.in}
\emailAdd{bdev@wustl.edu}
\emailAdd{arnabdasgupta@pitt.edu}
\emailAdd{devabrat@pragjyotishcollege.ac.in}

\abstract{We propose a minimal extension of the type-I seesaw model to realise leptogenesis from the coannihilation of dark sector particles. The type-I seesaw model is extended with a singlet fermion and two singlet scalars charged under a $Z_{2}$ symmetry. The $Z_{2}$-odd singlet scalar is the dark matter candidate. Here the usual type-I seesaw mechanism generates neutrino mass, and a net lepton asymmetry is generated from the coannihilation of the dark matter and the $Z_2$-odd singlet fermion. The $Z_{2}$-even singlet scalar is important in dark matter phenomenology. Successful leptogenesis is possible at TeV-scale, unlike the vanilla case. This minimal extension provides an elegant explanation of successful leptogenesis with direct connection to the dark matter abundance in the Universe. }


\makeatletter
\gdef\@fpheader{}
\makeatother

\begin{document}
\maketitle

\section{Introduction}
As suggested by various astrophysical and cosmological observations, the pressureless matter sector of the present Universe is dominated by a non-luminous, non-baryonic form of matter, known as dark matter (DM), comprising almost five times the density of ordinary matter or baryons~\cite{Planck:2018vyg, ParticleDataGroup:2024cfk, Cirelli:2024ssz}. More quantitatively, the present abundance of DM is quoted in terms of density parameter $\Omega_{\rm DM}$ and reduced Hubble constant $h$ = Hubble Parameter/$(100 \;\text{km} ~\text{s}^{-1} 
\text{Mpc}^{-1})$ as~\cite{Planck:2018vyg}
\begin{equation}
\Omega_{\text{DM}} h^2 = 0.120\pm 0.001
\label{dm_relic}
\end{equation}
\noindent at 68\% confidence level (CL). The presence of DM is still a mystery due to the fact that the Standard Model (SM) of particle physics does not have a suitable particle DM candidate. 

The observed matter-antimatter asymmetry in the visible or baryonic matter has led to another puzzle known as the baryon asymmetry of Universe (BAU)~\cite{ParticleDataGroup:2024cfk, Planck:2018vyg, Bodeker:2020ghk}. The observed BAU is quantified in terms of the baryon-to-photon number density ratio as~\cite{Planck:2018vyg} 
\begin{equation}
\eta_B = \frac{n_{B}-n_{\overline{B}}}{n_{\gamma}} = (6.12\pm 0.04) \times 10^{-10}, 
\label{etaBobs}
\end{equation} 
based on the cosmic microwave background (CMB) measurements, which agrees with the value extracted from the big bang nucleosynthesis (BBN) observations of  primordial elemental abundances~\cite{ParticleDataGroup:2024cfk}. The SM has, in principle, all the basic ingredients to satisfy the Sakharov conditions needed to generate a baryon asymmetry dynamically starting from a symmetric Universe~\cite{Sakharov:1967dj, Cline:2006ts}. However, the CP violation in the quark sector of the SM falls short by almost 10 orders of magnitude~\cite{Shaposhnikov:1987pf}, and moreover, the electroweak phase transition in the SM is not strongly first-order enough to produce the observed BAU~\cite{Kajantie:1996mn}.   

The lack of explanations in the SM to the observed DM and BAU has led to several beyond-the-standard model (BSM) proposals in the literature. Among them, the weakly-interacting-massive-particle (WIMP)~\cite{Kolb:1990vq, Jungman:1995df, Bertone:2004pz} paradigm for DM and baryogenesis~\cite{Weinberg:1979bt, Kolb:1979qa} or leptogenesis~\cite{Fukugita:1986hr} for BAU have been the most widely studied ones. In a typical WIMP framework, a particle having mass and interactions around the electroweak scale ballpark gives rise to the observed DM relic after freezing out from the thermal bath, a coincidence often referred to as the {\it WIMP Miracle}. Generic baryogenesis models, on the other hand, involve out-of-equilibrium decay of heavy new particles. In leptogenesis scenarios, a non-zero lepton asymmetry is first generated which later gets converted into the BAU via electroweak sphalerons~\cite{Kuzmin:1985mm}. An appealing feature of typical leptogenesis scenarios is their intimate connection to the seesaw mechanism~\cite{Minkowski:1977sc, Mohapatra:1979ia, Yanagida:1979as, GellMann:1980vs} for explaining the origin of neutrino mass and mixing, another experimental observation~\cite{ParticleDataGroup:2024cfk} which the SM fails to address.

While the BSM frameworks like WIMP and leptogenesis respectively can explain the observed DM and BAU independently, the striking similarity in the abundance of DM and baryons namely, $\Omega_{\rm DM} \approx 5\,\Omega_{B}$ may deserve further explanation and scrutiny. In the absence of any numerical coincidence, it is possible to explain similar order of magnitude abundances of DM and baryons by uniting their production mechanisms~\cite{Boucenna:2013wba}. While these common origin or cogenesis mechanisms are broadly classified into two categories namely, asymmetric dark matter (ADM)~\cite{Nussinov:1985xr, Davoudiasl:2012uw, Petraki:2013wwa, Zurek:2013wia,DuttaBanik:2020vfr, Barman:2021ost, Cui:2020dly, Borah:2024wos} and baryogenesis from annihilation of particles~\cite{Yoshimura:1978ex, Barr:1979wb, Baldes:2014gca, Gu:2009yx, Bhattacharya:2023kws} including DM annihilation~\cite{Chu:2021qwk, Cui:2011ab, Bernal:2012gv, Bernal:2013bga, Kumar:2013uca, Racker:2014uga, Dasgupta:2016odo, Borah:2018uci, Borah:2019epq, Dasgupta:2019lha, Mahanta:2022gsi}, there also exist other alternatives like Affleck-Dine~\cite{Affleck:1984fy} cogenesis~\cite{Roszkowski:2006kw, Seto:2007ym, Cheung:2011if, vonHarling:2012yn, Borah:2022qln, Borah:2023qag}, first-order phase transition (FOPT) related mass-gain mechanism~\cite{Huang:2022vkf,Dasgupta:2022isg, Borah:2022cdx, Borah:2023saq, Chun:2023ezg}, phase-separation mechanism~\cite{Arakawa:2024bkv}, bubble-filtering~\cite{Borah:2025wzl}, forbidden decay of DM~\cite{Borah:2023god} and conversion driven leptogenesis \cite{Heisig:2024mwr}. In the ADM scenario, the out-of-equilibrium decay of the same heavy particle is responsible for generating similar asymmetries in visible and dark sectors: $n_B-n_{\overline{B}} \sim \lvert n_{\rm DM}-n_{\overline{ \rm DM}} \rvert$. On the other hand, the other major class of cogenesis mechanism involving asymmetry from annihilations~\cite{Yoshimura:1978ex, Barr:1979wb, Baldes:2014gca}, one or more particles involved in the process eventually go out of thermal equilibrium to generate a net asymmetry\footnote{See Ref.~\cite{Chu:2021qwk} for a hybrid scenario where DM annihilates into metastable dark partners whose late decay produces baryon asymmetry.}. The so-called WIMPy baryogenesis~\cite{Cui:2011ab, Bernal:2012gv, Bernal:2013bga} belongs to this category, where a DM particle freezes out to generate its own relic abundance while simultaneously producing an asymmetry in the baryon sector. This idea has also been extended to leptogenesis, known as the WIMPy leptogenesis scenario~\cite{Kumar:2013uca, Racker:2014uga, Dasgupta:2016odo, Borah:2018uci, Borah:2019epq, Dasgupta:2019lha, Mahanta:2022gsi}.

In this work, we consider a WIMPy leptogenesis-inspired minimal scenario where lepton asymmetry is dominantly generated from coannihilation of DM and its heavier partners. While the impact of such coannihilations on DM abundance has been known for a long time~\cite{Griest:1990kh,Edsjo:1997bg}, the impact  on the production of baryon asymmetry via leptogenesis has not been studied before and this is our main focus here. We consider a minimal extension of the type-I seesaw model of neutrino mass~\cite{Minkowski:1977sc, GellMann:1980vs, Mohapatra:1979ia, Yanagida:1979as} by two $Z_2$-odd particles whose coannihilations provide the dominant source of lepton asymmetry. Another $Z_2$-even singlet scalar is introduced to ensure DM freeze-out after the freeze-out of the process responsible for generating lepton asymmetry. We solve the coupled Boltzmann equations for lepton asymmetry as well as abundances of all relevant particles to show that successful leptogenesis can take place at a low scale for DM mass $\sim \mathcal{O}(1)$ TeV. While the minimal model is also consistent with light neutrino data with a vanishing lightest neutrino mass, the TeV scale particle spectrum keeps the model testable at terrestrial experiments.

This paper is organised as follows. In Section~\ref{sec1}, we discuss our model. The details of DM and leptogenesis are discussed in Section~\ref{sec2}. This is followed by discussion of our numerical results in Section~\ref{sec3}. In Section~\ref{sec2a}, we briefly discuss the detection prospects of our model. Our conclusions are given in Section~\ref{sec4}. In Appendix~\ref{Appen2}, we give the relevant decay widths, while the calculation details of the CP-asymmetry are given in Appendix~\ref{Appen1}.

\section{The Model}
\label{sec1}
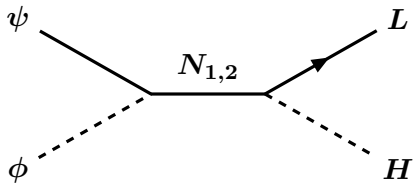
\begin{figure}
\begin{center}
\begin{tikzpicture}[scale=0.5]
    \centering
   \begin{feynman}
    \vertex (i1) at (-3.5, 2) {\(\boldsymbol{\psi}\)};
    \vertex (i2) at (-3.5, -2) {\(\boldsymbol{\phi}\)};
    
    \vertex (v1) at (0, 0);  
    \vertex (v2) at (3, 0); 

    \vertex (o1) at (6.5, 2) {\(\boldsymbol{L}\)};
    \vertex (o2) at (6.5, -2) {\(\boldsymbol{H}\)};
    
    \diagram* {
      (i1) -- [plain, very thick] (v1) -- [plain, very thick, edge label=\(\boldsymbol{N_{1,2}}\)] (v2) -- [fermion, very thick] (o1),
      (i2) -- [scalar,very thick] (v1),
      (v2) -- [scalar, very thick] (o2),
    };    
  \end{feynman}
\end{tikzpicture}
\end{center}
 \caption{DM coannihilation process responsible for creating the lepton asymmetry.}
    \label{fig1}
\end{figure}

We consider a minimal extension of the SM with three singlet fermions $N_{1,2}, \psi$ and two singlet scalars $\phi, \eta$. The new field content and their quantum numbers under the SM gauge symmetry augmented by a discrete $Z_2$ symmetry are shown in Table~\ref{tab1}. As we discuss below, $N_{1,2}$ generate light neutrino masses at tree-level while also playing the role of mediators in coannihilation of dark sector particles $\phi, \psi$; see Fig.~\ref{fig1}. The role of $Z_2$-even singlet scalar $\eta$ is to ensure the correct relic of the DM particle  $\phi$, details of which are given in the next section.

\begin{table}[h!]
		\small
		\begin{center}
			\begin{tabular}{||@{\hspace{0cm}}c@{\hspace{0cm}}|@{\hspace{0cm}}c@{\hspace{0cm}}|@{\hspace{0cm}}c@{\hspace{0cm}}|@{\hspace{0cm}}c@{\hspace{0cm}}||}
				\hline
				\hline
				\begin{tabular}{c}
					{\bf ~~~~ Symmetry~~~~}\\
					{\bf ~~~~Group~~~~}\\ 
					\hline
					$SU(3)$ \\
					\hline 
					$SU(2)_{L}$\\ 
					\hline
					$U(1)_{Y}$\\ 
					\hline
					$Z_2$\\ 
				
				\end{tabular}
				&
				&
				\begin{tabular}{c|c|c}
					\multicolumn{2}{c}{\bf Fermion Fields}\\
					\hline
					~~~$N_{1,2}$ & $\psi$ \\
					\hline
					$1$ & $1$\\
					\hline 
					$1$ & $1$\\
					\hline
					$0$ & $0$ \\
					\hline
					$+1$ & $-1$  \\
				\end{tabular}
				&
				\begin{tabular}{c|c|c}
					\multicolumn{2}{c}{\bf Scalar Field}\\
					\hline
					~~~$\phi$ & $\eta$ \\
					\hline
					$1$ & $1$\\
					\hline
					$1$ & $1$\\
					\hline
					$0$ & $0$\\
					\hline
					$-1$ & $1$ \\
		
				\end{tabular}\\
				\hline
				\hline
			\end{tabular}
			\caption{BSM fields,  along with their
				transformations under the SM gauge group and under a discrete $Z_2$ symmetry.}
			\label{tab1}
		\end{center}    
	\end{table}

The relevant fermionic Lagrangian can be written as
\begin{align}
   -\mathcal{L} \supset h_{\alpha i} \overline{L_\alpha} \tilde{H} N_i + y_{i} \overline{\psi^c} N_i \phi + \frac{m_{\psi}}{2} \overline{\psi^c} \psi + \frac{M_i}{2} \overline{N^c_i} N_i + {\rm h.c.},
\end{align}
where $i=1,2$. The unbroken $Z_2$ symmetry prevents $Z_2$-odd scalar $\phi$ from acquiring non-zero vacuum expectation value (VEV). While $\eta$ can acquire a non-zero VEV, we consider it to be negligible for simplicity. The doublet scalar field $H$ is parameterized as
\begin{align}
    H=\frac{1}{\sqrt{2}} \begin{pmatrix}
        0 \\
        h +v
    \end{pmatrix},
\end{align}
with $v=246$ GeV being the electroweak VEV. After electroweak symmetry breaking, neutrinos acquire a Dirac mass term $(M_D)_{\alpha i}= \frac{1}{\sqrt{2}} v h_{\alpha i}$. The type-I seesaw contribution to light neutrino mass in the seesaw limit $M_D \ll M_{i}$ is 
\begin{equation}
    M_\nu \simeq -M_D M^{-1} M^T_D\, , \quad {\rm or},\quad (M_\nu)_{\alpha\beta} \simeq -\frac{v^2}{2} \frac{h_{\alpha i} h_{\beta i}}{M_i}.
    \end{equation}
In order to incorporate the constraints from light neutrino masses, we use the Casas-Ibarra (CI) parametrisation~\cite{Casas:2001sr} for type-I seesaw given by
\begin{equation}
    h_{\alpha i} =\dfrac{1}{v} (U D_{\sqrt{M_{\nu}}} R^{\dagger} D_{\sqrt{M}}),
    \label{eq:CI}
\end{equation}
where $R$ is an arbitrary complex orthogonal matrix satisfying $RR^{T}=\mathbbm{1}$ and $U$ is the usual Pontecorvo-Maki-Nakagawa-Sakata (PMNS) mixing matrix which diagonalises the light neutrino mass matrix in a basis where charged lepton mass matrix is diagonal. For two right-handed neutrinos the diagonal matrix $D_{\sqrt{M}}$ is given by $D_{\sqrt{M}}={\rm diag}(\sqrt{M_{1}},\sqrt{M_{2}})$. The equivalent diagonal matrix for light neutrinos is $D_{\sqrt{M_{\nu}}}={\rm diag}(m_1, m_2, m_3)$. Since there are only two RHNs, the lightest neutrino remain massless, which is perfectly consistent with the current data. The $R$ matrix for 2 heavy neutrino scenario is given by~\cite{Ibarra:2003up}
\begin{equation}
    R = \begin{pmatrix}
        0 & \cos{z_1} & \pm \sin{z_1} \\
        0 & -\sin{z_1} & \pm \cos{z_1}
    \end{pmatrix}
\end{equation}
where $z_1=a+ib$ is a complex angle. The diagonal light neutrino mass matrix can be written as 
\begin{align}
    D_{\sqrt{M_\nu}} = \left\{ \begin{array}{cc}
    {\rm diag}\left(
    0,\sqrt{\Delta m^2_{\rm sol}}\, , \sqrt{\Delta m^2_{\rm sol}+\Delta m^2_{\rm atm}}\right) & {\rm (NH)} \\
    {\rm diag}\left(
    \sqrt{\Delta m^2_{\rm atm}-\Delta m^2_{\rm sol}}\, ,\sqrt{\Delta m^2_{\rm atm}}\, , 0\right) & {\rm (IH)}
    \end{array}
    \right.
\end{align}
for normal hierarchy (NH) and inverted hierarchy, respectively.  
The PMNS mixing parameters and mass squared differences are taken from Ref.~\cite{ParticleDataGroup:2024cfk}.


The $Z_2$-even singlet scalar $\eta$ couples to the $Z_2$-odd particles via the interactions 
\begin{equation}
    -\mathcal{L}_{\rm new} =y_\eta \eta \overline{\psi^c}\psi + \mu_{\eta\phi} \phi^2 \eta + \lambda_{\eta \phi} \eta^2 \phi^2.
\end{equation}
As we will discuss later, these interactions play a crucial role in generating the correct DM relic after thermal freeze-out. While one can also have similar couplings of $\eta$ to RHN $N_i$, we ignore them for simplicity. The complete scalar potential $V(H,\eta,\phi)$ can be written as
\begin{eqnarray}
   \nonumber
    V(H,\eta, \phi) & = & -\mu_{H}^{2}(H^{\dagger} H) + \frac{\lambda_{H}}{2}(H^{\dagger} H)^{2} + \frac{1}{2}m_{\phi}^{2}\phi^2 + \frac{\lambda_{\phi}}{2} \phi^4 + \frac{1}{2}\mu_{\eta}^{2}\eta^2 +\mu_{\eta\phi} \phi^2 \eta
    \nonumber \\ && 
+\: \frac{\lambda_{\eta}}{2} \eta^4 + \lambda_{H\phi}(H^{\dagger} H)\phi^2 +  \lambda_{H\eta} (H^\dagger H)\eta^2 +\lambda_{\eta\phi}\eta^2 \phi^2+\mu \eta (H^\dagger H).
\label{scalarpot}
\end{eqnarray}
Even if $\mu^2_\eta >0$, the scalar $\eta$ can acquire an induced VEV after electroweak symmetry breaking by virtue of the last term in Eq.~\eqref{scalarpot}. While a small $\mu$ term ensures a tiny VEV without any consequences to the discussion of leptogenesis and DM, it ensures that $\eta$ eventually decays into the SM particles due to its mixing with the SM Higgs boson.


\begin{figure}
\centering
\begin{tikzpicture}[scale=0.5]
  \begin{feynman}
    \vertex (i1) at (-3, 1.5) {\(\boldsymbol{\phi}\)};
    \vertex (i2) at (-3, -1.5) {\(\boldsymbol{\phi}\)};
    
    \vertex (v1) at (0, 1.5);  
    \vertex (v2) at (0, -1.5); 
    \vertex (c) at (0, 1.5);    
    
    \vertex (o1) at (3, 1.5) {\(\boldsymbol{\eta}\)};
    \vertex (o2) at (3, -1.5) {\(\boldsymbol{\eta}\)};
    
    \diagram* {
      (i1) -- [scalar, very thick] (v1) -- [scalar, very thick] (c),
      (i2) -- [scalar, very thick] (v2) -- [scalar, very thick,edge label'=\(\boldsymbol{\phi}\)] (c),
      (c) -- [scalar,very thick,edge label'=\(\boldsymbol{\eta}\)] (c),
      (c) -- [scalar,very thick] (v1) -- [scalar, very thick] (o1),
      (c) -- [scalar, very thick] (v2) -- [scalar, very thick] (o2),
    };
  \end{feynman}
\end{tikzpicture}
\begin{tikzpicture}
      \begin{feynman}
    \vertex (i1) at (-1.5, 0.9) {\(\boldsymbol{\phi}\)};
    \vertex (i2) at (-1.5, -0.9) {\(\boldsymbol{\phi}\)};
    
    \vertex (v1) at (0,0);  
    \vertex (v2) at (0, 0); 
    \vertex (c) at (0, 0);    
    
    \vertex (o1) at (1.5, 0.9) {\(\boldsymbol{\eta}\)};
    \vertex (o2) at (1.5, -0.9) {\(\boldsymbol{\eta}\)};
    
    \diagram* {
      (i1) -- [scalar, very thick] (v1) -- [scalar, very thick] (c),
      (i2) -- [scalar, very thick] (v2) -- [scalar, very thick,edge ] (c),
      (c) -- [scalar,very thick,edge ] (c),
      (c) -- [scalar,very thick] (v1) -- [scalar, very thick] (o1),
      (c) -- [scalar, very thick] (v2) -- [scalar, very thick] (o2),
    };
  \end{feynman}
\end{tikzpicture}
\begin{tikzpicture}[scale=0.5]
  \begin{feynman}
    \vertex (i1) at (-3, 1.5) {\(\boldsymbol{\psi}\)};
    \vertex (i2) at (-3, -1.5) {\(\boldsymbol{\psi}\)};
    
    \vertex (v1) at (0, 1.5);  
    \vertex (v2) at (0, -1.5); 
    \vertex (c) at (0, 1.5);    
    
    \vertex (o1) at (3, 1.5) {\(\boldsymbol{\eta}\)};
    \vertex (o2) at (3, -1.5) {\(\boldsymbol{\eta}\)};
    
    \diagram* {
      (i1) -- [plain, very thick] (v1) -- [scalar, very thick] (c),
      (i2) -- [plain, very thick] (v2) -- [plain, very thick,edge label'=\(\boldsymbol{\psi}\)] (c),
      (c) -- [plain,very thick,edge label'=\(\boldsymbol{\psi}\)] (c),
      (c) -- [plain,very thick] (v1) -- [scalar, very thick] (o1),
      (c) -- [plain, very thick] (v2) -- [scalar, very thick] (o2),
    };
  \end{feynman}
\end{tikzpicture}
\caption{Feynman diagrams for self-annihilation of $\phi$ and $\psi$ to $\eta$.}
\label{feyn1}
\end{figure}
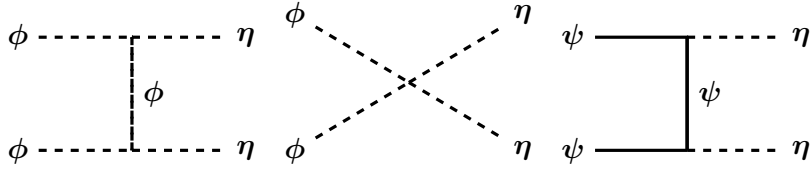

\begin{figure}
\centering
\begin{tikzpicture}[scale=0.5]
  \begin{feynman}
    \vertex (i1) at (-3, 1.5) {\(\boldsymbol{N_1}\)};
    \vertex (i2) at (-3, -1.5) {\(\boldsymbol{N_1}\)};
    
    \vertex (v1) at (0, 1.5);  
    \vertex (v2) at (0, -1.5); 
    \vertex (c) at (0, 1.5);    
    
    \vertex (o1) at (3, 1.5) {\(\boldsymbol{\phi}\)};
    \vertex (o2) at (3, -1.5) {\(\boldsymbol{\phi}\)};
    
    \diagram* {
      (i1) -- [plain, very thick] (v1) -- [scalar, very thick] (c),
      (i2) -- [plain, very thick] (v2) -- [plain, very thick,edge label'=\(\boldsymbol{\psi}\)] (c),
      (c) -- [plain,very thick,edge label'=\(\boldsymbol{\eta}\)] (c),
      (c) -- [plain,very thick] (v1) -- [scalar, very thick] (o1),
      (c) -- [plain, very thick] (v2) -- [scalar, very thick] (o2),
    };
  \end{feynman}
\end{tikzpicture}
\begin{tikzpicture}[scale=0.5]
  \begin{feynman}
    \vertex (i1) at (-3, 1.5) {\(\boldsymbol{N_1}\)};
    \vertex (i2) at (-3, -1.5) {\(\boldsymbol{N_1}\)};
    
    \vertex (v1) at (0, 1.5);  
    \vertex (v2) at (0, -1.5); 
    \vertex (c) at (0, 1.5);    
    
    \vertex (o1) at (3, 1.5) {\(\boldsymbol{\psi}\)};
    \vertex (o2) at (3, -1.5) {\(\boldsymbol{\psi}\)};
    
    \diagram* {
      (i1) -- [plain, very thick] (v1) -- [scalar, very thick] (c),
      (i2) -- [plain, very thick] (v2) -- [scalar, very thick,edge label'=\(\boldsymbol{\phi}\)] (c),
      (c) -- [scalar,very thick,edge label'=\(\boldsymbol{\eta}\)] (c),
      (c) -- [scalar,very thick] (v1) -- [plain, very thick] (o1),
      (c) -- [scalar, very thick] (v2) -- [plain, very thick] (o2),
    };
  \end{feynman}
\end{tikzpicture}
\caption{Feynman diagrams for annihilation of RHNs into $\phi, \psi$.}
\label{feyn2}
\end{figure}
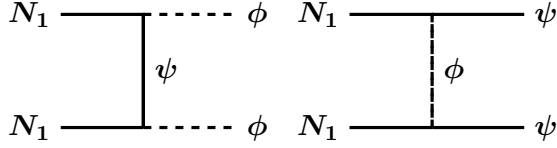

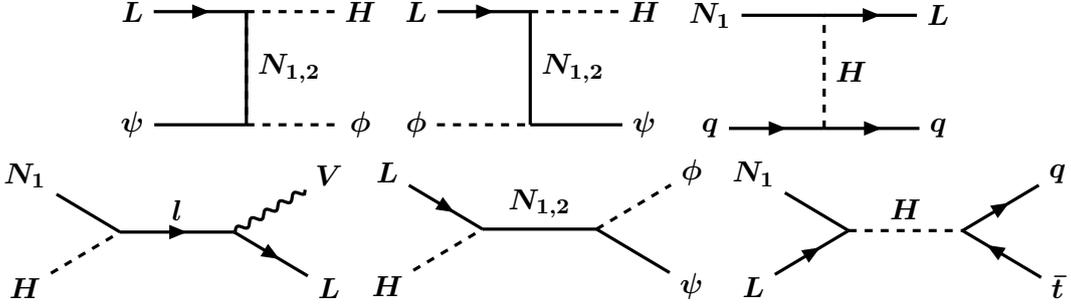
\begin{figure}
\centering
\begin{tikzpicture}[scale=0.5]
  \begin{feynman}
    \vertex (i1) at (-3, 1.5) {\(\boldsymbol{L}\)};
    \vertex (i2) at (-3, -1.5) {\(\boldsymbol{\psi}\)};
    
    \vertex (v1) at (0, 1.5);  
    \vertex (v2) at (0, -1.5); 
    \vertex (c) at (0, 1.5);    
    
    \vertex (o1) at (3, 1.5) {\(\boldsymbol{H}\)};
    \vertex (o2) at (3, -1.5) {\(\boldsymbol{\phi}\)};
    
    \diagram* {
      (i1) -- [fermion, very thick] (v1) -- [plain, very thick] (c),
      (i2) -- [plain, very thick] (v2) -- [plain, very thick,edge label'=\(\boldsymbol{N_{1,2}}\)] (c),
      (c) -- [scalar,very thick] (v1) -- [scalar, very thick] (o1),
      (c) -- [scalar, very thick] (v2) -- [scalar, very thick] (o2),
    };
  \end{feynman}
\end{tikzpicture}
\begin{tikzpicture}[scale=0.5]
  \begin{feynman}
    \vertex (i1) at (-3, 1.5) {\(\boldsymbol{L}\)};
    \vertex (i2) at (-3, -1.5) {\(\boldsymbol{\phi}\)};
    
    \vertex (v1) at (0, 1.5);  
    \vertex (v2) at (0, -1.5); 
    \vertex (c) at (0, 1.5);    
    
    \vertex (o1) at (3, 1.5) {\(\boldsymbol{H}\)};
    \vertex (o2) at (3, -1.5) {\(\boldsymbol{\psi}\)};
    
    \diagram* {
      (i1) -- [fermion, very thick] (v1) -- [scalar, very thick] (c),
      (i2) -- [scalar, very thick] (v2) -- [plain, very thick,edge label'=\(\boldsymbol{N_{1,2}}\)] (c),
       (v1) -- [scalar, very thick] (o1),
       (v2) -- [plain, very thick] (o2),
    };
  \end{feynman}
\end{tikzpicture}
\begin{tikzpicture}[scale=0.5]
  \begin{feynman}
    \vertex (i1) at (-3, 1.5) {\(\boldsymbol{N_{1}}\)};
    \vertex (i2) at (-3, -1.5) {\(\boldsymbol{q}\)};
    
    \vertex (v1) at (0, 1.5);  
    \vertex (v2) at (0, -1.5); 
    \vertex (c) at (0, 1.5);    
    
    \vertex (o1) at (3, 1.5) {\(\boldsymbol{L}\)};
    \vertex (o2) at (3, -1.5) {\(\boldsymbol{q}\)};
    
    \diagram* {
      (i1) -- [plain, very thick] (v1) -- [plain, very thick] (c),
      (i2) -- [fermion, very thick] (v2) -- [scalar, very thick,edge label'=\(\boldsymbol{H}\)] (c),
       (v1) -- [fermion, very thick] (o1),
       (v2) -- [fermion, very thick] (o2),
    };
  \end{feynman}
\end{tikzpicture}

\begin{tikzpicture}[scale=0.5]
  \begin{feynman}
    \vertex (i1) at (-2.5, 1.5) {\(\boldsymbol{N_1}\)};
    \vertex (i2) at (-2.5, -1.5) {\(\boldsymbol{H}\)};
    
    \vertex (v1) at (0, 0);  
    \vertex (v2) at (3, 0); 

    \vertex (o1) at (5.5, 1.5) {\(\boldsymbol{V}\)};
    \vertex (o2) at (5.5, -1.5) {\(\boldsymbol{L}\)};
    
    \diagram* {
      (i1) -- [plain, very thick] (v1) -- [fermion, very thick, edge label=\(\boldsymbol{l}\)] (v2) -- [photon, very thick] (o1),
      (i2) -- [scalar, very thick] (v1),
      (v2) -- [fermion, very thick] (o2),
    };    
\end{feynman}
\end{tikzpicture}
\begin{tikzpicture}[scale=0.5]
  \begin{feynman}
    \vertex (i1) at (-2.5, 1.5) {\(\boldsymbol{L}\)};
    \vertex (i2) at (-2.5, -1.5) {\(\boldsymbol{H}\)};
    
    \vertex (v1) at (0, 0);  
    \vertex (v2) at (3, 0); 

    \vertex (o1) at (5.5, 1.5) {\(\boldsymbol{\phi}\)};
    \vertex (o2) at (5.5, -1.5) {\(\boldsymbol{\psi}\)};
    
    \diagram* {
      (i1) -- [fermion, very thick] (v1) -- [plain, very thick, edge label=\(\boldsymbol{N_{1,2}}\)] (v2) -- [scalar, very thick] (o1),
      (i2) -- [scalar, very thick] (v1),
      (v2) -- [plain, very thick] (o2),
    };
    
  \end{feynman}
\end{tikzpicture}
\begin{tikzpicture}[scale=0.5]
  \begin{feynman}
    \vertex (i1) at (-2.5, 1.5) {\(\boldsymbol{N_1}\)};
    \vertex (i2) at (-2.5, -1.5) {\(\boldsymbol{L}\)};
    
    \vertex (v1) at (0, 0);  
    \vertex (v2) at (3, 0); 

    \vertex (o1) at (5.5, 1.5) {\(\boldsymbol{q}\)};
    \vertex (o2) at (5.5, -1.5) {\(\boldsymbol{\bar{t}}\)};
    
    \diagram* {
      (i1) -- [plain, very thick] (v1) -- [scalar, very thick, edge label=\(\boldsymbol{ H}\)] (v2) -- [fermion, very thick] (o1),
      (i2) -- [fermion, very thick] (v1),
      (v2) -- [anti fermion, very thick] (o2),
    };    
\end{feynman}
\end{tikzpicture}
\caption{Feynman diagrams for washout scattering processes with $\Delta L=1$.}
\label{feyn3}
\end{figure}

\begin{figure}
\centering
\begin{tikzpicture}[scale=0.5]
  \begin{feynman}
    \vertex (i1) at (-2.5, 1.5) {\(\boldsymbol{\bar{L}}\)};
    \vertex (i2) at (-2.5, -1.5) {\(\boldsymbol{H}\)};
    
    \vertex (v1) at (0, 0);  
    \vertex (v2) at (3, 0); 

    \vertex (o1) at (5.5, 1.5) {\(\boldsymbol{H^{\dagger}}\)};
    \vertex (o2) at (5.5, -1.5) {\(\boldsymbol{L}\)};
    
    \diagram* {
      (i1) -- [anti fermion, very thick] (v1) -- [ plain, very thick, edge label=\(\boldsymbol{N_{1,2}}\)] (v2) -- [scalar, very thick] (o1),
      (i2) -- [scalar, very thick] (v1),
      (v2) -- [ fermion, very thick] (o2),
    };
    
  \end{feynman}
\end{tikzpicture}
\begin{tikzpicture}[scale=0.5]
  \begin{feynman}
    \vertex (i1) at (-3, 1.5) {\(\boldsymbol{\bar{L}}\)};
    \vertex (i2) at (-3, -1.5) {\(\boldsymbol{H}\)};
    
    \vertex (v1) at (0, 1.5);  
    \vertex (v2) at (0, -1.5); 
    \vertex (c) at (0, 1.5);    
    
    \vertex (o1) at (3, 1.5) {\(\boldsymbol{H^{\dagger}}\)};
    \vertex (o2) at (3, -1.5) {\(\boldsymbol{L}\)};
    
    \diagram* {
      (i1) -- [anti fermion, very thick] (v1) -- [scalar, very thick] (c),
      (i2) -- [scalar, very thick] (v2) -- [plain, very thick,edge label'=\(\boldsymbol{N_{1,2}}\)] (c),
      (c) -- [scalar,very thick] (v1) -- [scalar, very thick] (o1),
      (c) -- [scalar, very thick] (v2) -- [fermion, very thick] (o2),
    };
  \end{feynman}
\end{tikzpicture}
\begin{tikzpicture}[scale=0.5]
  \begin{feynman}
    \vertex (i1) at (-3, 1.5) {\(\boldsymbol{H}\)};
    \vertex (i2) at (-3, -1.5) {\(\boldsymbol{H}\)};
    
    \vertex (v1) at (0, 1.5);  
    \vertex (v2) at (0, -1.5); 
    \vertex (c) at (0, 1.5);    
    
    \vertex (o1) at (3, 1.5) {\(\boldsymbol{L}\)};
    \vertex (o2) at (3, -1.5) {\(\boldsymbol{L}\)};
    
    \diagram* {
      (i1) -- [scalar, very thick] (v1) -- [scalar, very thick] (c),
      (i2) -- [scalar, very thick] (v2) -- [plain, very thick,edge label'=\(\boldsymbol{N_{1,2}}\)] (c),
      (c) -- [plain,very thick,edge label'=\(\boldsymbol{\psi}\)] (c),
      (c) -- [scalar,very thick] (v1) -- [fermion, very thick] (o1),
      (c) -- [scalar, very thick] (v2) -- [fermion, very thick] (o2),
    };
  \end{feynman}
\end{tikzpicture}
\caption{Feynman diagrams for washout scattering processes with $\Delta L=2$.}
\label{feyn4}
\end{figure}
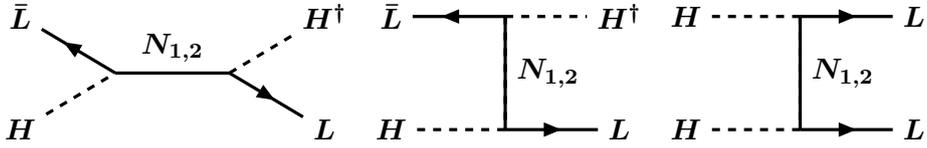

\section{Leptogenesis and Dark Matter}
\label{sec2}

There are two sources of CP violation in this model which can generate a non-zero lepton asymmetry. The first one is the decay of the lightest right-handed neutrino $N_{1}$, similar to leptogenesis in minimal type-I seesaw scenarios. The second one is the coannihilation of the dark sector particles $\phi$ and $\psi$, as shown in Fig.~\ref{fig1}. For $Z_2$-odd particles lighter than $N_1$, asymmetry is dominantly generated by coannihilation at a lower scale compared to $M_1$. The lighter of the $Z_2$-odd particles gives rise to DM relic after thermal freeze-out. Without any loss of generality, we consider scalar DM scenario here such that $m_{\phi} < m_{\psi}$. Although we show only the tree level diagrams in Fig.~\ref{fig1}, the propagators are resummed taking the radiative corrections into account, similar to Ref.~\cite{Borah:2020ivi} where three-body decay involving DM final state was considered instead of DM annihilation to source the lepton asymmetry. Additionally, we also have vertex corrections. In order to enhance the CP asymmetry, we consider the masses of RHNs $M_{1,2}$ to be of the same order of magnitude, although not necessarily quasi-degenerate as in resonant leptogenesis~\cite{Pilaftsis:2003gt, Dev:2017wwc} which keeps the vertex corrections contribution sub-leading compared to the self-energy corrections. The Feynman diagrams of other $2\to 2$ processes relevant for DM relic and lepton asymmetry are shown in Figs.~\ref{feyn1},~\ref{feyn2},~\ref{feyn3},~\ref{feyn4}. In addition to DM coannihilation process shown in Fig.~\ref{fig1}, we also have annihilation processes shown in Fig.~\ref{feyn1}. As pointed out in the original WIMPy baryogenesis work~\cite{Cui:2011ab}, it is crucial to ensure that the lepton number violating coannihilation processes freeze out before the freeze-out of DM with the latter being decided by the self-annihilation processes. While DM $\phi$ can freeze-out at later epochs due to SM Higgs-portal interactions too, this keeps $\phi$ mass constrained to the Higgs resonance $m_\phi \approx 62.5$ GeV~\cite{GAMBIT:2017gge} from the requirements of satisfying relic and direct detection constraints. Since we require DM mass to be a bit heavier for generating the required lepton asymmetry before the sphaleron decoupling epoch $T_{\rm sph} \sim 131.7$ GeV~\cite{DOnofrio:2014rug}, we consider the additional singlet scalar $\eta$ such that $\phi \phi \rightarrow \eta \eta$ can set the correct DM relic at a temperature below the freeze-out of DM coannihilation. As we discuss later, the scalar portal interactions of DM also keep its detection prospects alive at direct and indirect detection experiments.

The relevant decay widths and the details of the CP asymmetries required for leptogenesis are given in Appendices~\ref{Appen2} and~\ref{Appen1},  respectively. Unlike type-I seesaw leptogenesis where the parameters are tightly constrained from light neutrino data, which results in the Davidson-Ibarra bound on $M_1\gtrsim 10^9$ GeV~\cite{Davidson:2002qv, Buchmuller:2002rq},  our setup has additional parameters like $y_i$ relating dark sector particles with others, due to which it is possible to enhance the CP asymmetry for lower values of $M_{1,2}$ while being consistent with light neutrino mass. It should, however, be noted that the same additional Yukawa interaction can also enhance the CP asymmetry from two-body decay of RHNs via self-energy corrections. However, we choose the masses of $\phi, \psi$ such that their contribution to such self-energy corrections is negligible at one-loop level.

In order to track the evolution of lepton asymmetry, we need to solve the relevant Boltzmann equations. We write down the Boltzmann equations for the lightest RHN $N_{1}$, $Z_2$-odd singlet fermion $\psi$, DM $\phi$ and $B-L$ as follows.

\begin{align}
   \frac{dY_{N_{1}}}{dz}   & = -\dfrac{1}{s z\textbf{H}(z)} \Bigg[ \left( \dfrac{Y_{N_{1}}}{Y_{N_{1}}^{\rm eq}}-1 \right)\gamma_{N_{1}\to L H}  + \left(\dfrac{Y_{N_{1}}}{Y_{N_{1}}^{\rm eq}}- \dfrac{Y_{\phi}Y_{\psi}}{Y_{N_{1}}^{\rm eq}} \right) \gamma_{N_{1}\to \phi \psi}   \nonumber  \\  
         \displaybreak
   &  +\left(  \frac{Y_{N_{1}}}{Y_{N_{1}}^{\rm eq}}-1  \right) \Bigg[2 \gamma_{N_{1} H \to L V_{\mu}}+ 2\gamma_{N_{1}V_{\mu}\to L H}+ 4 \gamma_{N_{1}q\to L \bar{t}}+2\gamma_{N_{1}L\to \bar{t}q} \Bigg] \nonumber \\ 
   &  +  \left( \left( \frac{Y_{N_{1}}}{Y_{N_{1}}^{\rm eq}}\right)^{2}- \left( \frac{Y_{\phi}}{Y_{\phi}^{\rm eq}} \right)^{2} \right)   \gamma_{N_{1}N_{1}\to \phi \phi} + \left( \left( \frac{Y_{N_{1}}}{Y_{N_{1}}^{\rm eq}}\right)^{2}- \left( \frac{Y_{\psi}}{Y_{\psi}^{\rm eq}} \right)^{2} \right) \gamma_{N_{1}N_{1}\to \psi \psi}\Bigg], \nonumber \\ 
     \frac{d Y_{\psi}}{dz} &  = \frac{1}{sz\textbf{H}(z)} \Bigg[ \left( \frac{Y_{N_{1}}}{Y_{N_{1}}}-\frac{Y_{\phi} Y_{\psi}}{Y_{N_{1}}^{\rm eq}}  \right) \gamma_{N_{1}\to \phi \psi}  -\left(  \frac{Y_{\psi}}{Y_{\psi}^{\rm eq}}-\frac{Y_{\phi}Y_{L}^{\rm eq}}{Y_{\psi}^{\rm eq}} \right) \gamma_{\psi \to \phi \nu} \Theta(z-z_{\rm sph})   \nonumber \\ 
     & - \left( \frac{Y_{\psi}}{Y_{\psi}^{\rm eq}}-\frac{Y_{\phi}Y_{L}^{\rm eq}Y_{H}^{\rm eq}}{Y_{\psi}^{\rm eq}}   \right) \gamma_{\psi \to L H \phi} -2 \left( \frac{Y_{\phi}Y_{\psi}}{Y_{\phi}^{\rm eq}Y_{\psi}^{\rm eq}}-1  \right) \gamma_{\phi \psi \to L H}  \nonumber   \\ 
     & -2 \left( \frac{Y_{\psi}}{Y_{\psi}^{\rm eq}}-\frac{Y_{\phi}}{Y_{\phi}^{\rm eq}} \right)  \gamma_{L \psi \to H \phi} +\left( \left(  \frac{Y_{N_{1}}}{Y_{N_{1}}^{\rm eq}}\right)^{2}- \left( \frac{Y_{\psi}}{Y_{\psi}^{\rm eq}} \right)^{2} \right) \gamma_{N_{1}N_{1}\to \psi \psi } \Bigg], \nonumber \\ 
    \dfrac{dY_{\phi}}{dz} & = \frac{1}{sz\textbf{H}(z)}  \Bigg [ \left(  \frac{Y_{N_1}}{Y_{N_{1}}^{\rm eq}}-\frac{Y_{\phi}Y_{\psi}}{Y_{N_{1}}^{\rm eq}} \right) \gamma_{N_{1}\to \phi, \psi} +\left( \frac{Y_{\psi}}{Y_{\psi}^{\rm eq}}-\frac{Y_{\phi}Y_{L}^{\rm eq}}{Y_{\psi}^{\rm eq}}\right) \gamma_{\psi \to \phi \nu} \Theta(z-z_{\rm sph}) \nonumber \\ 
    & +\left(  \frac{Y_{\psi}}{Y_{\psi}^{\rm eq}}-\frac{Y_{\phi}Y_{L}^{\rm eq}Y_{H}^{\rm eq}}{Y_{\psi}^{\rm eq}}   \right) \gamma_{\psi \to L H \phi}- 2 \left( \frac{Y_{\phi}Y_{\psi}}{Y_{\phi}^{\rm eq}Y_{\psi}^{\rm eq}}-1  \right) \gamma_{\phi \psi \to L H}  - \left(  \left( \frac{Y_{\phi}}{Y_{\phi}^{\rm eq}}  \right)^{2}  -1\right)  \nonumber \\ 
    & \times \gamma_{\phi \phi \to \eta \eta} + 2 \left( \frac{Y_{\psi}}{Y_{\psi}^{\rm eq}}-\frac{Y_{\phi}}{Y_{\phi}^{\rm eq}} \right) \gamma_{L\psi \to H \phi} +\left( \left(  \frac{Y_{N_{1}}}{Y_{N_{1}^{\rm eq}}} \right)^{2}-\left( \frac{Y_{\phi}}{Y_{\phi}^{\rm eq}} \right)^{2}  \right) \gamma_{N_{1}N_{1}\to \phi \phi} \Bigg], \nonumber \\
    \frac{dY_{B-L}}{dz} & =-\frac{1}{sz\textbf{H}(z)}\Bigg [ 
 \epsilon_{1} \left( \frac{Y_{N_{1}}}{Y_{N_{1}}^{\rm eq}}-1  \right)\gamma_{N_{1}\to L H}+\left( \frac{Y_{\phi}Y_{\psi}}{Y_{\phi}^{\rm eq}Y_{\psi}^{\rm eq}}-1 \right) \gamma_{\phi \psi \to L H}^{\delta} +\frac{Y_{B-L}}{2Y_{L}^{\rm eq}}\gamma_{N_{1}\to L H}  \nonumber  \\ 
 & + \dfrac{Y_{B-L}}{Y_{L}^{\rm eq}} \bigg( 2 \gamma_{L H^{\dagger}\to \bar{L}H} +\frac{Y_{N_{1,2}}}{Y_{N_{1,2}}^{\rm eq}} \gamma_{\bar{L} N_{1,2}\to \bar{q}t} + 2 \gamma_{\bar{L}q\to N_{1,2}t}+2 \gamma_{LL \to H^{\dagger}H^{\dagger}}+ \gamma_{\bar{L}V_{\mu}\to H N_{1,2}} \nonumber \\ 
 & + \gamma_{L H \to \phi \psi}+\gamma_{L \phi \to H \psi}\bigg) \Bigg ].
\end{align}
\begin{figure}[h]
    \centering
    \includegraphics[scale=0.49]{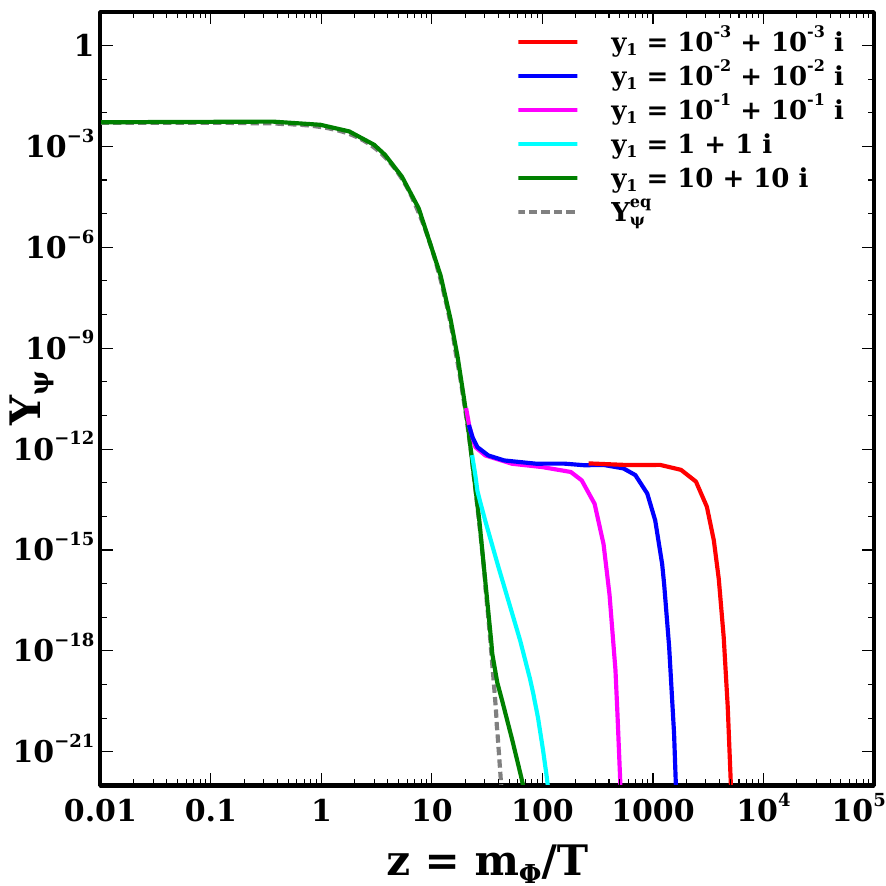}
    \includegraphics[scale=0.49]{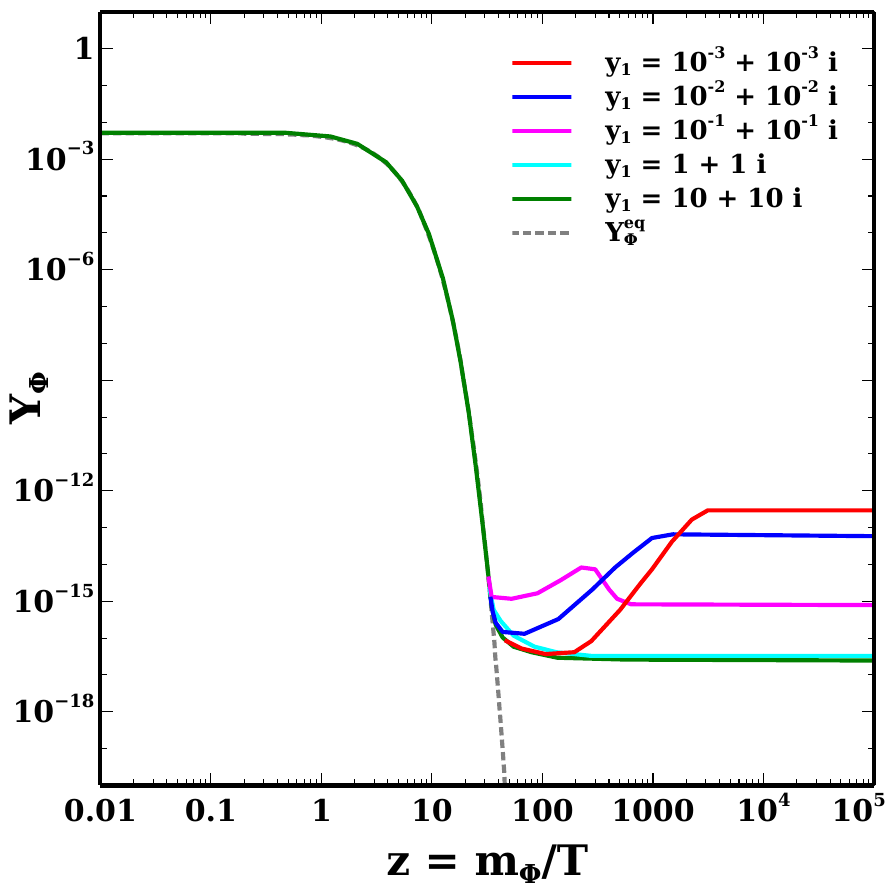}
    \includegraphics[scale=0.49]{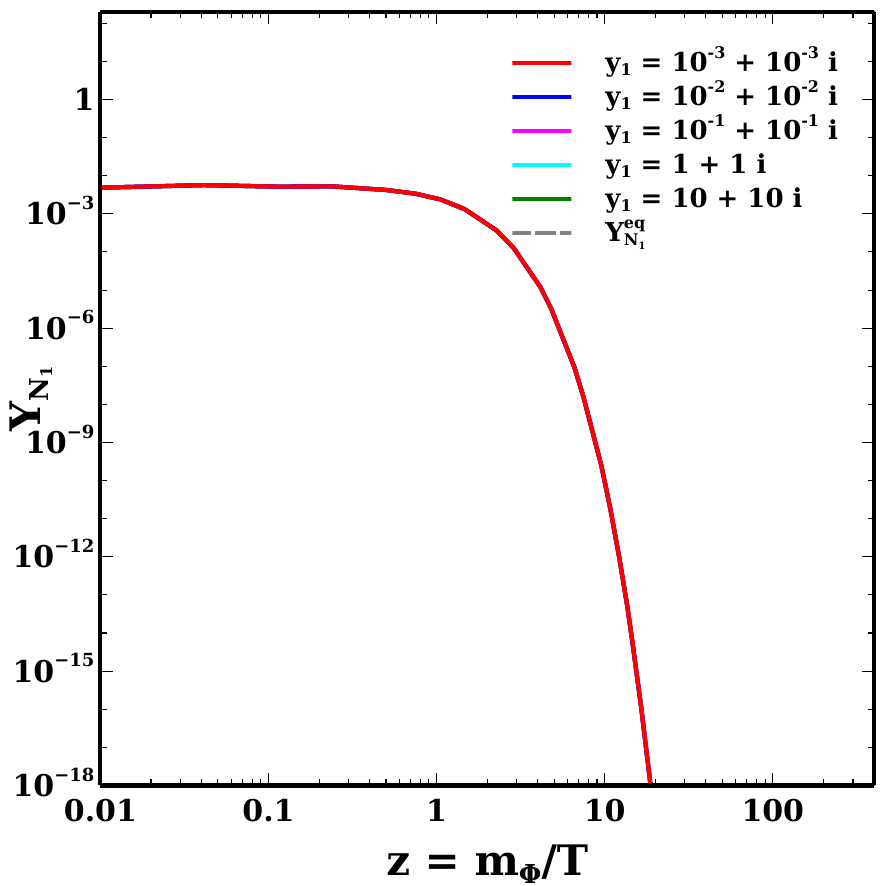}
    \includegraphics[scale=0.49]{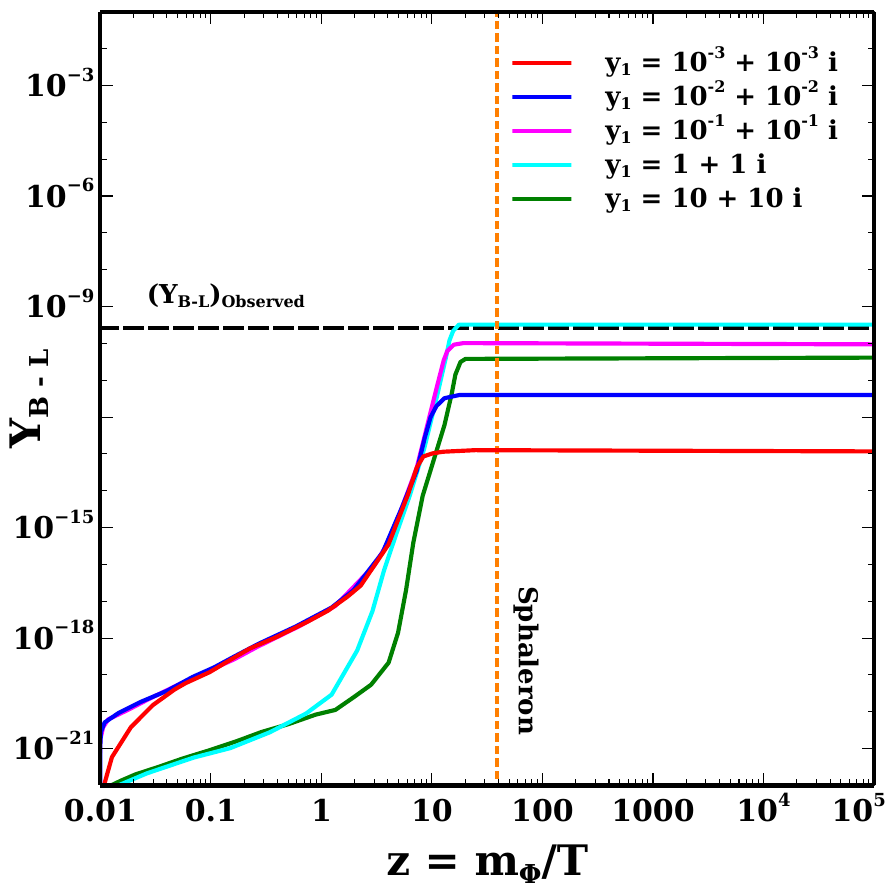}
    \caption{Evolution of comoving number densities of $\psi$ (top left), $\phi$ (top right),  $N_{1}$ (bottom left) and $B-L$ (bottom right) with $z=m_{\phi}/T$ for different values of Yukawa coupling $y_{1}=y_{2}$. The other relevant parameters are fixed at $m_{\phi}=5$ TeV, $m_{\psi}=6$ TeV, $M_{1}=11$ TeV, $M_{2}=M_{1}+100$ GeV, $y_\eta=0.1$ and $\lambda_{\eta \phi}=1$. In the bottom right plot, the horizontal line shows the value of $Y_{B-L}$ needed to reproduce the observed BAU [cf.~Eq.~\eqref{eq:3.4}]. The vertical line represents the sphaleron decoupling temperature below which a lepton asymmetry cannot be reprocessed into a baryon asymmetry.}
    \label{fig2}
\end{figure}
\begin{figure}
    \centering
     \includegraphics[scale=0.49]{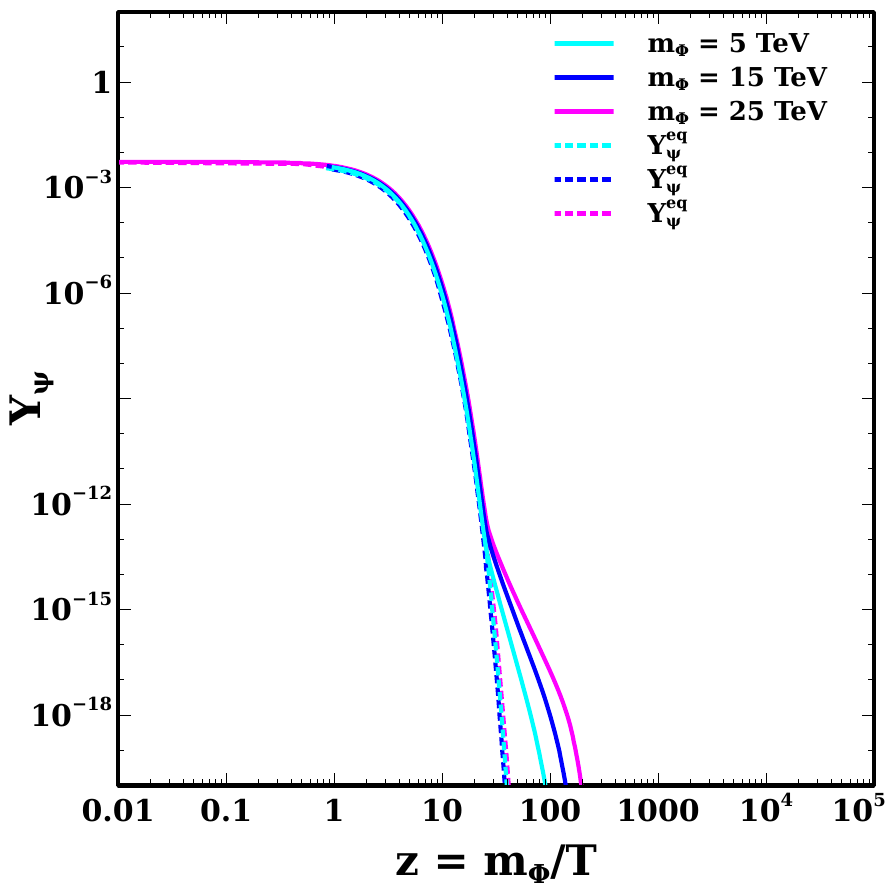}
    \includegraphics[scale=0.49]{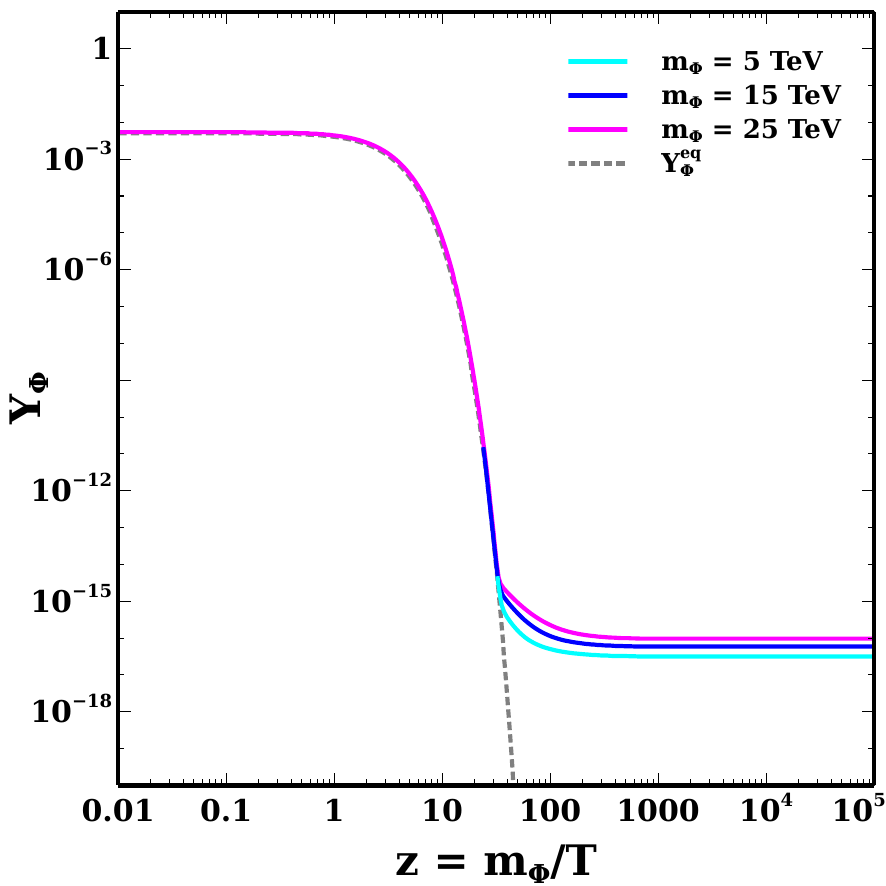}    \includegraphics[scale=0.49]{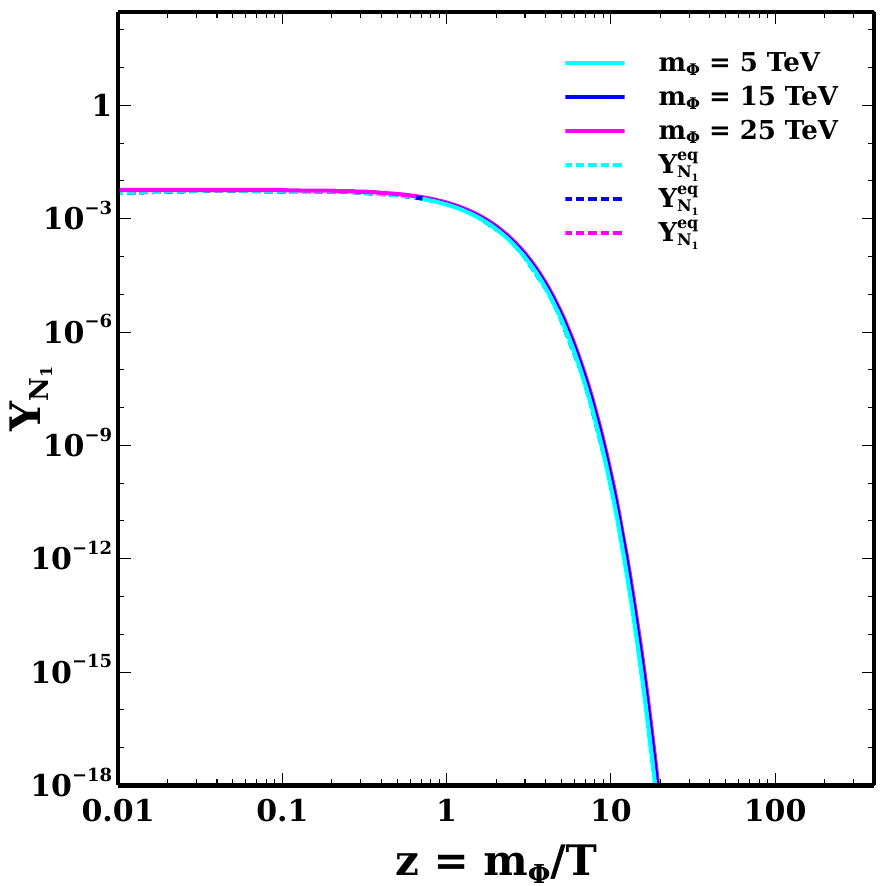}
\includegraphics[scale=0.49]{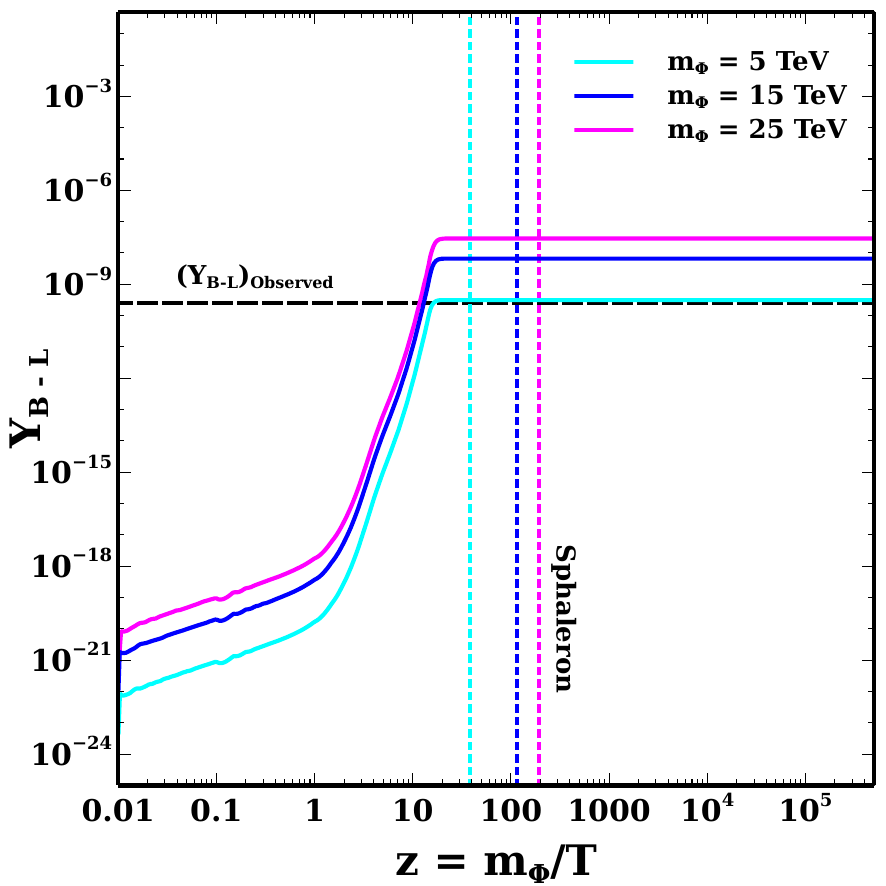}
\caption{Evolution of comoving number densities of $\psi$ (top left), $\phi$ (top right),  $N_{1}$ (bottom left) and $B-L$ (bottom right) with $z=m_{\phi}/T$ for different values of $m_{\phi}$ and $m_{\psi}$ keeping $\Delta m=m_{\psi}-m_{\phi}=1$ TeV fixed. The other relevant parameters are fixed at $M_{1}=m_{\phi}+m_{\psi}$, $M_{2}=M_{1}+100$ GeV, $y_1=y_2=1+1i$, $\lambda_{\eta \phi}=1$. The descriptions of the horizontal and vertical lines in bottom right plot remain same as in Fig. \ref{fig2}.}
    \label{fig3}
\end{figure}
In the above equations, $Y_i=n_i/s$ denotes the comoving number density of species $i$, with $s$ being the entropy density. Here $z=m_\phi/T$ and ${\bf H}(z)=\sqrt{8\pi^{3}g_{*}/90}\, m_{\phi}^{2}/(z^{2}M_{\rm Pl})$ is the Hubble expansion rate for the standard radiation dominated universe with $M_{\rm Pl} \simeq 1.22 \times 10^{19}$ GeV being the Planck mass and $g_{*}$ being the relativistic degrees of freedom. Here, $\gamma (A \rightarrow B)$ denotes the reaction rate for process $A \rightarrow B$ and is defined in Appendix~\ref{Appen2}.  The step function $\Theta$  ensures that the two-body decay $\psi \rightarrow \nu \phi$ occurs only after electroweak symmetry breaking, assumed to coincide with the sphaleron decoupling epoch $z_{\rm sph}=m_\phi/T_{\rm sph}$. In the Boltzmann equation for $B-L$, $\epsilon_1$ denotes the CP asymmetry from $N_1 \rightarrow L H$ decay while $\gamma_{\phi \psi \to L H}^{\delta}=n_{\psi}n_{\phi} \langle  \sigma v \rangle^{\delta}_{\phi \psi \to L H} $ includes the difference between reaction rates of $\phi \psi \to L H$ and $\phi \psi \to \bar{L} H^\dagger$ processes responsible for generating a net lepton asymmetry. The details of the thermally averaged cross-section difference $\langle  \sigma v \rangle^{\delta}_{\phi \psi \to L H}$ is given in Appendix~\ref{Appen1}. 

The lepton asymmetry at the sphaleron decoupling epoch $T_{\rm sph} \sim 131.7$ GeV~\cite{DOnofrio:2014rug}  gets converted into baryon asymmetry as~\cite{Harvey:1990qw}
\begin{align}
& Y_B (T_{\rm sph}) \simeq a_\text{Sph}\,Y_{B-L} (T_{\rm sph})=\frac{8\,N_F+4\,N_H}{22\,N_F+13\,N_H}\,Y_{B-L} (T_{\rm sph}) =\frac{28}{79} Y_{B-L} (T_{\rm sph})\,, 
\label{eq:sphaleron}
\end{align}
with $N_F=3\,,N_H=1$ being the fermion generations and the number of scalar doublets in our model respectively. The observational constraint on $\eta_B$ given in Eq.~\eqref{etaBobs} can be translated to $Y_{B} (T_{\rm sph}) $ as
\begin{equation}
    Y_{B} (T_{\rm sph}) = (8.69\pm 0.06) \times 10^{-11},
\end{equation} 
which can then be used in Eq.~\eqref{eq:sphaleron} to find 
\begin{equation}
    Y_{B-L} (T_{\rm sph}) =  (2.45\pm 0.02) \times 10^{-10}.
    \label{eq:3.4}
\end{equation}
This is the comoving density of $B-L$, referred to as $(Y_{B-L})_{\rm Observed}$ hereafter, that we need to reproduce at $T=T_{\rm sph}$ for successful baryogenesis via leptogenesis.


\begin{figure}
    \centering
     \includegraphics[scale=0.49]{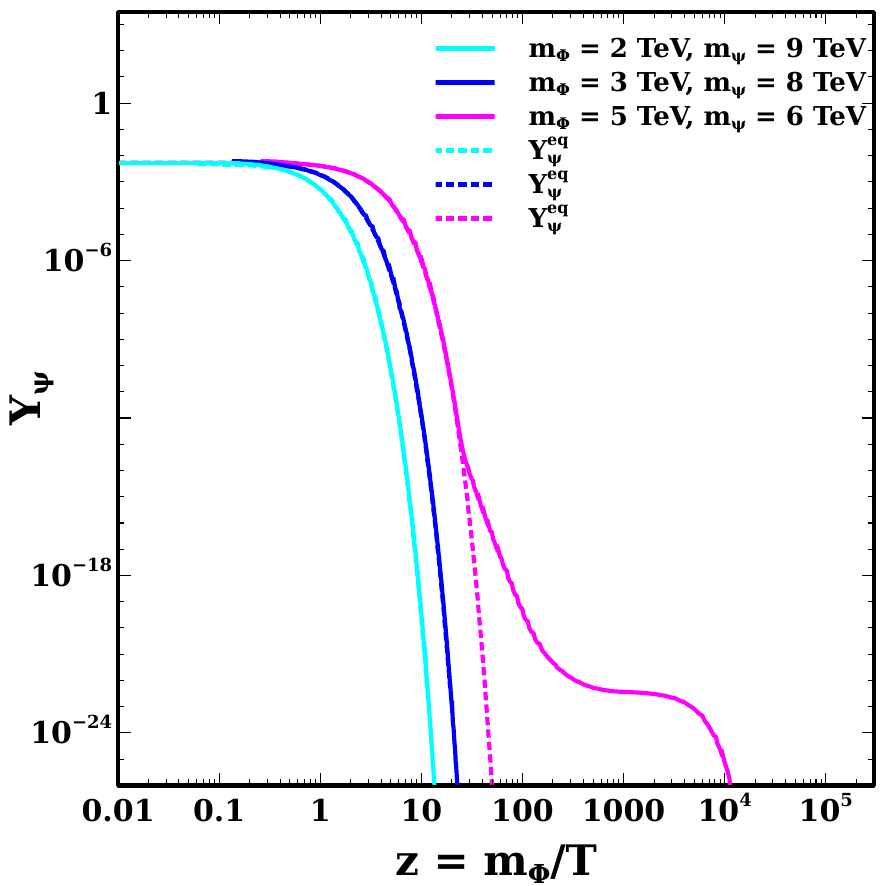}
    \includegraphics[scale=0.49]{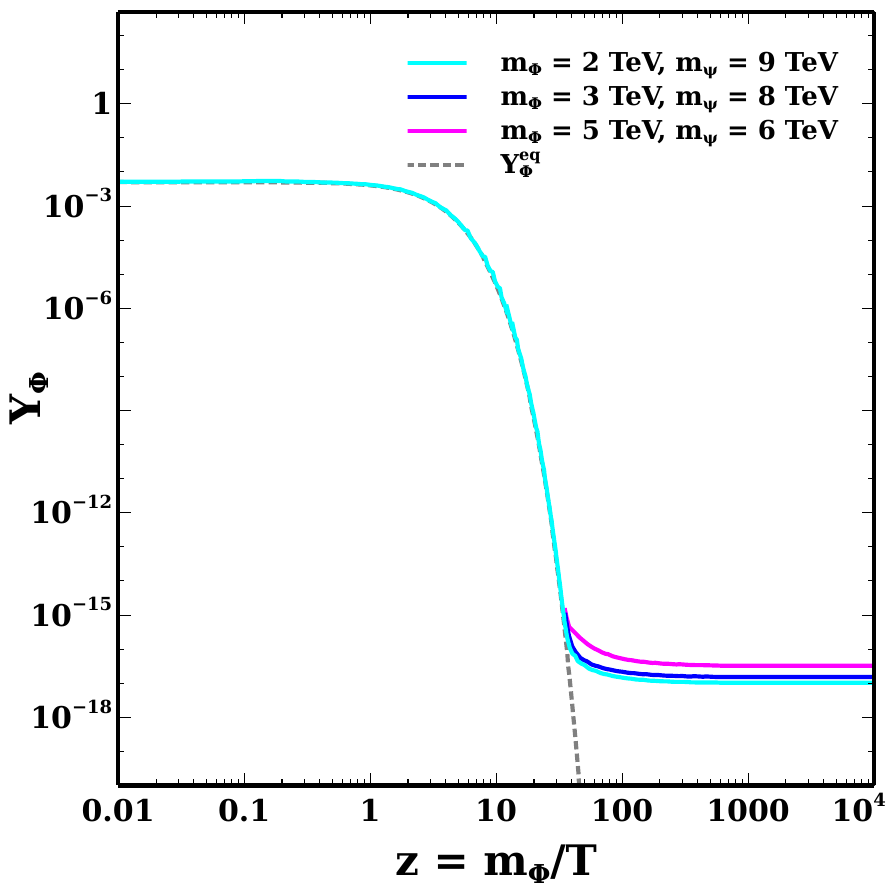}
    \includegraphics[scale=0.49]{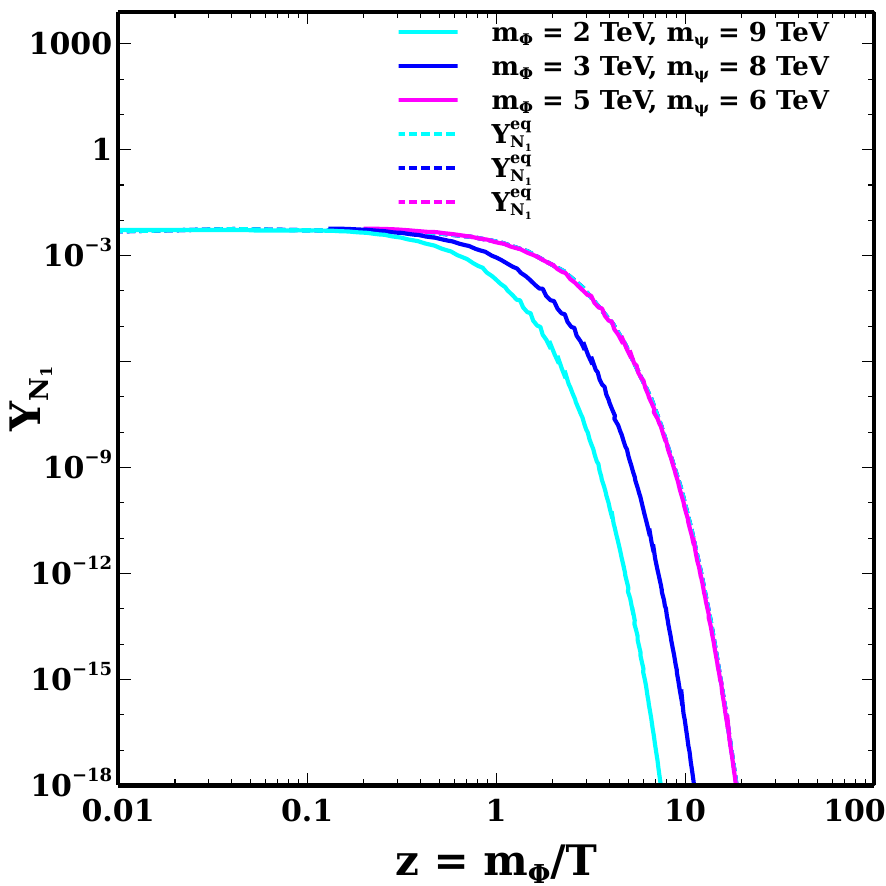}
    \includegraphics[scale=0.49]{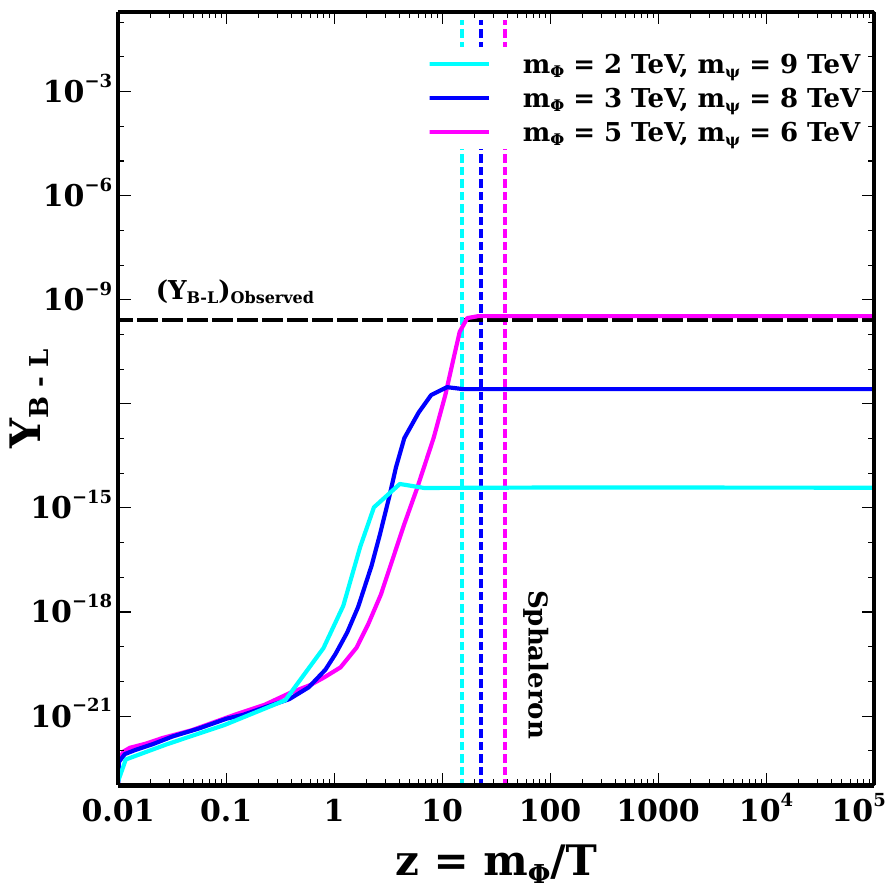}
    \caption{Evolution of comoving number densities of $\psi$ (top left), $\phi$ (top right),  $N_{1}$ (bottom left), and $B-L$ (bottom right)  with $z=m_{\phi}/T$ for different values of $m_{\phi}$ and $m_{\psi}$ keeping $m_{\phi} + m_{\psi} = M_{1}=11$ TeV.  The other relevant parameters are fixed at $y_1=y_2=1+1i$, $M_{2}=M_{1}+100$ GeV, $\lambda_{\eta \phi}=1$.The descriptions of the horizontal and vertical lines in bottom right plot remain same as in Fig. \ref{fig2}.}
    \label{fig5}
\end{figure}

\begin{figure}
    \centering
   \includegraphics[scale=0.49]{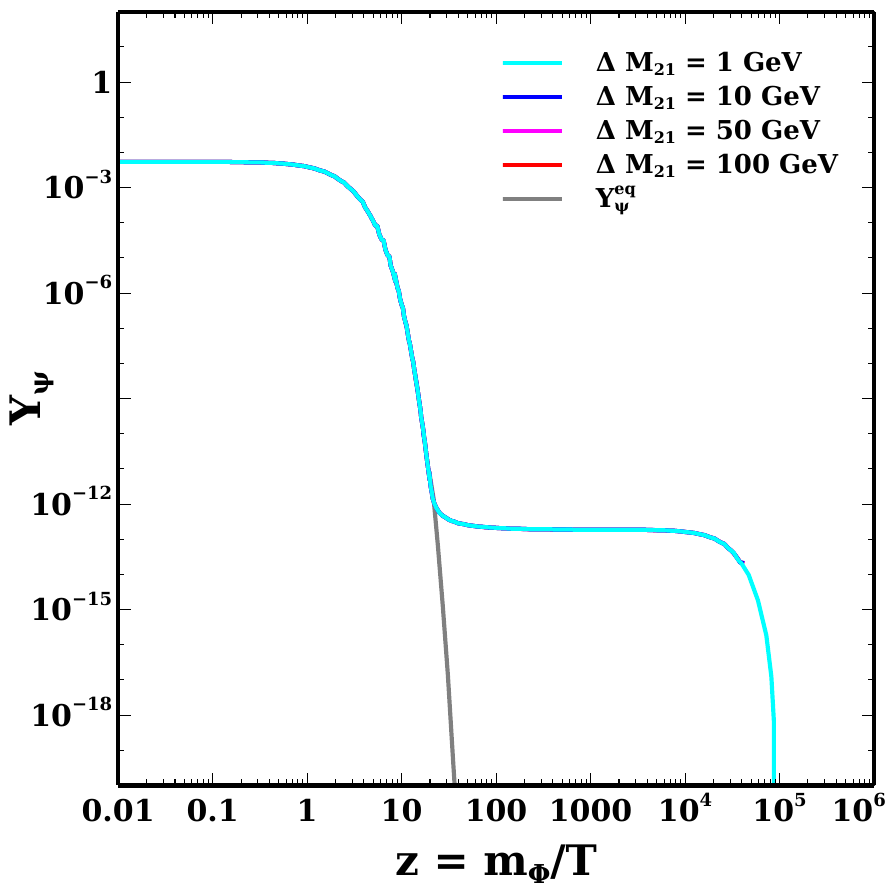}
   \includegraphics[scale=0.49]{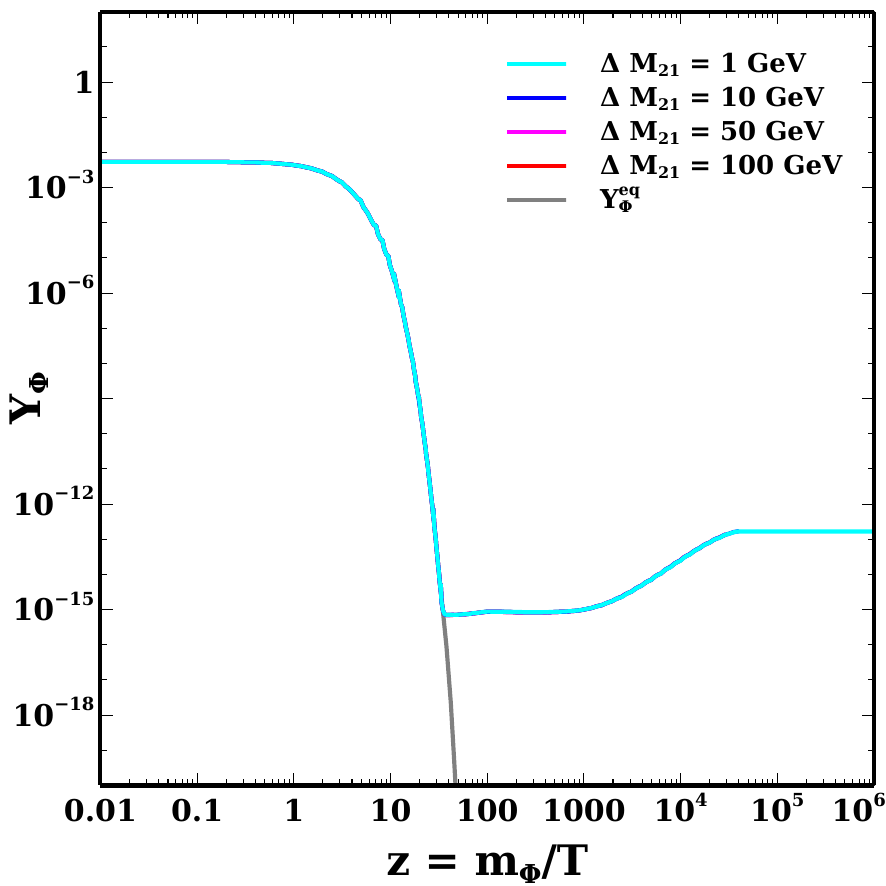}
   \includegraphics[scale=0.49]{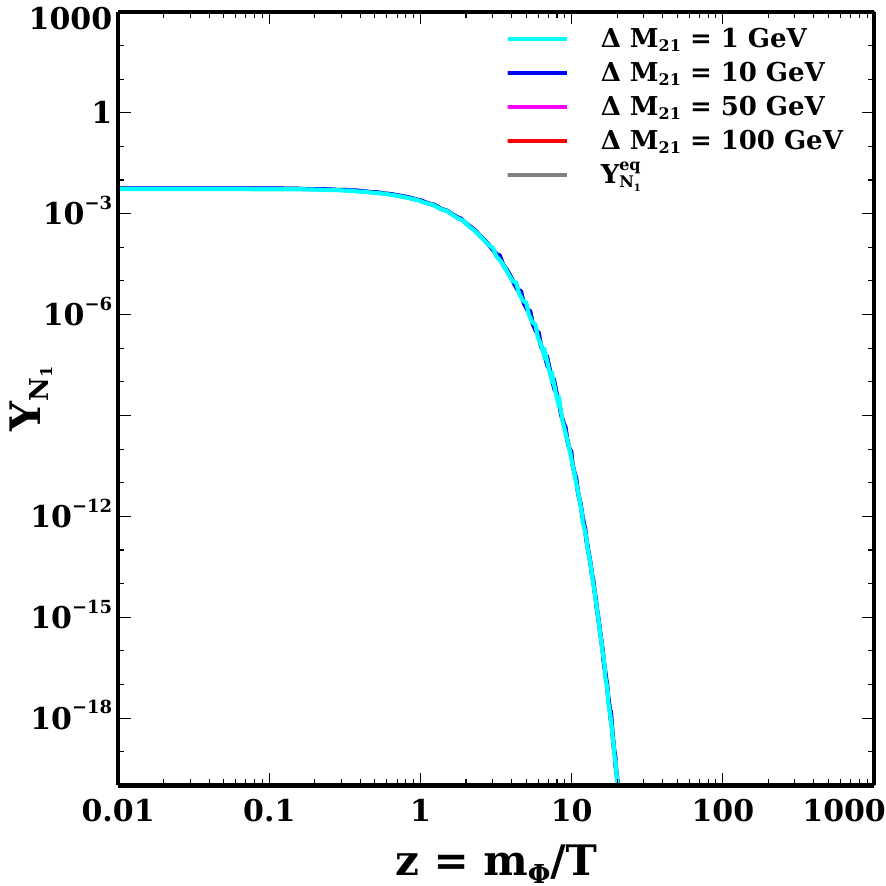}
   \includegraphics[scale=0.49]{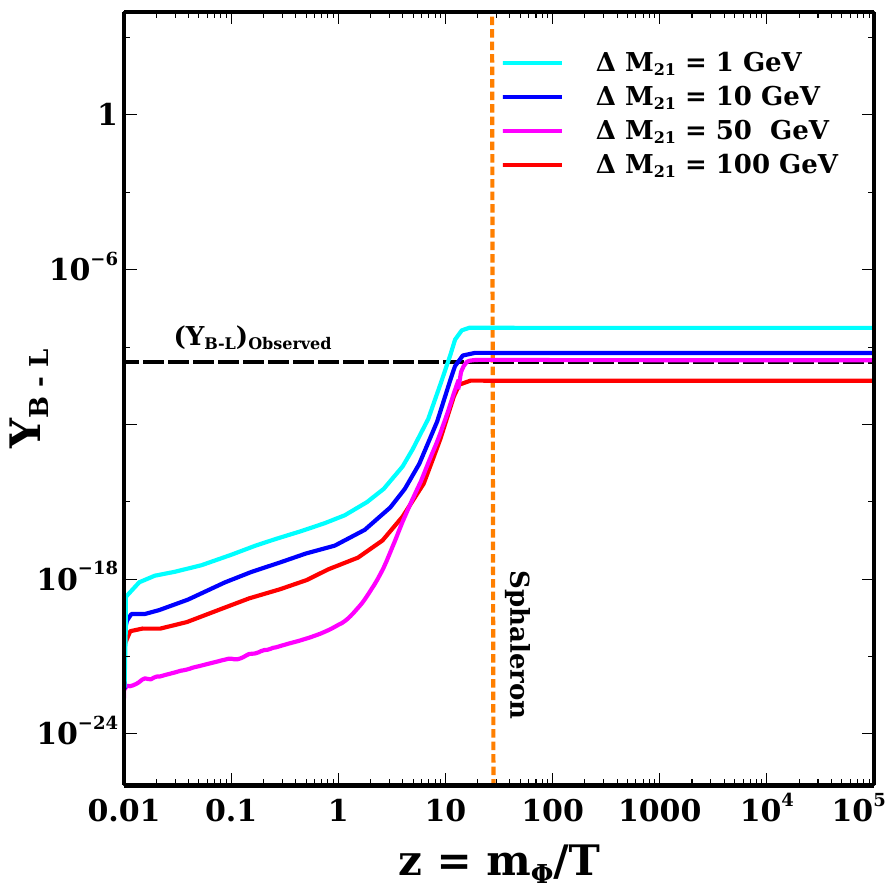}
    \caption{Evolution of comoving number densities of $\psi$ (top left), $\phi$ (top right),  $N_{1}$ (bottom left), and $B-L$ (bottom right) with $z=m_{\phi}/T$ for different value of mass splitting $\Delta M_{21} = M_{2}-M_{1}$. The relevant parameters are fixed at $m_{\phi}=3.5$ TeV, $m_{\psi}=4.5$ TeV, $M_{1}=8$ TeV, $y_{1}=y_{2}=0.08+0.08i$, $y_{\eta}=0.1$ and $\lambda_{\eta \phi}=1$. The descriptions of the horizontal and vertical lines in bottom right plot remain same as in Fig. \ref{fig2}.}
    \label{fig6}
\end{figure}

\begin{figure}
    \centering
    \includegraphics[scale=0.49]{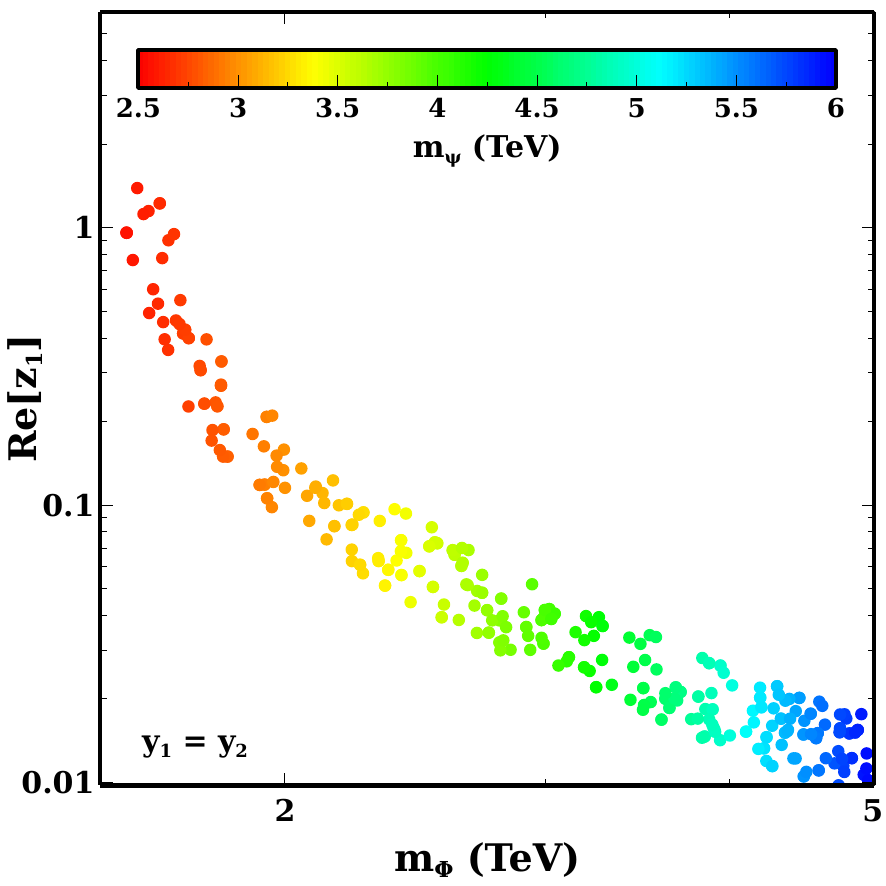}
    \includegraphics[scale=0.49]{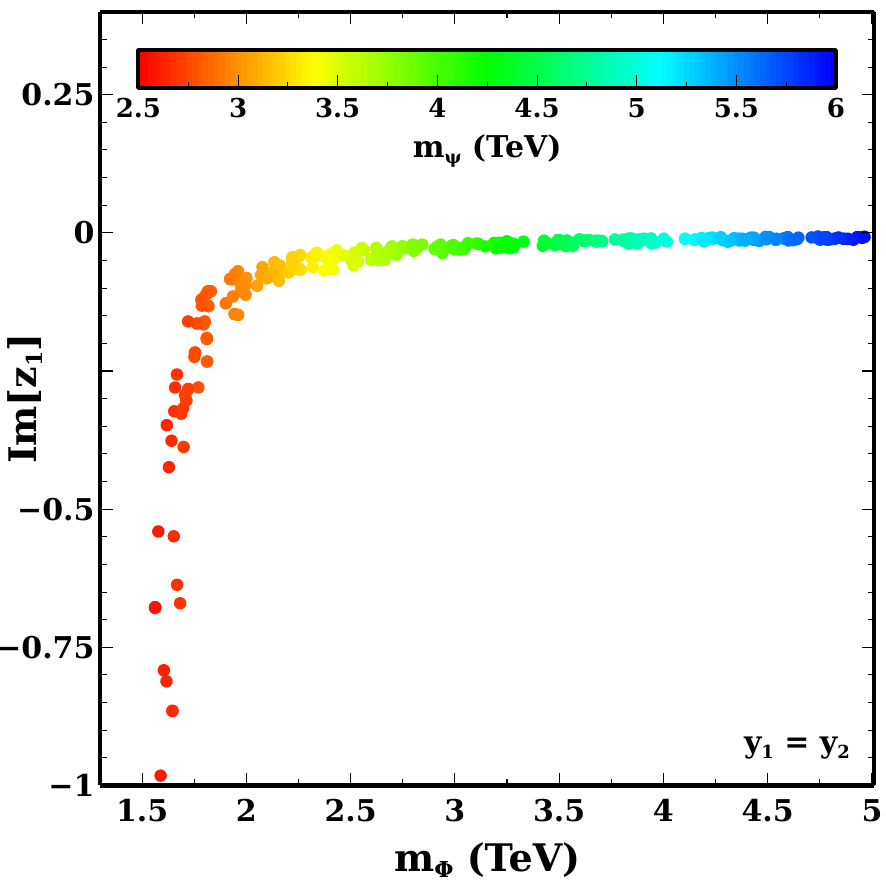}
    \caption{Parameter space consistent with the observed baryon asymmetry in $m_{\phi}-{\rm Re} [z_{1}]$ plane (left panel) and in $m_{\phi}-{\rm Im} [z_{1}]$ plane (right panel) for $y_1=y_2$. The values of $m_{\psi}$ are shown as color bar. Here we keep $M_{1}=m_{\psi}+m_{\phi}$ and $M_{2}=M_{1}+\Delta M_{21}$ and consider $10^{2} \,{\rm GeV} \lesssim \Delta M_{21}\lesssim 10^{3}\, {\rm GeV}$. }
    \label{fig66}
\end{figure}

\begin{figure}
    \centering
    \includegraphics[scale=0.49]{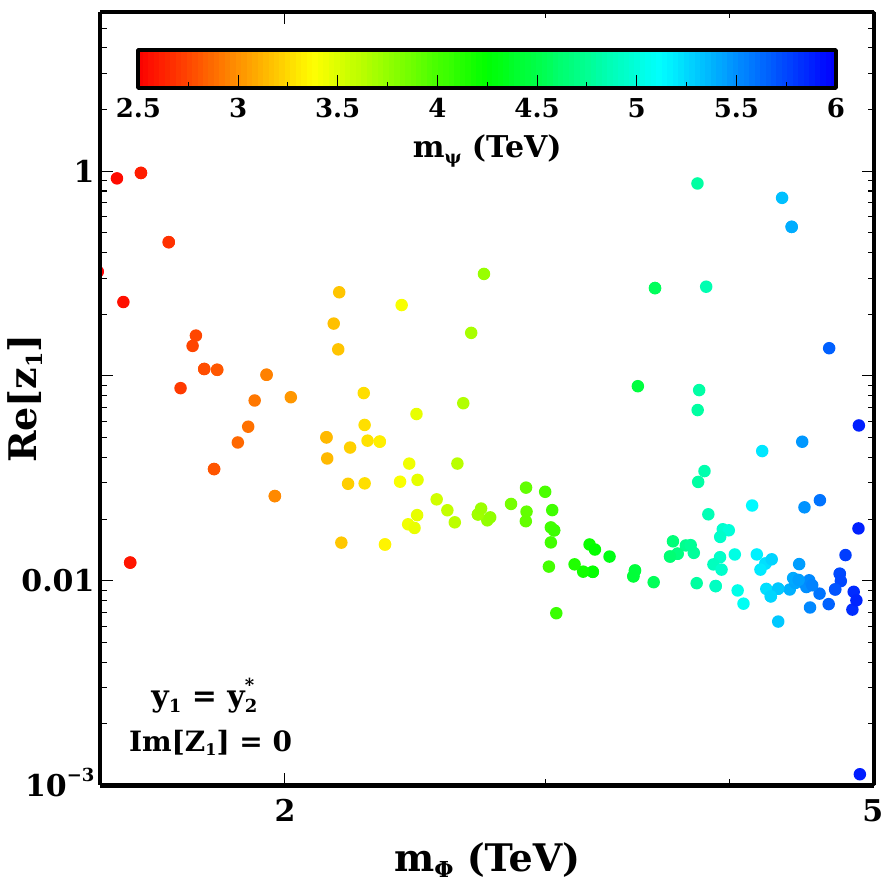}
    \caption{Parameter space consistent with the observed baryon asymmetry in $m_{\phi}-{\rm Re} [z_{1}]$ plane for $y_1=y^*_2$ and ${\rm Im}[z_1]=0$. The values of $m_{\psi}$ are shown as color bar. Here we keep $M_{1}=m_{\psi}+m_{\phi}$ and $M_{2}=M_{1}+\Delta M_{21}$ and consider $10^{2} \, {\rm GeV} \lesssim \Delta M_{21}\lesssim 10^{3} \, {\rm GeV}$. }
    \label{fig67}
\end{figure}

\begin{figure}
    \centering
    \includegraphics[scale=0.49]{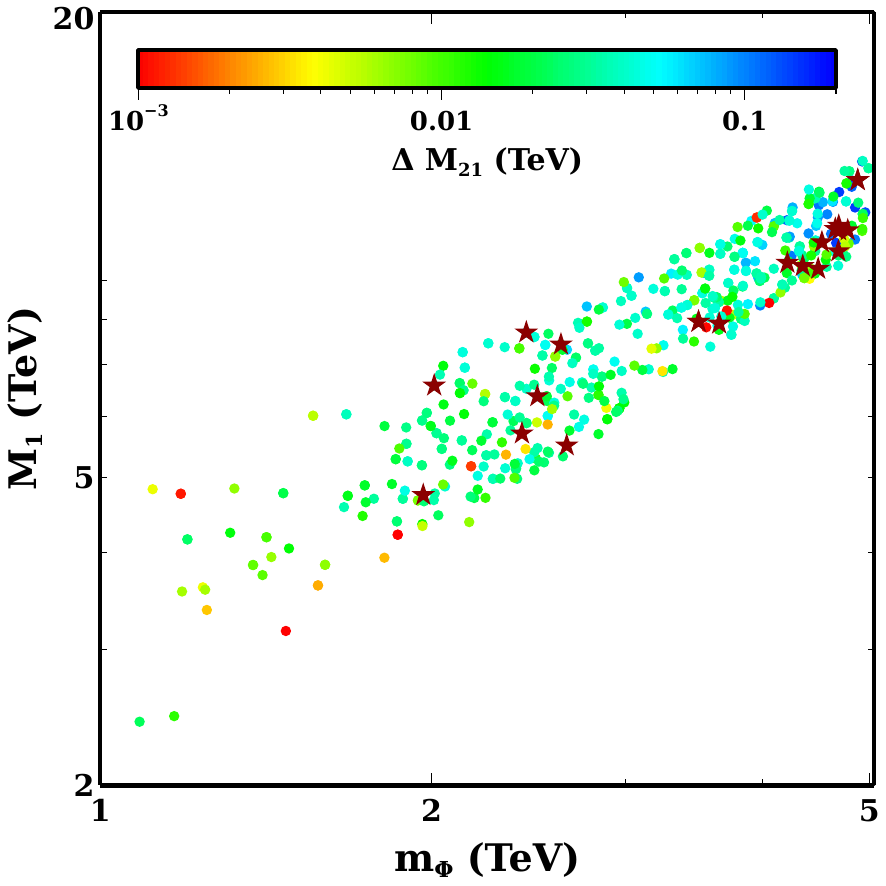}
    \includegraphics[scale=0.49]{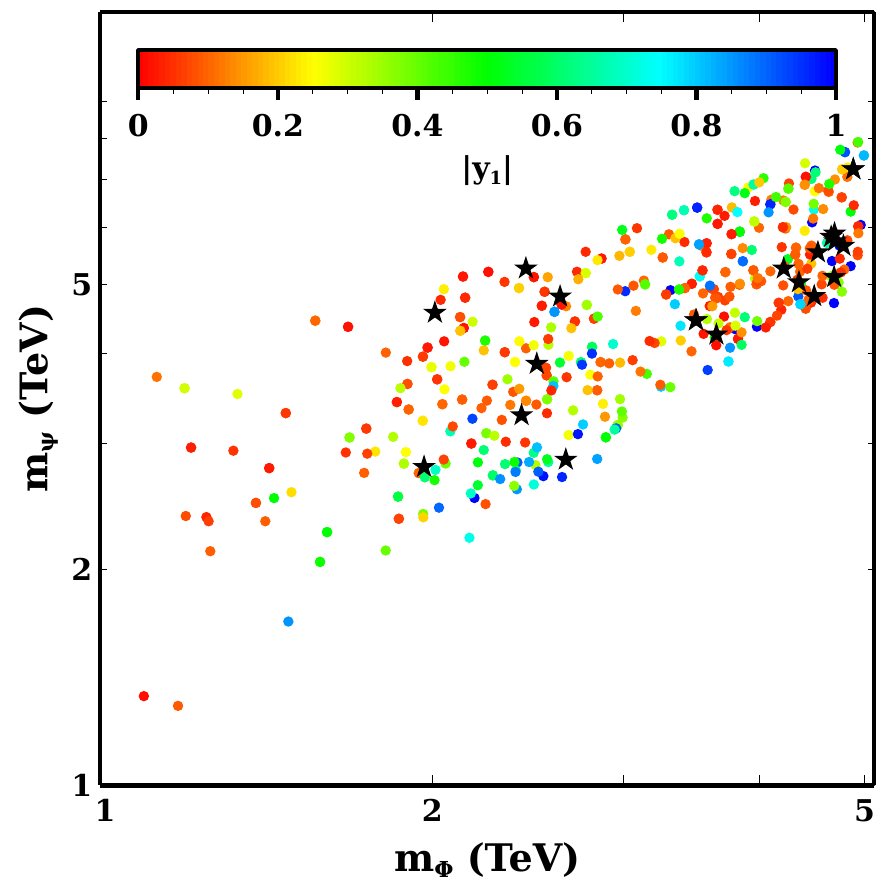}
    \caption{Parameter space consistent with the observed baryon asymmetry in $m_{\psi}-M_{1}$ plane with $\Delta M_{21}=M_{2}-M_{1}$ as color bar (left panel) and in $m_{\phi}-m_{\psi}$ plane with magnitude of Yukawa coupling $y_{1}=y_{2}$ as color bar (right panel). The points satisfying the correct baryon asymmetry and the observed DM relic are marked as $\star$. The other relevant parameters are fixed at $\lambda_{\eta \phi}=0.1$, $y_{\eta}=0.01$ and $\mu_{\eta\phi}=60$ GeV. }
    \label{fig7}
\end{figure}

\section{Results and Discussion}
\label{sec3}
We have implemented the model in \texttt{FeynRules}~~\cite{Alloul:2013bka} , and use \texttt{CalcHEP}~\cite{Belyaev:2012qa} and \texttt{micrOMEGAs}~\cite{Alguero:2023zol} for the purpose of numerical calculations. We then solve the coupled Boltzmann equations mentioned above using our own code to find the parameter space consistent with DM relic and baryon asymmetry of the Universe. Fig.~\ref{fig2} shows the evolution of comoving abundances of $\psi, \phi, N_1, B-L$ for different choices of Yukawa coupling $y_1$ which connects $\phi, \psi$ with RHN. For simplicity, we consider $y_1=y_2$ in our numerical calculations. As expected, smaller values of $y_1$ lead to a late decay of $\psi$ after freeze-out with consequently late enhancement of $\phi$ number density at late epochs. Since the CP asymmetry from coannihilation gets enhanced for larger $y_1$, we see the corresponding larger lepton asymmetry in the bottom right panel of Fig.~\ref{fig2}. However, larger $y_1$ can also enhance the washout processes and can lead to overall decrease in lepton asymmetry once it is increased beyond a threshold. As can be seen from the bottom right panel of Fig.~\ref{fig2}, final lepton asymmetry increases as $y_1$ is increased to $(1+i)$. If $y_1$ is increased further, the final lepton asymmetry decreases, as evident from the green colored contour. Fig.~\ref{fig3} shows the corresponding evolutions for different values of dark sector particle masses keeping their mass splitting fixed at 1 TeV. Lepton asymmetry starts to grow earlier for heavier $\psi, \phi$ as the coannihilation process goes out-of-equilibrium earlier. This leads to larger lepton asymmetry production for heavier $\psi, \phi$, a result similar to vanilla leptogenesis.
Fig.~\ref{fig5} shows the effect of different values of $\phi, \psi$ masses while keeping their sum equal to the lightest RHN mass. The heavier RHN mass is fixed at $M_2=M_1+100 \, {\rm GeV}$. The final asymmetry increases as $m_\psi-m_\phi$ decreases. This is expected due to larger coannihilation rate for reduced mass splitting. Fig.~\ref{fig6} shows the effect of RHN mass splitting on the evolution of comoving number densities. As expected, smaller mass splitting results in enhanced CP asymmetry resulting in larger lepton asymmetry.


As can be seen from the expression for CP asymmetry given in Appendix \ref{Appen1}, there are two sources of CP violation in our setup: one from the Dirac Yukawa coupling of neutrinos $h_{\alpha i}$ and other from the Yukawa coupling $y_i$ involving heavy RHNs and DM. As shown in Eq. \eqref{eq:CI}, the CP phases in Dirac Yukawa couplings can arise from a combination of CP phases in orthogonal matrix $R$ and the PMNS mixing matrix. Since $R$ is arbitrary, we can have the required CP violation for any values of low energy leptonic CP phase of the PMNS matrix. This makes typical leptogenesis scenarios insensitive to low energy CP violation in the leptonic sector \cite{Branco:2001pq, Pascoli:2006ie, Davidson:2007va}. In order to check the role of CP phases in $R$ matrix, we first perform a random scan over the complex angle $z_1$ which parametrizes the orthogonal matrix $R$. The magnitudes of the real and imaginary parts of $z_1$ are kept at values $\lesssim 1$. Such values of $z_1$ are consistent with the upper bound obtained from charged lepton flavor violation~\cite{BhupalDev:2014qbx}. Fig. \ref{fig66} shows the allowed values of the complex angle $z_1$ and its variation with $m_\phi, m_\psi$. We choose $y_1=y_2$ such that lepton asymmetry arises solely from the CP violating phases in Dirac Yukawa couplings. If we deviate from $y_1=y_2$, then CP phases in $y_i$ can also contribute to the generation of lepton asymmetry. To illustrate this we choose a particular case where $y_1=y^*_2$ and assume $z_1$ to be real such that lepton asymmetry is generated solely due to CP violating phases in $y_i$. The corresponding results are shown in Fig. \ref{fig67} with variation of ${\rm Re}[z_1]$ with dark sector particle masses. For both Fig. \ref{fig66} and Fig. \ref{fig67}, we choose $M_{1}=m_{\psi}+m_{\phi}$ and $M_{2}=M_{1}+\Delta M_{21}$ and consider $10^{2} \,{\rm GeV} \lesssim \Delta M_{21}\lesssim 10^{3}\, {\rm GeV}$.

After showing the allowed range of the complex angle $z_1$, we perform a numerical scan of physical masses and couplings to find the parameter space consistent with successful leptogenesis and dark matter relic density. For concreteness, we choose $y_{1}=y_{2}$ and $z_{1}=0.001+0.001i$ for this analysis. Fig.~\ref{fig7} shows the parameter space consistent with correct baryon asymmetry and DM relic abundance. The left panel shows the parameter space in the plane of DM mass $m_\phi$ and the mass of the lightest RHN mediating coannihilations with the color code indicating the mass splitting among RHNs. As DM mass increases, $M_{1,2}$ also increase in order to keep the coannihilation efficient. Similar pattern is also seen on the right panel plot where $m_\psi$ increases with DM mass while color code showing the Yukawa coupling $y_1$. In both the plots, the points marked as $\star$ satisfy the correct DM relic abundance as well. Clearly, we can have TeV scale leptogenesis without fine-tuning at the level of typical resonant leptogenesis~\cite{Pilaftsis:2003gt} where $\Delta M_{21} \sim \mathcal{O}(1)$ keV. As can be also seen from the left panel of Fig.~\ref{fig7}, decreasing RHN mass also decreases DM mass and therefore, the RHN mass, cannot be too much below the TeV scale because in that case, the lepton asymmetry from DM coannihilations would be generated after the sphaleron decoupling. 

\section{Detection Aspects}
\label{sec2a}
\begin{figure}[h!]
    \centering
    \includegraphics[scale=0.49]{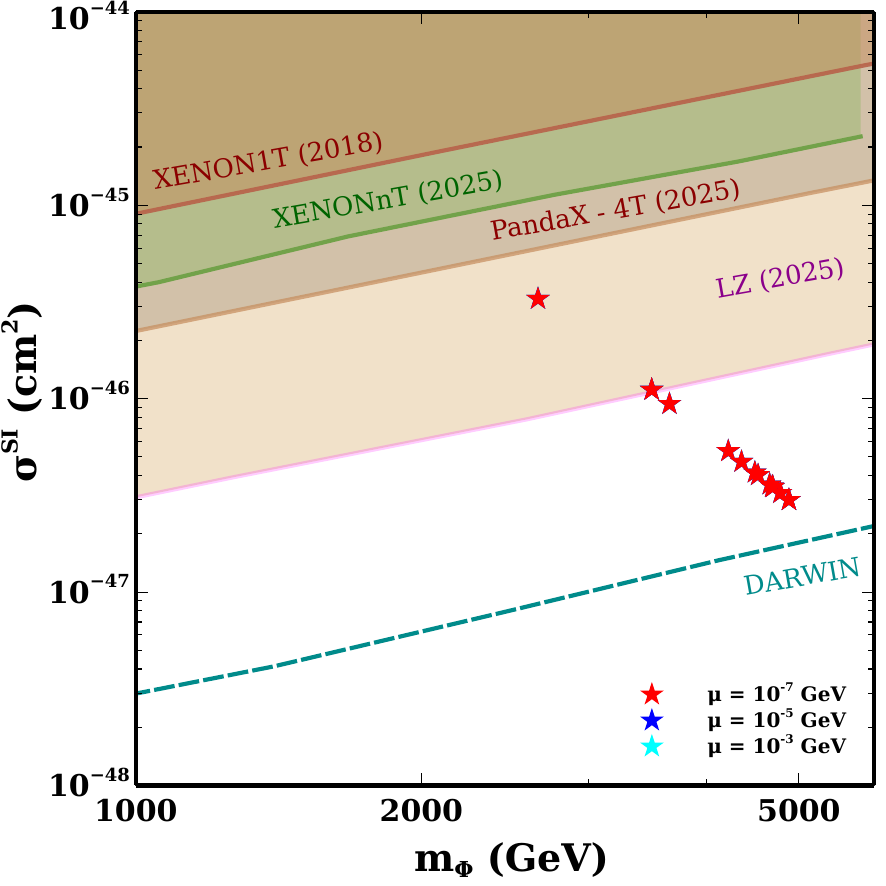}
    \caption{Spin-independent DM-nucleon scattering cross-section as a function of DM mass. The points shown here correspond to the ones consistent with successful leptogenesis. The shaded regions indicate the parameter space ruled out by existing constraints from \texttt{XENON1T}~\cite{XENON:2018voc}, \texttt{XENONnT}~\cite{XENON:2025vwd}, \texttt{PandaX-4T}~\cite{PandaX:2024qfu} and \texttt{LZ}~\cite{LZ:2024zvo} while the dashed contour indicates the future sensitivity from \texttt{DARWIN}~\cite{DARWIN:2016hyl}.}
    \label{fig:direct}
\end{figure}

Since the DM in our setup is of WIMP type, it has the usual direct and indirect detection prospects. Given that $\phi$ can couple to the SM Higgs $h$ both directly and via $\eta-h$ mixing, we have tree level Higgs mediated spin-independent DM-nucleon scattering which can be probed at direct detection experiments. 
Since the SM Higgs mixes with $\eta$ the effective Higgs to nucleon coupling become 
\begin{eqnarray}
g_{h_i nn} = \frac{f_{n}m_{n}}{v} \times
\begin{cases}
\cos\theta, & \text{for } h_1 \\
\sin\theta, & \text{for } h_2
\end{cases}
\end{eqnarray}
where $\theta$ is the mixing angle given by $\theta=\tan^{-1}\left( \frac{v\mu}{2\lambda_{H}v^{2}-m_{\eta}^{2}} \right)$ and $f_n=0.3$ is the Higgs-nucleon coupling. Similarly, DM $\phi$ couples to the mass eigenstates with couplings 
\begin{eqnarray}
\lambda_{h_i \phi \phi} =  
\begin{cases}
-\sin\theta \mu_{\eta \phi}+\cos\theta v \lambda_{H\phi} & \text{for } h_1 \, , \\
\cos\theta \mu_{\eta \phi}+\sin\theta v \lambda_{H\phi} & \text{for } h_2 \, .
\end{cases}
\end{eqnarray}
The effective DM-nucleon cross-section becomes 
\begin{equation}
    \sigma_{\rm SI}= \frac{1}{\pi} \left( \frac{\lambda_{h_{1}\phi\phi}g_{h_{1}nn}}{m_{h_{1}}^{2}}+\frac{\lambda_{h_{2}\phi\phi}g_{h_{2}nn}}{m_{h_{2}}^{2}}  \right)^{2} \mu_{r}^{2}. 
\end{equation}
where $\mu_r = m_\phi m_n/(m_\phi+m_n)$ is the DM-nucleon reduced mass. In order to simplify the analysis, we fix $m_\eta = 2 m_\phi, \mu_{\eta \phi}=60$ GeV, $\lambda_{H \phi}=10^{-4}, \lambda_H =0.129$ and calculate $\sigma_{\rm SI}$ for different $\mu$ and DM masses while also ensuring that the lightest mass eigenstate among $h_{1,2}$ correspond to the SM Higgs boson with mass $\sim 125$ GeV. This allows us to constrain the parameter space by using existing data from experiments like \texttt{LUX-ZEPLIN (LZ)} ~\cite{LZ:2024zvo}, \texttt{XENONnT}~\cite{XENON:2025vwd}, \texttt{PandaX-4T}~\cite{PandaX:2024qfu} with more scrutiny at future experiments like \texttt{DARWIN}~\cite{DARWIN:2016hyl}. In Fig.~\ref{fig:direct} we show the prediction for the spin-independent DM direct detection cross-section against the current experimental bounds and the future sensitivities by \texttt{DARWIN}. While variation in $\mu$ does not change $\sigma_{\rm SI}$ much, lighter DM masses lead to larger $\sigma_{\rm SI}$, already in tension with the {\tt LZ} bound. DM mass larger than $\sim 3$ TeV are currently allowed, but remains within sensitivity of future experiment \texttt{DARWIN}.

DM can also be probed at indirect detection experiments looking for DM annihilation into SM particles. Excess of gamma-rays, either monochromatic or diffuse, can be constrained from observations at such indirect detection experiments. While DM annihilation to monochromatic photons is loop-suppressed, tree-level DM annihilation into different charged particles can contribute to diffuse gamma-rays which can be constrained by existing data. While \texttt{Fermi-LAT}~\cite{Fermi-LAT:2017opo} and \texttt{HESS}~\cite{HESS:2022ygk} data provide stringent constraint at present, future experiments like \texttt{CTA}~\cite{CTA:2020qlo} and  \texttt{SWGO}~\cite{Viana:2019ucn} can probe such DM signatures even further.

\begin{figure}[h!]
    \centering
    \includegraphics[scale=0.49]{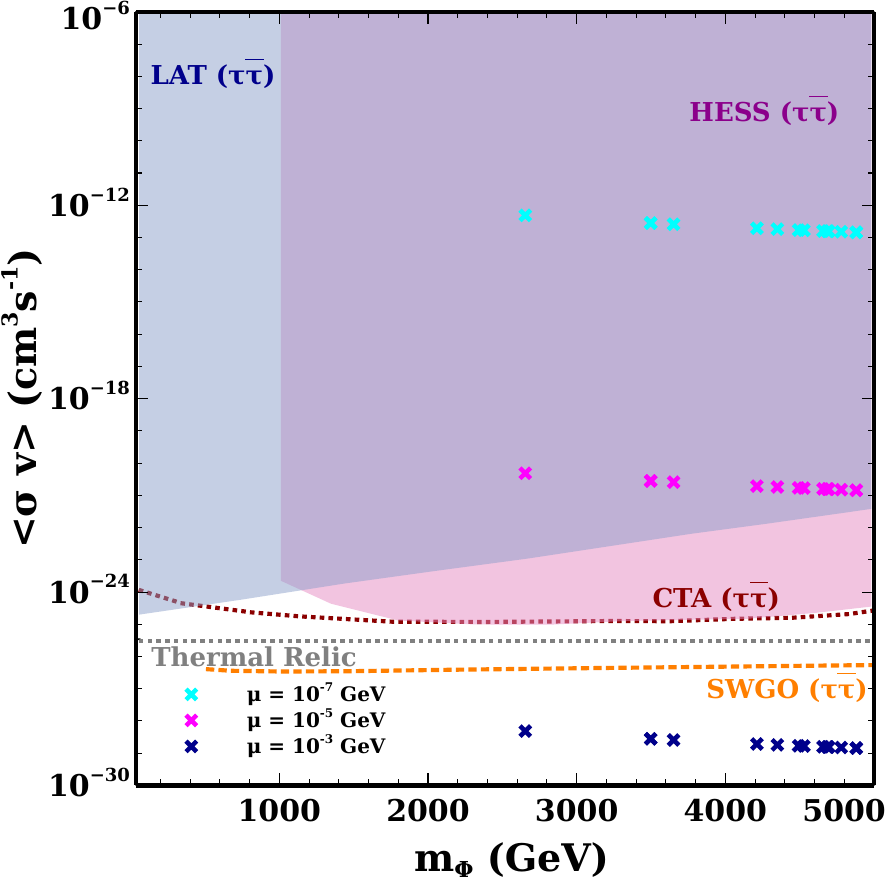}
    \includegraphics[scale=0.49]{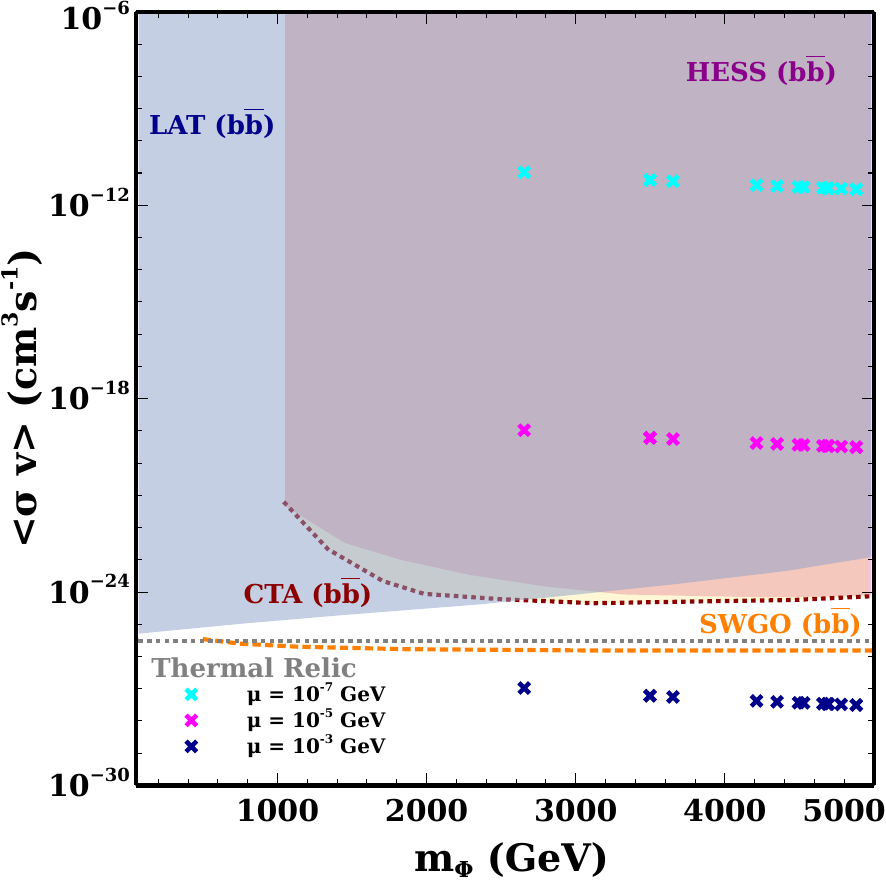}
     \includegraphics[scale=0.49]{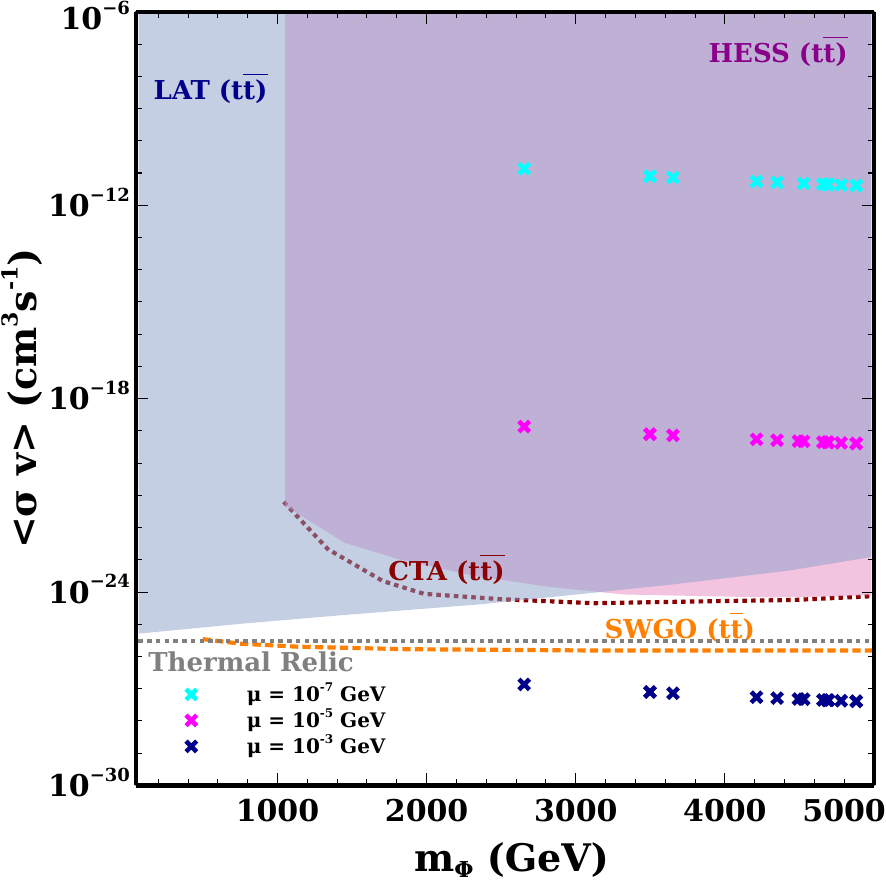}
      \includegraphics[scale=0.49]{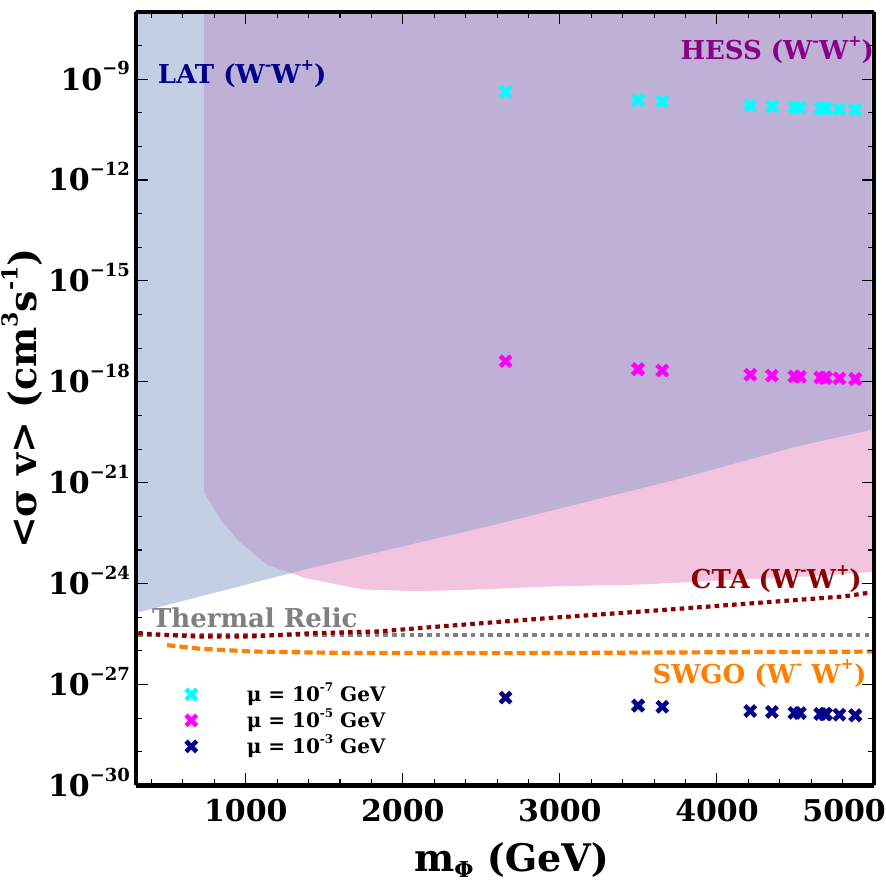}
    \caption{DM annihilation cross-section into different final states $\tau \bar{\tau}$ (top-left), $b \bar{b}$ (top-right), $t \bar{t}$ (bottom-left) and $W^- W^+$ (bottom-right), constrained from non-observation of secondary gamma-rays. The $\times$-shaped points of different color show the variation with respect to DM mass and $\mu$. The shaded regions are ruled out by existing constraints from \texttt{Fermi-LAT}~\cite{Fermi-LAT:2017opo} and \texttt{HESS}~\cite{HESS:2022ygk}. The maroon and orange colored contours correspond to future sensitivities from \texttt{CTA}~\cite{CTA:2020qlo} and \texttt{SWGO}~\cite{Viana:2019ucn}. The gray dotted contour denotes the typical cross-section for thermal DM required to satisfy the relic abundance~\cite{Steigman:2012nb}.}
    \label{fig:indirect}
\end{figure}

For indirect detection search, we consider the process $\phi \phi \to f \bar{f}$ with charged fermion final states assuming $f \equiv \tau, b, t$ due to their larger Yukawa coupling to the SM Higgs. In the non-relativistic limit, the corresponding thermal averaged cross-section can be written as 
\begin{equation}
    \langle \sigma v  \rangle = \frac{N_{c}m_{f}^{2}}{128\pi m_{\phi}^{2} } \left( 1-\frac{m_{f}^{2}}{m_{\phi}^{2}} \right)^{3/2} \left|  \sum_{i=1}^{2} \frac{\lambda_{h_{i}\phi \phi}g_{h_{i}f \bar{f}}}{4m_{\phi}^{2}-m_{h_{i}}^{2}+ i m_{h_{i}}\Gamma_{i}}  \right|^{2} \, ,
\end{equation} 
where we have ignored the interference term. In the above expression, $g_{h_{i}f \bar{f}}$ are the couplings of $h_{i}$ with fermion $f$ given by 
\begin{eqnarray}
g_{h_i f \bar{f}} = \frac{m_{f}}{v} \times 
\begin{cases}
\cos\theta, & \text{for } h_1 \\
\sin\theta, & \text{for } h_2.
\end{cases}
\end{eqnarray}
We also consider the annihilation process $\phi \phi \to W^{+} W^{-}$ whose thermal averaged cross-section can be written as 
\begin{equation}
    \langle \sigma v  \rangle = \frac{1}{8\pi m_{\phi}^{2} } \left( 1-\frac{m_{W}^{2}}{m_{\phi}^{2}} \right)^{1/2} \left|  \sum_{i=1}^{2} \frac{\lambda_{h_{i}\phi \phi}g_{h_{i}W^{+} W^{-}}}{4m_{\phi}^{2}-m_{h_{i}}^{2}+ i m_{h_{i}}\Gamma_{i}}  \right|^{2}F_{W}(m_{\phi}),
\end{equation} 
where the interference term is ignored once again. The form of $F_{W}(m_{\phi})$ is given as
\begin{equation}
 F_{W}(m_{\phi}) = \Bigg[1-\frac{m_{W}^{2}}{m_{\phi}^{2}}+\frac{3m_{W}^{4}}{4m_{\phi}^{4}}\Bigg]   \, ,
\end{equation}
and $g_{h_{i}W^{+} W^{-}}$ are the couplings of $h_{i}$ with $W^{+},  W^{-}$ bosons given by 
\begin{eqnarray}
g_{h_{i}W^{+} W^{-}} = \frac{2 m_{W}^{2}}{v} \times 
\begin{cases}
\cos\theta & \text{for } h_1 \, ,\\
\sin\theta & \text{for } h_2\, .
\end{cases}
\end{eqnarray}
In Fig.~\ref{fig:indirect} we show the predictions for DM indirect detection rates together with existing constraints and future sensitivities by considering $\tau \bar{\tau}$ (top-left), $b \bar{b}$ (top-right), $t \bar{t}$ (bottom-left) and $W^- W^+$ (bottom-right) final states respectively. For smaller $\mu$, the physical mass of the heavier scalar among $h_{1,2}$ remains almost same as the bare mass $m_\eta =2m_\phi$ leading to a resonantly enhanced cross-section. With increase in $\mu$, the physical mass of the heavier scalar changes from the resonance limit leading to a smaller cross-section. Therefore, in the resonant limit $m_\eta =2m_\phi$, smaller values of $\mu$ can be ruled out by existing constraints from indirect detection experiments like \texttt{Fermi-LAT}~\cite{Fermi-LAT:2017opo} and \texttt{HESS}~\cite{HESS:2022ygk} while future experiments like \texttt{CTA}~\cite{CTA:2020qlo}, \texttt{SWGO}~\cite{Viana:2019ucn}\footnote{We used similar sensitivity line for $b \bar{b}$ and $t \bar{t}$ final states as the latter was not available.} can probe larger values of $\mu$ or equivalently, DM annihilations away from the resonant limit. It is also worth mentioning that indirect detection experiments typically put bound on a specific final state considering $100\%$ branching ratio. However, in our model, DM can annihilate into different final states and hence the bounds shown in Fig.~\ref{fig:indirect} are of conservative nature. We also note that DM detection prospects in this section uses the resonant condition $m_\eta=2 m_\phi$ in order to enhance indirect detection prospects whereas the analysis of leptogenesis and DM in section~\ref{sec2} used $m_\eta \ll m_\phi$. The mass of $\eta$ and its coupling to DM $\phi$ are introduced only to satisfy the relic of DM which freezes out at a temperature below the freeze-out of lepton asymmetry production process. Therefore, tuning the parameters related to DM freeze-out does not change the leptogenesis satisfying parameter space significantly.

In addition to the detection prospects of DM in our model, RHNs with masses in the TeV range can also have promising experimental signatures~\cite{Abdullahi:2022jlv}. 
Although TeV-scale RHNs in the minimal type-I seesaw, as considered in our work, are usually difficult to be produced at terrestrial experiments like the LHC~\cite{CMS:2018iaf, ATLAS:2024rzi} due to  their suppressed Yukawa interactions, there exist model constructions with specific textures of Dirac and Majorana neutrinos where larger Yukawa couplings are allowed~\cite{Pilaftsis:1991ug, Kersten:2007vk}, thereby enhancing their collider prospects. Moreover, in the presence of additional interactions like gauged $U(1)_X$~\cite{FileviezPerez:2009hdc, Chauhan:2021xus, Das:2021esm,Liu:2021akf} or $SU(2)_L\times SU(2)_R$~\cite{Keung:1983uu,Chen:2013foz,Nemevsek:2018bbt}, production of RHNs can be further enhanced leading to interesting collider signatures.

\section{Conclusions}
\label{sec4}
We have proposed a minimal leptogenesis scenario where lepton asymmetry is generated from DM coannihilation. While leptogenesis from DM annihilation has been known for a long time in the context of WIMPy leptogenesis, we consider coannihilation among DM and its heavier partners to be the primary source of leptogenesis. Unlike typical WIMPy leptogenesis models, our proposed idea can be realised in a very minimal extension of the type-I seesaw model of light neutrino mass. In addition to two heavy right handed neutrinos of type-I seesaw model, we have two $Z_2$-odd coannihilating dark sector particles $\phi, \psi$ and an additional $Z_2$-even scalar $\eta$, all singlets under the SM gauge symmetry. While coannihilations occur via heavy RHNs, the $Z_2$-even scalar singlet helps in getting the correct scalar DM relic by facilitating $\phi \phi \rightarrow \eta \eta$ annihilation. While RHNs by themselves can contribute to the origin of lepton asymmetry, in our setup the asymmetry is generated at TeV scale primarily from DM coannihilations. In fact, in presence of additional CP violating phases associated with DM couplings to RHNs, it is possible to have non-zero CP violation in DM coannihilations even when minimal type-I seesaw CP violation remains vanishing at one-loop level. The DM in our setup is WIMP type and hence can have usual direct and indirect detection aspects via scalar portal interactions with the SM that can be tested in the upcoming experiments.

\section*{Acknowledgments}
The work of D.B. is supported by the Science and Engineering Research Board (SERB), Government of India grant MTR/2022/000575. The work of B.D. was partly supported by the US Department of Energy under grant No.~DE-SC0017987 and by a Humboldt Fellowship from the Alexander von Humboldt Foundation. B.D. thanks the High Energy Physics group at IIT Guwahati for local hospitality where a part of this work was done. S.A. and B.D. also thank  the organizers of PPC 2024 at IIT Hyderabad for local hospitality where a part of this work was done.

\appendix

\section{Relevant decay widths}
\label{Appen2}

\begin{figure}[!t]
    \centering
\includegraphics[width=0.45\linewidth]{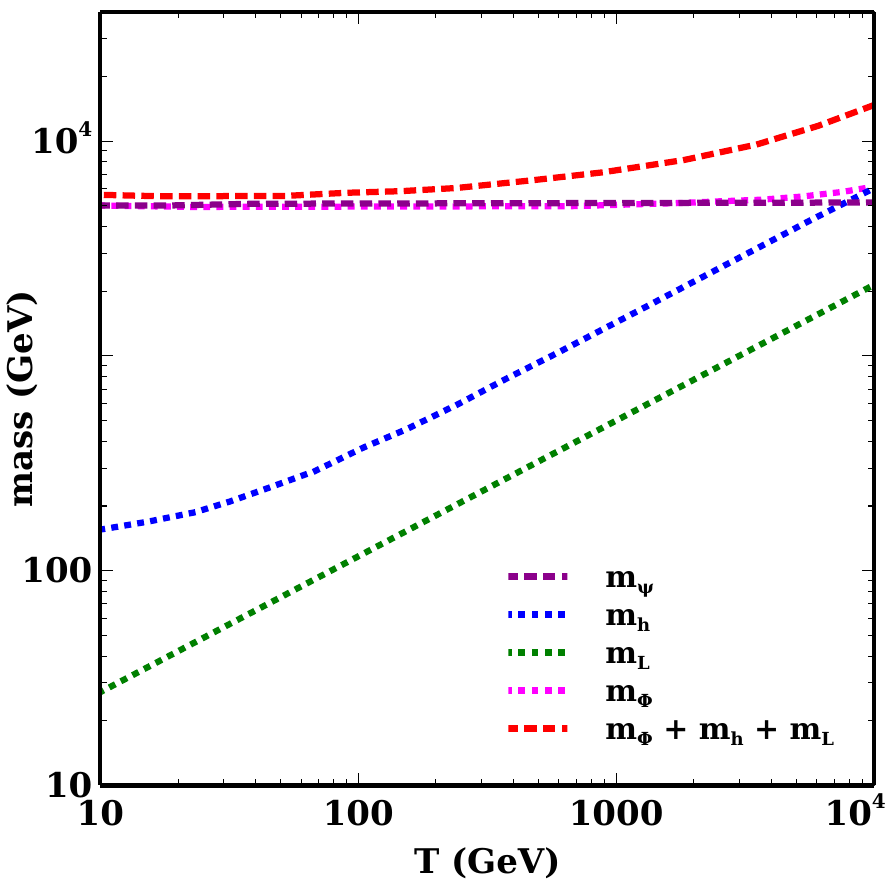}
\includegraphics[width=0.45\linewidth]{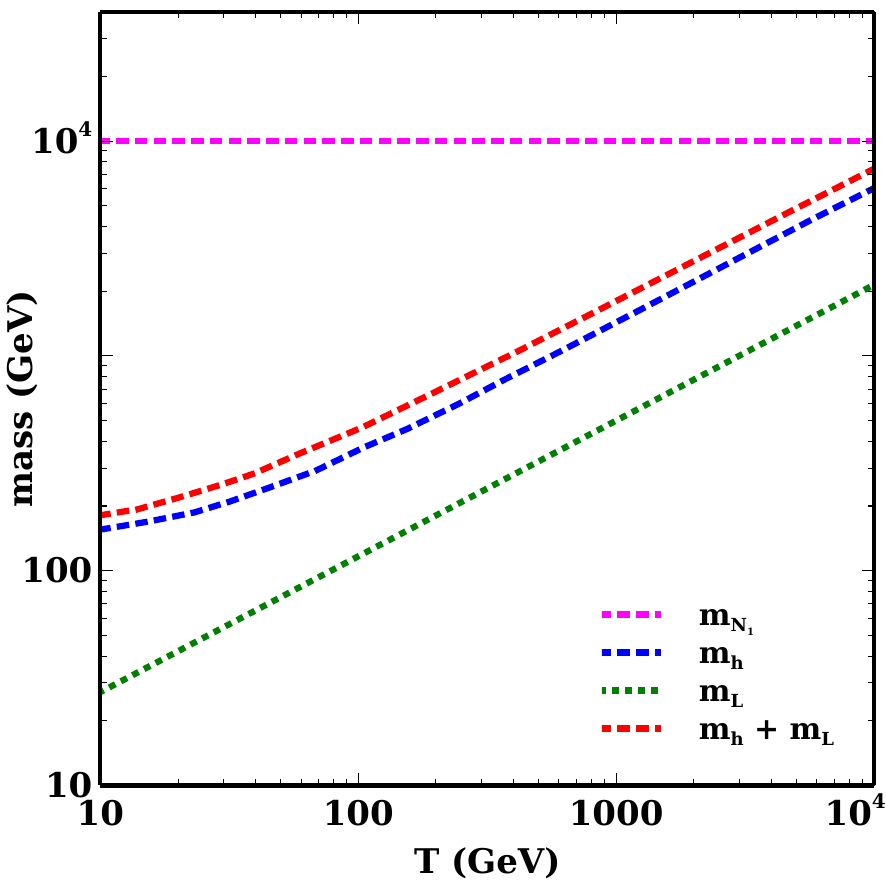}
    \caption{Thermal masses for $\psi$, $\phi$, Higgs and SM lepton (left panel plot) and $N_{1}$, Higgs and SM lepton (right panel plot). The other relevant parameters are fixed at $m_{\phi}=5$ TeV, $m_{\psi}=5.1$ TeV, $M_{1} = 11$ TeV, $\lambda_{\eta \phi}=1$ and $\lambda_{h \phi}=10^{-3}$. }
    \label{fig:app0}
\end{figure}

The decay widths for the process $N_{1}\rightarrow L H$ and $N_{1}\rightarrow \phi \psi$ are written as 
\begin{align}
    \Gamma_{(N_1 \rightarrow L H)} =& \frac{M_{1}}{8 \pi} (h^{\dagger}h)_{11} \left(1-\frac{m_{H}^{2}}{M_{1}^{2}}\right)^{2} \, , \\
    \Gamma_{(N_1 \rightarrow \phi \psi)} =&   \frac{(m_{\psi}^2 - 2 m_{\psi} M_1 + M_1^2 - m_\phi^2)\sqrt{m_{\psi}^4 + (M_{1}^{2} - m_{\phi}^{2})^{2}-2 m_{\psi}^{2}(M_{1}^{2} + m_{\phi}^{2})}y_{1}^{2}}{4 \pi M_{1}^{3}} \, .
 \end{align}
The decay width $\Gamma_{(\psi \rightarrow \phi \nu_{i})}$ is given by
\begin{equation}
    \Gamma_{(\psi \rightarrow \phi \nu_{i})} =  \frac{m_{\psi}}{8 \pi} \sum_{j=1}^{2} (y_{j}V_{ij})(y_{j}V_{ij})^{*}\bigg(1-\frac{m_{\phi}^{2}}{m_{\Psi}^{2}}\bigg)^{2} \, .
    \end{equation}
 Here $V=M_{D}M^{-1}$ is the mixing matrix between active and sterile neutrinos. The total mass of SM Higgs after taking finite temperature contribution is 
\begin{equation}
    m^{2}_{h}(T) \to m_{h}^{2} + \Pi_{h}(T) \, ,
\end{equation}
where 
\begin{equation}
    \Pi_{h}(T) = T^{2}\bigg(\frac{3}{16}g^{2}+\frac{1}{16}g'^{2}+\frac{1}{4}\lambda_{H} + \frac{1}{12}\lambda_{H\phi}+ \frac{1}{4}y_{t}^{2}+\frac{1}{4}y_{b}^{2}+\frac{1}{12}y_{\tau}^{2}\bigg) \, ,
\end{equation}
and that of SM leptons is~\cite{Giudice:2003jh}
\begin{equation}
    m^{2}_{L}(T) \to m_{L}^{2} + \frac{1}{2}\Pi_{L}(T) \, ,
\end{equation}
where 
\begin{equation}
    \Pi_{L}(T) = T^{2}\bigg(\frac{1}{16}g'^{2}+\frac{3}{16}g^{2}\bigg) \, .
\end{equation}
Similarly, the thermal mass of $\phi$ is given as 
\begin{equation}
    m^{2}_{\phi}(T) \to m_{\phi}^{2} + \Pi_{\phi}(T) \, ,
\end{equation}
where 
\begin{equation}
    \Pi_{\phi}(T) = T^{2}\bigg(\frac{1}{4}\lambda_{\phi} + \frac{1}{3}\lambda_{H\phi}+ \frac{1}{12}\lambda_{\eta\phi}\bigg) \, .
\end{equation}
Fig.~\ref{fig:app0} shows the finite-temperature masses of $\phi, \psi, L, H, N_1$ for a fixed choice of other parameters. As we can see from the left panel plot, the three-body decay $\psi \rightarrow \phi L H$ remains forbidden at high temperatures. This ensures the absence of any additional source of lepton asymmetry from three-body decay of $\psi$~\cite{Borah:2020ivi}. The usual two-body decay of $N_1$ is however allowed, as in the case of type-I seesaw leptogenesis.

The thermal averaged cross-section is defined as~\cite{Gondolo:1990dk}
\begin{equation}
    \langle \sigma v \rangle_{ij \rightarrow kl} = \frac{1}{8Tm^2_i m^2_j \kappa_2 (z_i) \kappa_2 (z_j)} \int^{\infty}_{(m_i+m_j)^2} ds \frac{\lambda (s, m^2_i, m^2_j)}{\sqrt{s}} \kappa_1 (\sqrt{s}/T) \sigma \, ,
\end{equation}
with $z_i=m_i/T$ and $\lambda (s, m^2_i, m^2_j)=[s-(m_i+m_j)^2][s-(m_i-m_j)^2]$. Here $\sigma$ is the cross-section for the process $i+j \to k+l$ with $\kappa_{i}$ being the modified Bessel functions of order $i$. The thermally-averaged reaction densities for the $2\to 2$ scattering processes are defined as 
\begin{eqnarray}
    \gamma_{ij\to kl } = \frac{ T}{64 \pi^{4}} \int_{(m_{i}+m_{j})^{2}}^{\infty} ds \sqrt{s}\kappa_{1} \left( \sqrt{s}/T \right) \hat{\sigma}(s),
\end{eqnarray}
where $\hat{\sigma}(s)$ is the reduced cross-section for the process and is given by 
\begin{equation}
\hat{\sigma} (s)= 8 \left[ (p_{i}.p_{j})^{2}-m_{i}^{2}m_{j}^{2} \right] \sigma (s).
\end{equation}
The reaction rate density $\gamma_{ij \to kl}$ is related to $\langle  \sigma v \rangle_{ij\to kl}$ as
\begin{equation}
\gamma_{ij \to kl}=n_{i}^{\rm eq}n_{j}^{\rm eq}\langle  \sigma v \rangle_{ij \to kl} \, ,
\end{equation}
where $n_{i}^{\rm eq}$ and $n_{j}^{\rm eq}$ are the equilibrium number densities of species $i$ and $j$ respectively. For decay $A\to B$ the reaction densities are defined as 
\begin{equation}
    \gamma_{A\to B}=n_{A}^{\rm eq} \frac{\kappa_{1}\left( m_{A}/T \right)}{\kappa_{2} \left(  m_{B}/T \right)}\Gamma_{A\to B}.
\end{equation}


\section{Calculation of CP Asymmetry}
\label{Appen1}
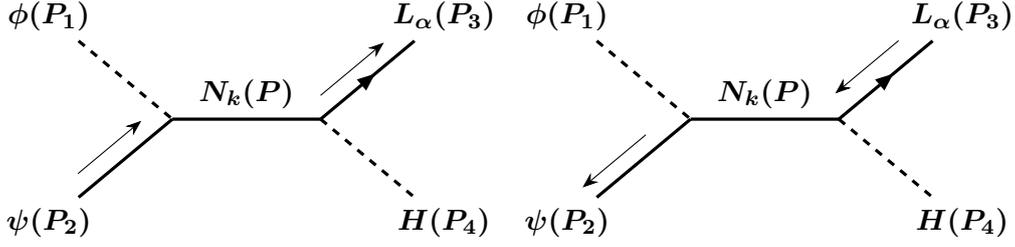
\begin{figure}
    \centering
    \begin{tikzpicture}[scale=0.65]
  \begin{feynman}
    \vertex (i1) at (-2.5, 2.1) {\(\boldsymbol{\phi(P_1)}\)};
    \vertex (i2) at (-2.5, -2.1) {\(\boldsymbol{\psi(P_2)}\)};
    
    \vertex (v1) at (0, 0);  
    \vertex (v2) at (3, 0); 

    \vertex (o1) at (5.5, 2.1) {\(\boldsymbol{L_{\alpha}(P_3)}\)};
    \vertex (o2) at (5.5, -2.1) {\(\boldsymbol{H(P_4)}\)};
    
    \diagram* {
      (i1) -- [scalar, very thick] (v1) -- [plain, very thick, edge label=\(\boldsymbol{N_{k}(P)}\)] (v2) -- [fermion, very thick,momentum={[arrow shorten=0.15,xshift=0.2pt,yshift=-2pt]}] (o1),
      (i2) -- [plain, very thick,momentum={[arrow shorten=0.15,xshift=0pt,yshift=-2pt]}] (v1),
      (v2) -- [scalar, very thick] (o2),
    };
 
  \end{feynman}
\end{tikzpicture}
    \begin{tikzpicture}[scale=0.65]
  \begin{feynman}
    \vertex (i1) at (-2.5, 2.1) {\(\boldsymbol{\phi(P_1)}\)};
    \vertex (i2) at (-2.5, -2.1) {\(\boldsymbol{\psi(P_2)}\)};
    
    \vertex (v1) at (0, 0);  
    \vertex (v2) at (3, 0); 

    \vertex (o1) at (5.5, 2.1) {\(\boldsymbol{L_{\alpha}(P_3)}\)};
    \vertex (o2) at (5.5, -2.1) {\(\boldsymbol{H(P_4)}\)};
    
    \diagram* {
      (i1) -- [scalar, very thick] (v1) -- [plain, very thick, edge label=\(\boldsymbol{N_{k}(P)}\)] (v2) -- [fermion, very thick, reversed momentum={[arrow shorten=0.15,xshift=-2pt,yshift=-2pt]}] (o1),
      (i2) -- [plain, very thick, reversed momentum={[arrow shorten=0.15,xshift=-6pt,yshift=-8pt]}] (v1),
      (v2) -- [scalar, very thick] (o2),
    };
 
  \end{feynman}
\end{tikzpicture}
    \caption{Feynman diagrams of the process $\phi \psi \to LH$.}
    \label{f1}
\end{figure}

To calculate the CP asymmetry we consider the coannihilation of $\phi$ and $\psi$ as shown Fig.~\ref{f1}. The amplitude for the left panel plot of Fig.~\ref{f1} can be written as~\cite{Denner:1992vza}
\begin{eqnarray}
 \mathcal{M}_{1} &= -  y_{k} h_{\alpha k} [\bar{u}(P_{3})S(P)u(P_{2})] .
\end{eqnarray}
The amplitude for the right panel plot of Fig.~\ref{f1} can be written as
\begin{eqnarray}
  \mathcal{M}_{2} &=&   -  y_{k} h_{\alpha k}[\bar{u(P_{3})}S(-P)u(P_{2})] .
\end{eqnarray}

\noindent Total amplitude for the process $\phi \psi \to L H$ mediated by $N_{k}$ is
\begin{equation}
\begin{aligned}
    \mathcal{M}_k &= \mathcal{M}_1 + \mathcal{M}_2
    = -  y_{k} h_{\alpha k}\bar{u}(P_3)\bigg[\frac{2 M_k}{s - M_{k}^{2}+ i M_k\Gamma_k}\bigg]u(P_2) \, .
\end{aligned}
\end{equation}
Similarly, the amplitude for the same process mediated by $N_{m}$
\begin{equation}
\begin{aligned}
    \mathcal{M}_m = -  y_{m} h_{\alpha m}\bar{u}(P_3)\bigg[\frac{2 M_m}{s - M_{m}^{2}+ i M_m\Gamma_m}\bigg]u(P_2)  \, .
\end{aligned}
\end{equation}
The total amplitude for the process $\phi \psi \to L H$ gets contribution from the $N_{k}$ and $N_{m}$ mediated diagrams and is given by 
\begin{eqnarray}
      \mathcal{M} &=&  \mathcal{M}_k +  \mathcal{M}_m \, .
\end{eqnarray}
The asymmetry parameter at the amplitude level is given by 
\begin{eqnarray}
     \delta &=& |\mathcal{M}|^2 - |\overline{\mathcal{M}}|^2 \nonumber \\
    &=& -4 \text{Im}[C_1C^{*}_2]\text{Im}[\mathfrak{M}_k\mathfrak{M}^{*}_m]|\mathcal{W}|^{2} \nonumber \\
    &=& -4 \text{Im}[y_k h_{\alpha k}y^{*}_{m}h_{\alpha m}^{*}] \text{Im}[\mathfrak{M}_k\mathfrak{M}^{*}_m]|\mathcal{W}|^{2} \, . 
\end{eqnarray}
Here, $\mathfrak{M}_{k,m}$ are the contribution of the propagator to the amplitudes $\mathcal{M}_{k,m}$ and $\mathcal{W}$ is the contribution of the wavefunction contribution to the amplitudes. The imaginary part coming from the propagator can be calculated as  
\begin{equation}
\begin{aligned}
    \text{Im}[\mathfrak{M}_k\mathfrak{M}^{*}_m] &= 
   4M_kM_m \times \frac{M_m\Gamma_m (s - M^{2}_k)-  M_k\Gamma_k (s - M^{2}_m)}{((s-M^{2}_k)^2+ M^{2}_k\Gamma_k^2)((s-M^{2}_m)^2+ M^{2}_m\Gamma_m^2)}.
\end{aligned}    
\end{equation}
The wavefunction contribution to the amplitude is 
\begin{equation}
\begin{aligned}
    \mathcal{W} &= \bar{u}(P_3)u(P_2), \\
    \overline{\mathcal{W}} &= \bar{u}(P_2)u(P_3).\\
    |\mathcal{W}|^2 &= \frac{1}{2}  \sum_{\alpha,\beta} (\bar{u}_{\alpha}(P_2)u_{\alpha}(P_3))\bar{u}_{\beta}(P_3)u_\beta(P_2)
    =  (P_3 . P_2).
\end{aligned}    
\end{equation}
The asymmetry at the amplitude level is calculated to be
\begin{equation}
\begin{aligned}
    \delta = -16 \times \text{Im}[y_k h_{\alpha k}y^{*}_{m}h_{\alpha m}^{*}] M_k M_m \frac{M_m\Gamma_m  (s - M^{2}_k)-  M_k\Gamma_k (s - M^{2}_m)}{((s-M^{2}_k)^2+ M^{2}_k\Gamma_k^2)((s-M^{2}_m)^2+ M^{2}_m\Gamma_m^2)} (P_3 . P_2).  
\end{aligned}    
\end{equation}
In the CM frame 
the asymmetry parameter $\delta$ is written in terms of the $s$ parameter as
\begin{equation}
\begin{aligned}
    \delta &= -16 \times \text{Im}[y_k h_{\alpha k}y^{*}_{m}h_{\alpha m}^{*}] M_k M_m \frac{M_m\Gamma_m  (s - M^{2}_k)-  M_k\Gamma_k (s - M^{2}_m)}{((s-M^{2}_k)^2+ M^{2}_k\Gamma_k^2)((s-M^{2}_m)^2+ M^{2}_m\Gamma_m^2)} \\ 
    &\times \frac{\sqrt{s}}{2}\bigg[\sqrt{\frac{[s- (m_\phi+m_\psi)^2][s- (m_\phi-m_\psi)^2]}{4 s}} + \sqrt{m^{2}_{\psi} + \frac{[s- (m_\phi+m_\psi)^2][s- (m_\phi-m_\psi)^2]}{4 s}}\bigg]  .
    \label{B10}
\end{aligned}
\end{equation}

\noindent  Here we define the important quantity used in the Boltzmann equations $\langle \sigma v \rangle^{\delta}_{\phi \psi \to L H}$. It is the difference of thermally averaged cross-section of the process $\phi \psi \to L H$ and its Hermitian conjugate, and is given by 
\begin{eqnarray}
    \langle  \sigma v \rangle_{\phi \psi \to L H}^{\delta}  & = & \dfrac{1}{2Tm_{\phi}^{2} \kappa_{2}(m_{\phi}/T)m_{\psi}^{2} \kappa_{2}(m_{\psi}/T)  } \int_{s_{\rm in}}^{\infty} \int_{-1}^{1} \dfrac{1}{32\pi} \dfrac{\delta}{\sqrt{s}} \nonumber  \\ & & \times p_{\phi \psi} p_{LH} \kappa_{1}\left( \sqrt{s}/T  \right)  ds \, d(\cos \theta)   . 
\end{eqnarray}
The function $p_{i j }$ and $s_{\rm in}$ are given by
\begin{eqnarray}
    p_{ij} & = & \dfrac{1}{2}\sqrt{\lambda(s,m_{i}^{2},m_{j}^{2})/s}, \\
     s_{\rm in} & = & {\rm max} \left[  (m_{\phi}+m_{\psi})^{2},(m_{L}+m_{H})^{2}  \right].
\end{eqnarray}

\begin{figure}
    \centering
        \begin{tikzpicture}[scale=0.64]
  \begin{feynman}
    \vertex (i1) at (-2.5, 2.1) {\(\boldsymbol{\phi(P_1)}\)};
    \vertex (i2) at (-2.5, -2.1) {\(\boldsymbol{\psi(P_2)}\)};
    
    \vertex (v1) at (0, 0);  
    \vertex (v2) at (3, 0); 

    \vertex (o1) at (5.5, 2.1) {\(\boldsymbol{L_{\alpha}(P_3)}\)};
    \vertex (o2) at (5.5, -2.1) {\(\boldsymbol{H(P_4)}\)};
    
    \diagram* {
      (i1) -- [scalar, very thick] (v1) -- [plain, very thick, edge label=\(\boldsymbol{N_{k}(P)}\)] (v2) -- [fermion, very thick,momentum={[arrow shorten=0.15,xshift=0.2pt,yshift=-2pt]}] (o1),
      (i2) -- [plain, very thick,momentum={[arrow shorten=0.15,xshift=0pt,yshift=-2pt]}] (v1),
      (v2) -- [scalar, very thick] (o2),
    };
 
  \end{feynman}
\end{tikzpicture}
    \begin{tikzpicture}[scale=0.6]
  \begin{feynman}
    \vertex (i1) at (-2.5, 2.1) {\(\boldsymbol{\phi (P_1)}\)};
    \vertex (i2) at (-2.5, -2.1) {\(\boldsymbol{\psi (P_2)}\)};
    \vertex (v1) at (0, 0);
    \vertex (v2) at (3, 0);
    \vertex (o11) at (5.5, 1);
    \vertex (o22) at (5.5, -1);
    \vertex (o1) at (8.5, 2.5) {\(\boldsymbol{L_{\alpha} (P_3)}\)};
    \vertex (o2) at (8.5, -2.5) {\(\boldsymbol{H (P_4)}\)};
    
    \diagram* {
      (i1) -- [scalar, very thick] (v1),
      (i2) -- [plain, very thick, momentum={[arrow shorten=0.15, yshift=-2pt]}] (v1),
      (v1) -- [plain, very thick, edge label=\(\boldsymbol{N_{k^{'}}(P)}\)] (v2),
      (v2) -- [scalar, very thick, edge label=\(\boldsymbol{H}\),momentum={[arrow shorten=0.3, xshift=5pt,yshift=-3pt]}] (o11),
      (o11) -- [fermion, very thick, momentum={[arrow shorten=0.15, yshift=-2pt]}] (o1),
      
      (v2) -- [anti fermion, very thick, edge label'=\(\boldsymbol{L}\)] (o22),
      (o22) -- [momentum={[arrow shorten=0.3, yshift=4pt]}] (v2),
      (o22) -- [scalar, very thick, momentum={[arrow shorten=0.15, yshift=-16pt]}] (o2),
      
      (o11) -- [plain, thick, edge label'=\(\boldsymbol{N_m}\), momentum={[arrow shorten=0.15, yshift=-2pt]}] (o22),
    };
  \end{feynman}
\end{tikzpicture}
    \caption{CP asymmetry from vertex correction.}
    \label{fig11}
\end{figure}
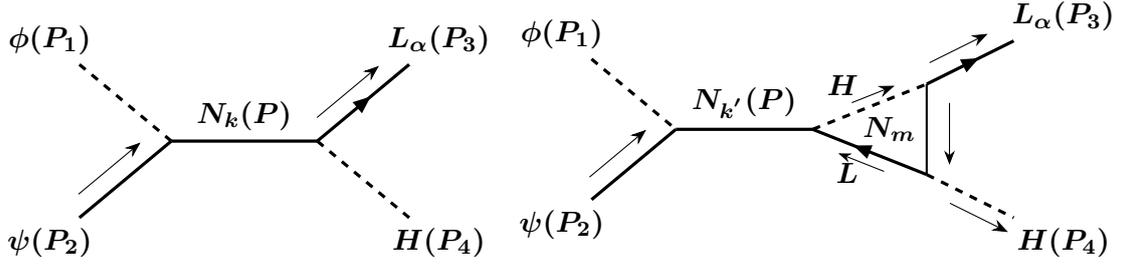

Similarly, the CP asymmetry from the vertex correction (as shown in Fig.~\ref{fig11}) can be found as 
\begin{eqnarray}
    \delta &=& -\dfrac{1}{8\pi} \sum_{k,k^{'}=1}^{2}\dfrac{1}{(s-M_{k}^{2})(s-M_{k^{'}}^{2})}M_{k}m_{\psi} \sum_{\alpha,\beta} \sum_{m\neq k^{'}}{\rm Im}\left[  h_{\alpha m} h_{\beta m} h_{\beta k^{'}}h^{*}_{\alpha k} y_{k} y_{k^{'}}\right] \nonumber \\ & &  \sqrt{s}  \bigg[ \sqrt{m_{\psi}^{2}+p_{\phi \psi}^{2}} ln \left(\dfrac{1+z_{m}}{z_{m}} \right) + p_{\psi}  \cos \theta \left( 4-(1+4z_{m})ln \left( \dfrac{1+z_{m}}{z_{m}} \right) \right)    \nonumber \\ & &  -\pi  p_{\phi \psi}  \left( 1+2z_{m}-2\sqrt{z_{m}(1+z_{m})}  \right) \sin \theta -2 M_{k^{'}}\sqrt{z_{m}} \left( 1+z_{m} ln \left( \frac{1+z_{m}}{z_{m}}  \right)  \right) \bigg]. \nonumber \\
\end{eqnarray}
Here, the dimensionless parameter $z$ is defined as $z_{m}=M_{m}^{2}/s$. 



\begin{figure}[t!]
    \centering
\includegraphics[scale=0.49]{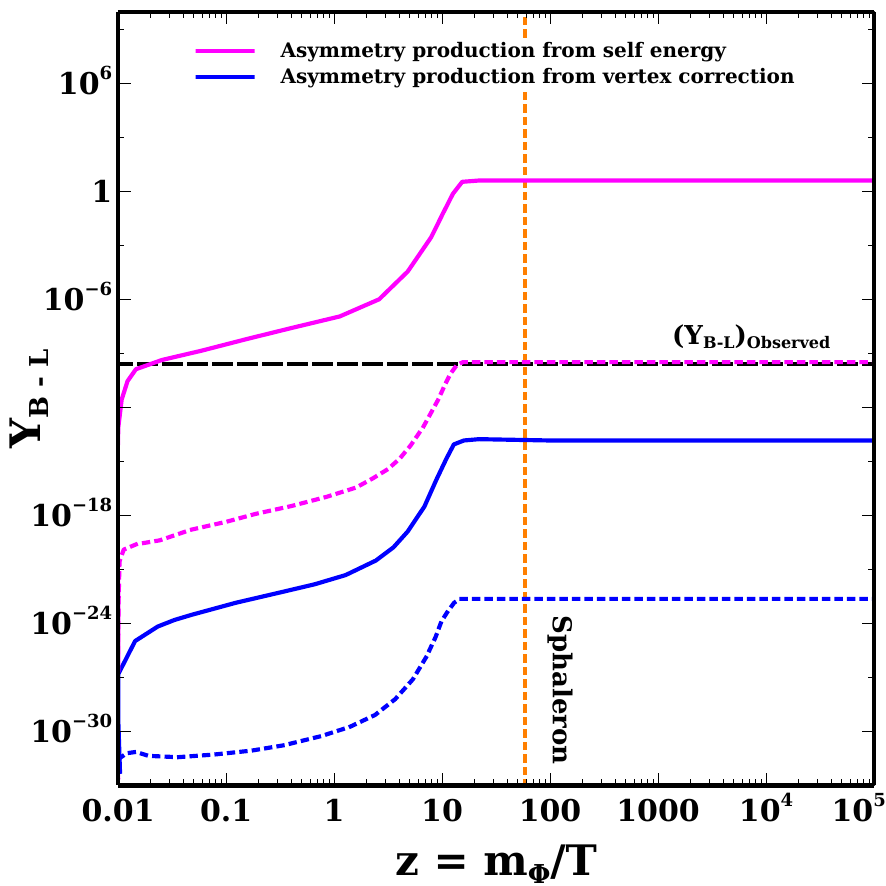}
    \caption{Comparison of the $B-L$ asymmetry generated from the self energy and vertex contributions to the scattering $\psi \phi \to L H$. Here $M_{1}=m_{\phi}+m_{\psi}$ and $M_{2}=M_{1}+0.1$ keV and for the dashed lines $M_{2}=M_{1}+1$ GeV. The relevant parameters are fixed at $M_{1}=8$ TeV, $m_{\phi}=3.5$ TeV, $m_{\psi}=4.5$ TeV, $y_{1}=y_{2}=0.06+0.06i$, $\lambda_{\eta \phi}=1$ and $y_{\eta}=0.1$.  }
    \label{fig:app1}
\end{figure}

Fig.~\ref{fig:app1} shows the comparison between the CP asymmetry contributions from self-energy correction and vertex corrections. Clearly, in the resonant limit, the self-energy correction dominates. Even for mass splitting larger than typical resonant limits for TeV scale leptogenesis, the self-energy correction dominates. 

\begin{figure}[h]
    \centering
    \includegraphics[scale=0.5]{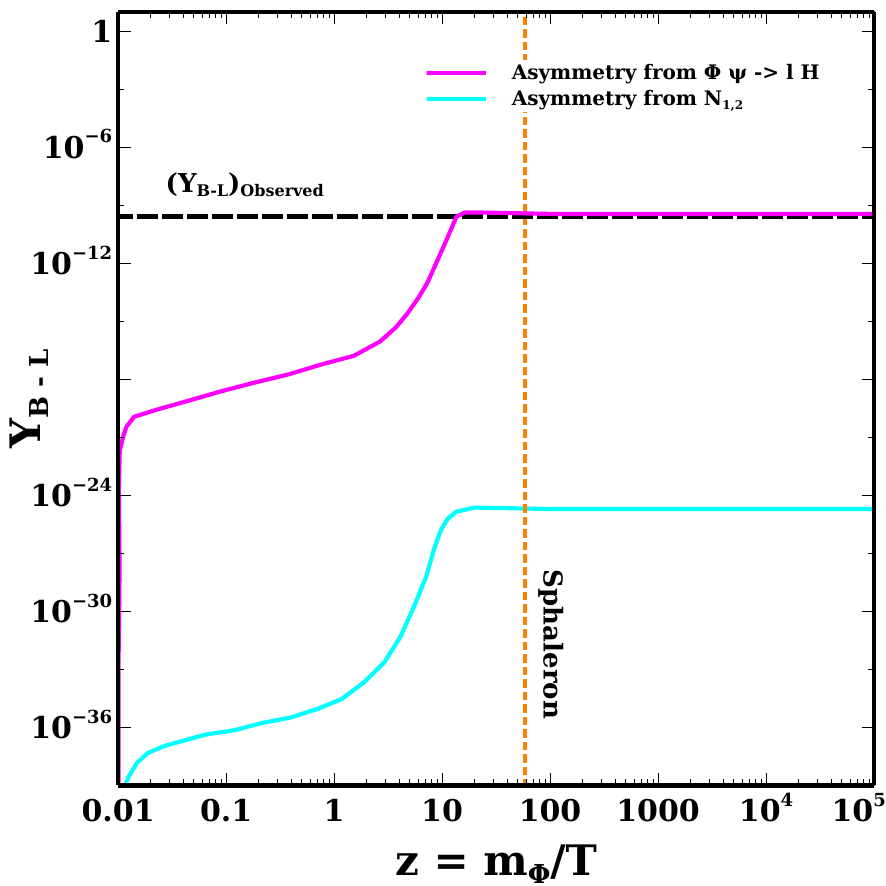}
    \caption{Comparison plot of the $B-L$ asymmetry generated from the decay of $N_{1,2}$ and the scattering $\psi \phi \to L H$. The relevant parameters are fixed at $m_{\phi}=3.5$ TeV, $m_{\psi}=4.5$ TeV, $M_{1}=8$ TeV, $M_{2}=M_{1}+1$ GeV, $y_{1}=y_{2}=0.06+0.06i$, $\lambda_{\eta \phi}=1$ and $y_{\eta}=0.1$.}
    \label{fig:app2}
\end{figure}

Finally, the CP asymmetry from heavy RHN $N_i$ decay is given by~\cite{Luty:1992un, Flanz:1994yx, Covi:1996wh, Pilaftsis:2003gt}
\begin{align}
    \epsilon_{i \alpha} & = \dfrac{\Gamma_{(N_{i}\to \sum_{\alpha} L_{\alpha} H )}-\Gamma_{(N_{i}\to \sum_{\alpha} L_{\alpha}^{c}H)}}{\Gamma_{(N_{i}\to \sum_{\alpha}L_{\alpha}H)}+\Gamma_{(N_{i}\to \sum_{i}L_{\alpha}^{c}H)}}  \\ 
    & = \frac{1}{8 \pi (h^{\dagger}h)_{ii}} \sum_{j\neq i} \bigg [ \mathcal{F} \left( \frac{M^2_j}{M^2_i} \right) {\rm Im} [ h^*_{\alpha i} h_{\alpha j} (h^{\dagger} h)_{ij}] + \mathcal{G} \left( \frac{M^2_j}{M^2_i} \right) {\rm Im} [ h^*_{\alpha i} h_{\alpha j} (h^{\dagger} h)_{ij}^*]]\, , \label{eqn:CPasy}
\end{align}
where $\mathcal{F}(x)$ and $\mathcal{G}(x)$ are the loop functions defined as
\begin{align}
    \mathcal{F}(x)= \sqrt{x}\left[1+\frac{1}{1-x}+(1+x){\rm ln}\left(\frac{x}{1+x}\right)\right]\,, \qquad  \mathcal{G}(x)= \frac{1}{1-x}.
\end{align}
Ignoring the effect of lepton flavors~\cite{Abada:2006fw,Nardi:2006fx,Abada:2006ea,Blanchet:2006be, Dev:2014pfm, Dev:2017trv}, Eq.~\eqref{eqn:CPasy} simplifies to
\begin{align}
    \epsilon_i =\sum_\alpha \epsilon_{i\alpha} = \frac{1}{8 \pi (h^{\dagger}h)_{ii}}\sum_{j\neq i} \mathcal{F} \left( \frac{M^2_j}{M^2_i} \right) {\rm Im} [ (h^{\dagger} h)_{ij}^2]. \label{eqn:CPasynoflav}
\end{align}
In the resonant limit $|M_j-M_i| \sim \Gamma_i$, one can use the general formula for CP asymmetry dominantly sourced by self-energy correction~\cite{Pilaftsis:2003gt}:
\begin{eqnarray}
\epsilon_{i} = \sum_{i\neq j}\dfrac{{\rm Im}[(h^{\dagger}h)_{ij}^{2}]}{(h^{\dagger}h)_{ii}(h^{\dagger}h)_{jj}}\dfrac{(M_{i}^{2}-M_{j}^{2})M_{i}\Gamma_{j}}{(M_{i}^{2}-M_{j}^{2})^{2}+M_{i}^{2}\Gamma_{j}^{2}}. \label{eq:asymmparameter}
\end{eqnarray}
Fig.~\ref{fig:app2} shows the comparison between lepton asymmetry originating from RHN decay and that from DM coannihilations. Even in the resonant limit, the lepton asymmetry from RHN decay gets washed out due to lighter $\phi, N$ with large Yukawa coupling $y_1$. For non-resonant $N_1, N_2$, lepton asymmetry can still be generated at low scale from DM coannihilation as opposed to the type-I seesaw leptogenesis where the scale has a lower bound $M_1 \gtrsim 10^9$ GeV~\cite{Davidson:2002qv}.

\bibliographystyle{JHEP}
\bibliography{ref}

@article{Minkowski:1977sc,
    author = "Minkowski, Peter",
    title = "{$\mu \to e\gamma$ at a Rate of One Out of $10^{9}$ Muon Decays?}",
    reportNumber = "Print-77-0182 (BERN)",
    doi = "10.1016/0370-2693(77)90435-X",
    journal = "Phys. Lett. B",
    volume = "67",
    pages = "421--428",
    year = "1977"
}

@article{Denner:1992vza,
    author = "Denner, Ansgar and Eck, H. and Hahn, O. and Kublbeck, J.",
    title = "{Feynman rules for fermion number violating interactions}",
    reportNumber = "CERN-TH-6549-92",
    doi = "10.1016/0550-3213(92)90169-C",
    journal = "Nucl. Phys. B",
    volume = "387",
    pages = "467--481",
    year = "1992"
}

@article{Dev:2017wwc,
    author = "Dev, P. S. Bhupal and Garny, Mathias and Klaric, Juraj and Millington, Peter and Teresi, Daniele",
    title = "{Resonant enhancement in leptogenesis}",
    eprint = "1711.02863",
    archivePrefix = "arXiv",
    primaryClass = "hep-ph",
    reportNumber = "TUM-HEP-1110-17",
    doi = "10.1142/S0217751X18420034",
    journal = "Int. J. Mod. Phys. A",
    volume = "33",
    pages = "1842003",
    year = "2018"
}

@article{Chen:2013foz,
    author = "Chen, Chien-Yi and Dev, P. S. Bhupal and Mohapatra, R. N.",
    title = "{Probing Heavy-Light Neutrino Mixing in Left-Right Seesaw Models at the LHC}",
    eprint = "1306.2342",
    archivePrefix = "arXiv",
    primaryClass = "hep-ph",
    reportNumber = "MAN-HEP-2013-09",
    doi = "10.1103/PhysRevD.88.033014",
    journal = "Phys. Rev. D",
    volume = "88",
    pages = "033014",
    year = "2013"
}

@article{Yanagida:1979as,
    author = "Yanagida, Tsutomu",
    editor = "Sawada, Osamu and Sugamoto, Akio",
    title = "{Horizontal gauge symmetry and masses of neutrinos}",
    reportNumber = "KEK-79-18-95",
    journal = "Conf. Proc. C",
    volume = "7902131",
    pages = "95--99",
    year = "1979"
}

@article{Casas:2001sr,
    author = "Casas, J. A. and Ibarra, A.",
    title = "{Oscillating neutrinos and $\mu \to e, \gamma$}",
    eprint = "hep-ph/0103065",
    archivePrefix = "arXiv",
    reportNumber = "IEM-FT-211-01, OUTP-01-11P, IFT-UAM-CSIC-01-08",
    doi = "10.1016/S0550-3213(01)00475-8",
    journal = "Nucl. Phys. B",
    volume = "618",
    pages = "171--204",
    year = "2001"
}

@article{Giudice:2003jh,
    author = "Giudice, G. F. and Notari, A. and Raidal, M. and Riotto, A. and Strumia, A.",
    title = "{Towards a complete theory of thermal leptogenesis in the SM and MSSM}",
    eprint = "hep-ph/0310123",
    archivePrefix = "arXiv",
    reportNumber = "IFUP-TH-2003-37, CERN-TH-2003-240",
    doi = "10.1016/j.nuclphysb.2004.02.019",
    journal = "Nucl. Phys. B",
    volume = "685",
    pages = "89--149",
    year = "2004"
}

@article{Borah:2025wzl,
    author = "Borah, Debasish and Saha, Indrajit",
    title = "{Filtered cogenesis of PBH dark matter and baryons}",
    eprint = "2502.12248",
    archivePrefix = "arXiv",
    primaryClass = "hep-ph",
    month = "2",
    year = "2025"
}

@article{Cirelli:2024ssz,
    author = "Cirelli, Marco and Strumia, Alessandro and Zupan, Jure",
    title = "{Dark Matter}",
    eprint = "2406.01705",
    archivePrefix = "arXiv",
    primaryClass = "hep-ph",
    month = "6",
    year = "2024"
}

@article{Edsjo:1997bg,
    author = "Edsjo, Joakim and Gondolo, Paolo",
    title = "{Neutralino relic density including coannihilations}",
    eprint = "hep-ph/9704361",
    archivePrefix = "arXiv",
    reportNumber = "UUITP-11-97, MPI-PHT-97-27",
    doi = "10.1103/PhysRevD.56.1879",
    journal = "Phys. Rev. D",
    volume = "56",
    pages = "1879--1894",
    year = "1997"
}

@article{Ibarra:2003up,
    author = "Ibarra, A. and Ross, Graham G.",
    title = "{Neutrino phenomenology: The Case of two right-handed neutrinos}",
    eprint = "hep-ph/0312138",
    archivePrefix = "arXiv",
    reportNumber = "CERN-TH-2003-294, OUTP-0333P",
    doi = "10.1016/j.physletb.2004.04.037",
    journal = "Phys. Lett. B",
    volume = "591",
    pages = "285--296",
    year = "2004"
}

@article{Harvey:1990qw,
    author = "Harvey, Jeffrey A. and Turner, Michael S.",
    title = "{Cosmological baryon and lepton number in the presence of electroweak fermion number violation}",
    reportNumber = "FERMILAB-PUB-90-049-A, EFI-90-33",
    doi = "10.1103/PhysRevD.42.3344",
    journal = "Phys. Rev. D",
    volume = "42",
    pages = "3344--3349",
    year = "1990"
}

@article{Alguero:2023zol,
    author = "Alguero, G. and Belanger, G. and Boudjema, F. and Chakraborti, S. and Goudelis, A. and Kraml, S. and Mjallal, A. and Pukhov, A.",
    title = "{micrOMEGAs 6.0: N-component dark matter}",
    eprint = "2312.14894",
    archivePrefix = "arXiv",
    primaryClass = "hep-ph",
    doi = "10.1016/j.cpc.2024.109133",
    journal = "Comput. Phys. Commun.",
    volume = "299",
    pages = "109133",
    year = "2024"
}

@article{Luty:1992un,
    author = "Luty, M. A.",
    title = "{Baryogenesis via leptogenesis}",
    doi = "10.1103/PhysRevD.45.455",
    journal = "Phys. Rev. D",
    volume = "45",
    pages = "455--465",
    year = "1992"
}

@article{Flanz:1994yx,
    author = "Flanz, Marion and Paschos, Emmanuel A. and Sarkar, Utpal",
    title = "{Baryogenesis from a lepton asymmetric universe}",
    eprint = "hep-ph/9411366",
    archivePrefix = "arXiv",
    reportNumber = "DO-TH-94-15",
    doi = "10.1016/0370-2693(94)01555-Q",
    journal = "Phys. Lett. B",
    volume = "345",
    pages = "248--252",
    year = "1995",
    note = "[Erratum: Phys.Lett.B 384, 487--487 (1996), Erratum: Phys.Lett.B 382, 447--447 (1996)]"
}

@article{Covi:1996wh,
    author = "Covi, Laura and Roulet, Esteban and Vissani, Francesco",
    title = "{CP violating decays in leptogenesis scenarios}",
    eprint = "hep-ph/9605319",
    archivePrefix = "arXiv",
    reportNumber = "SISSA-66-96-EP, IC-96-73",
    doi = "10.1016/0370-2693(96)00817-9",
    journal = "Phys. Lett. B",
    volume = "384",
    pages = "169--174",
    year = "1996"
}

@article{Borah:2022cdx,
    author = "Borah, Debasish and Dasgupta, Arnab and Saha, Indrajit",
    title = "{Leptogenesis and dark matter through relativistic bubble walls with observable gravitational waves}",
    eprint = "2207.14226",
    archivePrefix = "arXiv",
    primaryClass = "hep-ph",
    doi = "10.1007/JHEP11(2022)136",
    journal = "JHEP",
    volume = "11",
    pages = "136",
    year = "2022"
}

@article{Jungman:1995df,
    author = "Jungman, Gerard and Kamionkowski, Marc and Griest, Kim",
    title = "{Supersymmetric dark matter}",
    eprint = "hep-ph/9506380",
    archivePrefix = "arXiv",
    reportNumber = "SU-4240-605, UCSD-PTH-95-02, IASSNS-HEP-95-14, CU-TP-677",
    doi = "10.1016/0370-1573(95)00058-5",
    journal = "Phys. Rept.",
    volume = "267",
    pages = "195--373",
    year = "1996"
}

@article{Bertone:2004pz,
    author = "Bertone, Gianfranco and Hooper, Dan and Silk, Joseph",
    title = "{Particle dark matter: Evidence, candidates and constraints}",
    eprint = "hep-ph/0404175",
    archivePrefix = "arXiv",
    reportNumber = "FERMILAB-PUB-04-047-A",
    doi = "10.1016/j.physrep.2004.08.031",
    journal = "Phys. Rept.",
    volume = "405",
    pages = "279--390",
    year = "2005"
}

@article{Planck:2018vyg,
    author = "Aghanim, N. and others",
    collaboration = "Planck",
    title = "{Planck 2018 results. VI. Cosmological parameters}",
    eprint = "1807.06209",
    archivePrefix = "arXiv",
    primaryClass = "astro-ph.CO",
    doi = "10.1051/0004-6361/201833910",
    journal = "Astron. Astrophys.",
    volume = "641",
    pages = "A6",
    year = "2020",
    note = "[Erratum: Astron.Astrophys. 652, C4 (2021)]"
}

@article{XENON:2018voc,
    author = "Aprile, E. and others",
    collaboration = "XENON",
    title = "{Dark Matter Search Results from a One Ton-Year Exposure of XENON1T}",
    eprint = "1805.12562",
    archivePrefix = "arXiv",
    primaryClass = "astro-ph.CO",
    doi = "10.1103/PhysRevLett.121.111302",
    journal = "Phys. Rev. Lett.",
    volume = "121",
    number = "11",
    pages = "111302",
    year = "2018"
}

@article{GAMBIT:2017gge,
    author = "Athron, Peter and others",
    collaboration = "GAMBIT",
    title = "{Status of the scalar singlet dark matter model}",
    eprint = "1705.07931",
    archivePrefix = "arXiv",
    primaryClass = "hep-ph",
    reportNumber = "COEPP-MN-17-10, CERN-TH-2017-170, NORDITA 2017-079, CoEPP-MN-17-10, gambit-physics, gambit-physics-2017",
    doi = "10.1140/epjc/s10052-017-5113-1",
    journal = "Eur. Phys. J. C",
    volume = "77",
    number = "8",
    pages = "568",
    year = "2017"
}

@article{DOnofrio:2014rug,
    author = "D'Onofrio, Michela and Rummukainen, Kari and Tranberg, Anders",
    title = "{Sphaleron Rate in the Minimal Standard Model}",
    eprint = "1404.3565",
    archivePrefix = "arXiv",
    primaryClass = "hep-ph",
    doi = "10.1103/PhysRevLett.113.141602",
    journal = "Phys. Rev. Lett.",
    volume = "113",
    number = "14",
    pages = "141602",
    year = "2014"
}

@article{Fermi-LAT:2017opo,
    author = "Ackermann, M. and others",
    collaboration = "Fermi-LAT",
    title = "{The Fermi Galactic Center GeV Excess and Implications for Dark Matter}",
    eprint = "1704.03910",
    archivePrefix = "arXiv",
    primaryClass = "astro-ph.HE",
    doi = "10.3847/1538-4357/aa6cab",
    journal = "Astrophys. J.",
    volume = "840",
    number = "1",
    pages = "43",
    year = "2017"
}

@article{HESS:2022ygk,
    author = "Abdalla, H. and others",
    collaboration = "H.E.S.S.",
    title = "{Search for Dark Matter Annihilation Signals in the H.E.S.S. Inner Galaxy Survey}",
    eprint = "2207.10471",
    archivePrefix = "arXiv",
    primaryClass = "astro-ph.HE",
    doi = "10.1103/PhysRevLett.129.111101",
    journal = "Phys. Rev. Lett.",
    volume = "129",
    number = "11",
    pages = "111101",
    year = "2022"
}

@article{CTA:2020qlo,
    author = "Acharyya, A. and others",
    collaboration = "CTA",
    title = "{Sensitivity of the Cherenkov Telescope Array to a dark matter signal from the Galactic centre}",
    eprint = "2007.16129",
    archivePrefix = "arXiv",
    primaryClass = "astro-ph.HE",
    doi = "10.1088/1475-7516/2021/01/057",
    journal = "JCAP",
    volume = "01",
    pages = "057",
    year = "2021"
}

@article{DARWIN:2016hyl,
    author = "Aalbers, J. and others",
    collaboration = "DARWIN",
    title = "{DARWIN: towards the ultimate dark matter detector}",
    eprint = "1606.07001",
    archivePrefix = "arXiv",
    primaryClass = "astro-ph.IM",
    doi = "10.1088/1475-7516/2016/11/017",
    journal = "JCAP",
    volume = "11",
    pages = "017",
    year = "2016"
}

@article{FileviezPerez:2009hdc,
    author = "Fileviez Perez, Pavel and Han, Tao and Li, Tong",
    title = "{Testability of Type I Seesaw at the CERN LHC: Revealing the Existence of the B-L Symmetry}",
    eprint = "0907.4186",
    archivePrefix = "arXiv",
    primaryClass = "hep-ph",
    reportNumber = "MADPH-09-1534",
    doi = "10.1103/PhysRevD.80.073015",
    journal = "Phys. Rev. D",
    volume = "80",
    pages = "073015",
    year = "2009"
}

@article{Liu:2021akf,
    author = "Liu, Wei and Xie, Ke-Pan and Yi, Zihan",
    title = "{Testing leptogenesis at the LHC and future muon colliders: A Z' scenario}",
    eprint = "2109.15087",
    archivePrefix = "arXiv",
    primaryClass = "hep-ph",
    doi = "10.1103/PhysRevD.105.095034",
    journal = "Phys. Rev. D",
    volume = "105",
    number = "9",
    pages = "095034",
    year = "2022"
}

@article{Chauhan:2021xus,
    author = "Chauhan, Garv and Dev, P. S. Bhupal",
    title = "{Interplay between resonant leptogenesis, neutrinoless double beta decay and collider signals in a model with flavor and CP symmetries}",
    eprint = "2112.09710",
    archivePrefix = "arXiv",
    primaryClass = "hep-ph",
    doi = "10.1016/j.nuclphysb.2022.116058",
    journal = "Nucl. Phys. B",
    volume = "986",
    pages = "116058",
    year = "2023"
}

@article{Abdullahi:2022jlv,
    author = "Abdullahi, Asli M. and others",
    title = "{The present and future status of heavy neutral leptons}",
    eprint = "2203.08039",
    archivePrefix = "arXiv",
    primaryClass = "hep-ph",
    reportNumber = "FERMILAB-CONF-22-184-T-V",
    doi = "10.1088/1361-6471/ac98f9",
    journal = "J. Phys. G",
    volume = "50",
    number = "2",
    pages = "020501",
    year = "2023"
}

@article{LZ:2024zvo,
    author = "Aalbers, J. and others",
    collaboration = "LZ",
    title = "{Dark Matter Search Results from 4.2{\,}{\,}Tonne-Years of Exposure of the LUX-ZEPLIN (LZ) Experiment}",
    eprint = "2410.17036",
    archivePrefix = "arXiv",
    primaryClass = "hep-ex",
    reportNumber = "FERMILAB-PUB-24-0796-V",
    doi = "10.1103/4dyc-z8zf",
    journal = "Phys. Rev. Lett.",
    volume = "135",
    number = "1",
    pages = "011802",
    year = "2025"
}

@article{PandaX:2024qfu,
    author = "Bo, Zihao and others",
    collaboration = "PandaX",
    title = "{Dark Matter Search Results from 1.54{\,}{\,}Tonne{\textperiodcentered}Year Exposure of PandaX-4T}",
    eprint = "2408.00664",
    archivePrefix = "arXiv",
    primaryClass = "hep-ex",
    doi = "10.1103/PhysRevLett.134.011805",
    journal = "Phys. Rev. Lett.",
    volume = "134",
    number = "1",
    pages = "011805",
    year = "2025"
}

@article{Affleck:1984fy,
    author = "Affleck, Ian and Dine, Michael",
    title = "{A New Mechanism for Baryogenesis}",
    reportNumber = "Print-84-0574 (PRINCETON)",
    doi = "10.1016/0550-3213(85)90021-5",
    journal = "Nucl. Phys. B",
    volume = "249",
    pages = "361--380",
    year = "1985"
}

@article{GellMann:1980vs,
    author = "Gell-Mann, Murray and Ramond, Pierre and Slansky, Richard",
    title = "{Complex Spinors and Unified Theories}",
    eprint = "1306.4669",
    archivePrefix = "arXiv",
    primaryClass = "hep-th",
    reportNumber = "PRINT-80-0576",
    journal = "Conf. Proc. C",
    volume = "790927",
    pages = "315--321",
    year = "1979"
}

@article{Mohapatra:1979ia,
    author = "Mohapatra, Rabindra N. and Senjanovic, Goran",
    title = "{Neutrino Mass and Spontaneous Parity Nonconservation}",
    reportNumber = "MDDP-TR-80-060, MDDP-PP-80-105, CCNY-HEP-79-10",
    doi = "10.1103/PhysRevLett.44.912",
    journal = "Phys. Rev. Lett.",
    volume = "44",
    pages = "912",
    year = "1980"
}

@article{Griest:1990kh,
    author = "Griest, Kim and Seckel, David",
    title = "{Three exceptions in the calculation of relic abundances}",
    reportNumber = "CFPA-TH-90-001A, BA-90-79",
    doi = "10.1103/PhysRevD.43.3191",
    journal = "Phys. Rev. D",
    volume = "43",
    pages = "3191--3203",
    year = "1991"
}

@article{Alloul:2013bka,
    author = "Alloul, Adam and Christensen, Neil D. and Degrande, Céline and Duhr, Claude and Fuks, Benjamin",
    title = "{FeynRules  2.0 - A complete toolbox for tree-level phenomenology}",
    eprint = "1310.1921",
    archivePrefix = "arXiv",
    primaryClass = "hep-ph",
    reportNumber = "CERN-PH-TH-2013-239, MCNET-13-14, IPPP-13-71, DCPT-13-142, PITT-PACC-1308",
    doi = "10.1016/j.cpc.2014.04.012",
    journal = "Comput. Phys. Commun.",
    volume = "185",
    pages = "2250--2300",
    year = "2014"
}

@article{Gondolo:1990dk,
      author         = "Gondolo, Paolo and Gelmini, Graciela",
      title          = "{Cosmic abundances of stable particles: Improved
                        analysis}",
      journal        = "Nucl. Phys.",
      volume         = "B360",
      year           = "1991",
      pages          = "145-179",
      doi            = "10.1016/0550-3213(91)90438-4",
      reportNumber   = "UCLA-90-TEP-68",
      SLACcitation   = "%%CITATION = NUPHA,B360,145;%%"
}

@article{ParticleDataGroup:2024cfk,
    author = "Navas, S. and others",
    collaboration = "Particle Data Group",
    title = "{Review of particle physics}",
    doi = "10.1103/PhysRevD.110.030001",
    journal = "Phys. Rev. D",
    volume = "110",
    number = "3",
    pages = "030001",
    year = "2024"
}

@inproceedings{Cline:2006ts,
    author = "Cline, James M.",
    title = "{Baryogenesis}",
    booktitle = "{Les Houches Summer School - Session 86: Particle Physics and Cosmology: The Fabric of Spacetime}",
    eprint = "hep-ph/0609145",
    archivePrefix = "arXiv",
    month = "9",
    year = "2006"
}

@article{Shaposhnikov:1987pf,
    author = "Shaposhnikov, M. E.",
    title = "{Structure of the High Temperature Gauge Ground State and Electroweak Production of the Baryon Asymmetry}",
    reportNumber = "NBI-HE-87-36",
    doi = "10.1016/0550-3213(88)90373-2",
    journal = "Nucl. Phys. B",
    volume = "299",
    pages = "797--817",
    year = "1988"
}

@article{Kajantie:1996mn,
    author = "Kajantie, K. and Laine, M. and Rummukainen, K. and Shaposhnikov, Mikhail E.",
    title = "{Is there a~ hot electroweak phase transition at $m_H \gtrsim m_W$?}",
    eprint = "hep-ph/9605288",
    archivePrefix = "arXiv",
    reportNumber = "CERN-TH-96-126, HD-THEP-96-15, IUHET-333",
    doi = "10.1103/PhysRevLett.77.2887",
    journal = "Phys. Rev. Lett.",
    volume = "77",
    pages = "2887--2890",
    year = "1996"
}

@article{CMS:2018iaf,
    author = "Sirunyan, Albert M and others",
    collaboration = "CMS",
    title = "{Search for heavy neutral leptons in events with three charged leptons in proton-proton collisions at $\sqrt{s} =$ 13 TeV}",
    eprint = "1802.02965",
    archivePrefix = "arXiv",
    primaryClass = "hep-ex",
    reportNumber = "CMS-EXO-17-012, CERN-EP-2018-006",
    doi = "10.1103/PhysRevLett.120.221801",
    journal = "Phys. Rev. Lett.",
    volume = "120",
    number = "22",
    pages = "221801",
    year = "2018"
}

@article{ATLAS:2024rzi,
    author = "Aad, Georges and others",
    collaboration = "ATLAS",
    title = "{Search for heavy Majorana neutrinos in e{\ensuremath{\pm}}e{\ensuremath{\pm}} and e{\ensuremath{\pm}}{\ensuremath{\mu}}{\ensuremath{\pm}} final states via WW scattering in pp collisions at s=13 TeV with the ATLAS detector}",
    eprint = "2403.15016",
    archivePrefix = "arXiv",
    primaryClass = "hep-ex",
    reportNumber = "CERN-EP-2024-083",
    doi = "10.1016/j.physletb.2024.138865",
    journal = "Phys. Lett. B",
    volume = "856",
    pages = "138865",
    year = "2024"
}

@article{Gu:2009yx,
    author = "Gu, Pei-Hong and Sarkar, Utpal",
    title = "{Annihilating Leptogenesis}",
    eprint = "0903.3473",
    archivePrefix = "arXiv",
    primaryClass = "hep-ph",
    doi = "10.1016/j.physletb.2009.07.029",
    journal = "Phys. Lett. B",
    volume = "679",
    pages = "118--121",
    year = "2009"
}

@article{Bodeker:2020ghk,
    author = "Bodeker, Dietrich and Buchmuller, Wilfried",
    title = "{Baryogenesis from the weak scale to the grand unification scale}",
    eprint = "2009.07294",
    archivePrefix = "arXiv",
    primaryClass = "hep-ph",
    reportNumber = "DESY 20-141, DESY-20-141",
    doi = "10.1103/RevModPhys.93.035004",
    journal = "Rev. Mod. Phys.",
    volume = "93",
    number = "3",
    pages = "035004",
    year = "2021"
}

@article{Pilaftsis:2003gt,
    author = "Pilaftsis, Apostolos and Underwood, Thomas E. J.",
    title = "{Resonant leptogenesis}",
    eprint = "hep-ph/0309342",
    archivePrefix = "arXiv",
    reportNumber = "MC-TH-2003-09",
    doi = "10.1016/j.nuclphysb.2004.05.029",
    journal = "Nucl. Phys. B",
    volume = "692",
    pages = "303--345",
    year = "2004"
}

@article{Abada:2006fw,
    author = "Abada, Asmaa and Davidson, Sacha and Josse-Michaux, Francois-Xavier and Losada, Marta and Riotto, Antonio",
    title = "{Flavor issues in leptogenesis}",
    eprint = "hep-ph/0601083",
    archivePrefix = "arXiv",
    reportNumber = "CERN-PH-TH-2006-001, DNI-UAN-06-03, LPT-ORSAY-06-03, LYCEN-2006-03",
    doi = "10.1088/1475-7516/2006/04/004",
    journal = "JCAP",
    volume = "04",
    pages = "004",
    year = "2006"
}

@article{Abada:2006ea,
    author = "Abada, A. and Davidson, S. and Ibarra, A. and Josse-Michaux, F. -X. and Losada, M. and Riotto, A.",
    title = "{Flavour Matters in Leptogenesis}",
    eprint = "hep-ph/0605281",
    archivePrefix = "arXiv",
    reportNumber = "CERN-PH-TH-2006-093, DNI-UAN-06-97FT, IFT-UAM-CSIC-06-23, LPT-ORSAY-06-21, LYCEN-2006-07",
    doi = "10.1088/1126-6708/2006/09/010",
    journal = "JHEP",
    volume = "09",
    pages = "010",
    year = "2006"
}

@article{Nardi:2006fx,
    author = "Nardi, Enrico and Nir, Yosef and Roulet, Esteban and Racker, Juan",
    title = "{The Importance of flavor in leptogenesis}",
    eprint = "hep-ph/0601084",
    archivePrefix = "arXiv",
    doi = "10.1088/1126-6708/2006/01/164",
    journal = "JHEP",
    volume = "01",
    pages = "164",
    year = "2006"
}

@article{Blanchet:2006be,
    author = "Blanchet, Steve and Di Bari, Pasquale",
    title = "{Flavor effects on leptogenesis predictions}",
    eprint = "hep-ph/0607330",
    archivePrefix = "arXiv",
    doi = "10.1088/1475-7516/2007/03/018",
    journal = "JCAP",
    volume = "03",
    pages = "018",
    year = "2007"
}

@article{XENON:2025vwd,
    author = "Aprile, E. and others",
    collaboration = "XENON",
    title = "{WIMP Dark Matter Search using a 3.1 tonne $\times$ year Exposure of the XENONnT Experiment}",
    eprint = "2502.18005",
    archivePrefix = "arXiv",
    primaryClass = "hep-ex",
    month = "2",
    year = "2025"
}

@article{Belyaev:2012qa,
    author = "Belyaev, Alexander and Christensen, Neil D. and Pukhov, Alexander",
    title = "{CalcHEP 3.4 for collider physics within and beyond the Standard Model}",
    eprint = "1207.6082",
    archivePrefix = "arXiv",
    primaryClass = "hep-ph",
    reportNumber = "PITT-PACC-1209",
    doi = "10.1016/j.cpc.2013.01.014",
    journal = "Comput. Phys. Commun.",
    volume = "184",
    pages = "1729--1769",
    year = "2013"
}

@article{Bhattacharya:2023kws,
    author = "Bhattacharya, Subhaditya and Mondal, Niloy and Roshan, Rishav and Vatsyayan, Drona",
    title = "{Leptogenesis, dark matter and gravitational waves from discrete symmetry breaking}",
    eprint = "2312.15053",
    archivePrefix = "arXiv",
    primaryClass = "hep-ph",
    doi = "10.1088/1475-7516/2024/06/029",
    journal = "JCAP",
    volume = "06",
    pages = "029",
    year = "2024"
}

@article{Kolb:1990vq,
      author         = "Kolb, Edward W. and Turner, Michael S.",
      title          = "{The Early Universe}",
      journal        = "Front. Phys.",
      volume         = "69",
      year           = "1990",
      pages          = "1-547",
      SLACcitation   = "%%CITATION = FRPHA,69,1;%%"
}

@article{Keung:1983uu,
    author = "Keung, Wai-Yee and Senjanovic, Goran",
    title = "{Majorana Neutrinos and the Production of the Right-handed Charged Gauge Boson}",
    reportNumber = "BNL-32872",
    doi = "10.1103/PhysRevLett.50.1427",
    journal = "Phys. Rev. Lett.",
    volume = "50",
    pages = "1427",
    year = "1983"
}

@article{Nemevsek:2018bbt,
    author = "Nemev{\v{s}}ek, Miha and Nesti, Fabrizio and Popara, Goran",
    title = "{Keung-Senjanovi{\'c} process at the LHC: From lepton number violation to displaced vertices to invisible decays}",
    eprint = "1801.05813",
    archivePrefix = "arXiv",
    primaryClass = "hep-ph",
    doi = "10.1103/PhysRevD.97.115018",
    journal = "Phys. Rev. D",
    volume = "97",
    number = "11",
    pages = "115018",
    year = "2018"
}

@article{Sakharov:1967dj,
      author         = "Sakharov, A. D.",
      title          = "{Violation of CP Invariance, C asymmetry, and baryon
                        asymmetry of the universe}",
      journal        = "Pisma Zh. Eksp. Teor. Fiz.",
      volume         = "5",
      year           = "1967",
      pages          = "32-35",
      doi            = "10.1070/PU1991v034n05ABEH002497",
      note           = "[Usp. Fiz. Nauk161,no.5,61(1991)]",
      SLACcitation   = "%%CITATION = ZFPRA,5,32;%%"
}

@article{Weinberg:1979bt,
      author         = "Weinberg, Steven",
      title          = "{Cosmological Production of Baryons}",
      journal        = "Phys. Rev. Lett.",
      volume         = "42",
      year           = "1979",
      pages          = "850-853",
      doi            = "10.1103/PhysRevLett.42.850",
      reportNumber   = "HUTP-78/A040",
      SLACcitation   = "%%CITATION = PRLTA,42,850;%%"
}

@article{Kolb:1979qa,
      author         = "Kolb, Edward W. and Wolfram, Stephen",
      title          = "{Baryon Number Generation in the Early Universe}",
      journal        = "Nucl. Phys.",
      volume         = "B172",
      year           = "1980",
      pages          = "224",
      doi            = "10.1016/0550-3213(80)90167-4,
                        10.1016/0550-3213(82)90012-8",
      note           = "[Erratum: Nucl. Phys.B195,542(1982)]",
      reportNumber   = "Print-79-0956 (CAL TECH), OAP-579, CALT-68-754",
      SLACcitation   = "%%CITATION = NUPHA,B172,224;%%"
}

@article{Fukugita:1986hr,
      author         = "Fukugita, M. and Yanagida, T.",
      title          = "{Baryogenesis Without Grand Unification}",
      journal        = "Phys. Lett.",
      volume         = "B174",
      year           = "1986",
      pages          = "45-47",
      doi            = "10.1016/0370-2693(86)91126-3",
      reportNumber   = "RIFP-641",
      SLACcitation   = "%%CITATION = PHLTA,B174,45;%%"
}

@article{Kuzmin:1985mm,
      author         = "Kuzmin, V. A. and Rubakov, V. A. and Shaposhnikov, M. E.",
      title          = "{On the Anomalous Electroweak Baryon Number
                        Nonconservation in the Early Universe}",
      journal        = "Phys. Lett.",
      volume         = "155B",
      year           = "1985",
      pages          = "36",
      doi            = "10.1016/0370-2693(85)91028-7",
      reportNumber   = "IC/85/8",
      SLACcitation   = "%%CITATION = PHLTA,155B,36;%%"
}

@article{Davidson:2002qv,
      author         = "Davidson, Sacha and Ibarra, Alejandro",
      title          = "{A Lower bound on the right-handed neutrino mass from
                        leptogenesis}",
      journal        = "Phys. Lett.",
      volume         = "B535",
      year           = "2002",
      pages          = "25-32",
      doi            = "10.1016/S0370-2693(02)01735-5",
      eprint         = "hep-ph/0202239",
      archivePrefix  = "arXiv",
      primaryClass   = "hep-ph",
      reportNumber   = "OUTP-02-10P, IPPP-02-16, DCPT-02-32",
      SLACcitation   = "%%CITATION = HEP-PH/0202239;%%"
}

@article{DuttaBanik:2020vfr,
    author = "Dutta Banik, Amit and Roshan, Rishav and Sil, Arunansu",
    title = "{Neutrino mass and asymmetric dark matter: study with inert Higgs doublet and high scale validity}",
    eprint = "2011.04371",
    archivePrefix = "arXiv",
    primaryClass = "hep-ph",
    month = "11",
    year = "2020"
}

@article{Nussinov:1985xr,
    author = "Nussinov, S.",
    title = "{TECHNOCOSMOLOGY: COULD A TECHNIBARYON EXCESS PROVIDE A 'NATURAL' MISSING MASS CANDIDATE?}",
    reportNumber = "CLNS-85/703",
    doi = "10.1016/0370-2693(85)90689-6",
    journal = "Phys. Lett. B",
    volume = "165",
    pages = "55--58",
    year = "1985"
}

@article{Davoudiasl:2012uw,
    author = "Davoudiasl, Hooman and Mohapatra, Rabindra N.",
    title = "{On Relating the Genesis of Cosmic Baryons and Dark Matter}",
    eprint = "1203.1247",
    archivePrefix = "arXiv",
    primaryClass = "hep-ph",
    doi = "10.1088/1367-2630/14/9/095011",
    journal = "New J. Phys.",
    volume = "14",
    pages = "095011",
    year = "2012"
}

@article{Petraki:2013wwa,
    author = "Petraki, Kalliopi and Volkas, Raymond R.",
    title = "{Review of asymmetric dark matter}",
    eprint = "1305.4939",
    archivePrefix = "arXiv",
    primaryClass = "hep-ph",
    reportNumber = "NIKHEF-2013-016",
    doi = "10.1142/S0217751X13300287",
    journal = "Int. J. Mod. Phys. A",
    volume = "28",
    pages = "1330028",
    year = "2013"
}

@article{Zurek:2013wia,
    author = "Zurek, Kathryn M.",
    title = "{Asymmetric Dark Matter: Theories, Signatures, and Constraints}",
    eprint = "1308.0338",
    archivePrefix = "arXiv",
    primaryClass = "hep-ph",
    doi = "10.1016/j.physrep.2013.12.001",
    journal = "Phys. Rept.",
    volume = "537",
    pages = "91--121",
    year = "2014"
}

@article{Yoshimura:1978ex,
    author = "Yoshimura, Motohiko",
    title = "{Unified Gauge Theories and the Baryon Number of the Universe}",
    reportNumber = "TU/78/179",
    doi = "10.1103/PhysRevLett.41.281",
    journal = "Phys. Rev. Lett.",
    volume = "41",
    pages = "281--284",
    year = "1978",
    note = "[Erratum: Phys.Rev.Lett. 42, 746 (1979)]"
}

@article{Barr:1979wb,
    author = "Barr, Stephen M.",
    title = "{Comments on Unitarity and the Possible Origins of the Baryon Asymmetry of the Universe}",
    reportNumber = "UPR-0122T",
    doi = "10.1103/PhysRevD.19.3803",
    journal = "Phys. Rev. D",
    volume = "19",
    pages = "3803",
    year = "1979"
}

@article{Baldes:2014gca,
    author = "Baldes, Iason and Bell, Nicole F. and Petraki, Kalliopi and Volkas, Raymond R.",
    title = "{Particle-antiparticle asymmetries from annihilations}",
    eprint = "1407.4566",
    archivePrefix = "arXiv",
    primaryClass = "hep-ph",
    doi = "10.1103/PhysRevLett.113.181601",
    journal = "Phys. Rev. Lett.",
    volume = "113",
    number = "18",
    pages = "181601",
    year = "2014"
}

@article{Cui:2011ab,
    author = "Cui, Yanou and Randall, Lisa and Shuve, Brian",
    title = "{A WIMPy Baryogenesis Miracle}",
    eprint = "1112.2704",
    archivePrefix = "arXiv",
    primaryClass = "hep-ph",
    doi = "10.1007/JHEP04(2012)075",
    journal = "JHEP",
    volume = "04",
    pages = "075",
    year = "2012"
}

@article{Bernal:2012gv,
    author = "Bernal, Nicolas and Josse-Michaux, Francois-Xavier and Ubaldi, Lorenzo",
    title = "{Phenomenology of WIMPy baryogenesis models}",
    eprint = "1210.0094",
    archivePrefix = "arXiv",
    primaryClass = "hep-ph",
    reportNumber = "BONN-TH-2012-24, CFTP-12-014",
    doi = "10.1088/1475-7516/2013/01/034",
    journal = "JCAP",
    volume = "01",
    pages = "034",
    year = "2013"
}

@article{Bernal:2013bga,
    author = "Bernal, Nicol\'as and Colucci, Stefano and Josse-Michaux, Fran\c{c}ois-Xavier and Racker, J. and Ubaldi, Lorenzo",
    title = "{On baryogenesis from dark matter annihilation}",
    eprint = "1307.6878",
    archivePrefix = "arXiv",
    primaryClass = "hep-ph",
    reportNumber = "BONN-TH-2013-11, CETUP2013-010, CFTP-13-01, IFIC-13-49",
    doi = "10.1088/1475-7516/2013/10/035",
    journal = "JCAP",
    volume = "10",
    pages = "035",
    year = "2013"
}

@article{Kumar:2013uca,
    author = "Kumar, Jason and Stengel, Patrick",
    title = "{WIMPy Leptogenesis With Absorptive Final State Interactions}",
    eprint = "1309.1145",
    archivePrefix = "arXiv",
    primaryClass = "hep-ph",
    reportNumber = "UH-511-1216-13, CETUP2013-015",
    doi = "10.1103/PhysRevD.89.055016",
    journal = "Phys. Rev. D",
    volume = "89",
    number = "5",
    pages = "055016",
    year = "2014"
}

@article{Racker:2014uga,
    author = "Racker, J. and Rius, N.",
    title = "{Helicitogenesis: WIMPy baryogenesis with sterile neutrinos and other realizations}",
    eprint = "1406.6105",
    archivePrefix = "arXiv",
    primaryClass = "hep-ph",
    reportNumber = "IFIC-14-42, FTUV-14-0419",
    doi = "10.1007/JHEP11(2014)163",
    journal = "JHEP",
    volume = "11",
    pages = "163",
    year = "2014"
}

@article{Dasgupta:2016odo,
    author = "Dasgupta, Arnab and Hati, Chandan and Patra, Sudhanwa and Sarkar, Utpal",
    title = "{A minimal model of TeV scale WIMPy leptogenesis}",
    eprint = "1605.01292",
    archivePrefix = "arXiv",
    primaryClass = "hep-ph",
    month = "5",
    year = "2016"
}

@article{Borah:2018uci,
    author = "Borah, Debasish and Dasgupta, Arnab and Kang, Sin Kyu",
    title = "{TeV Scale Leptogenesis via Dark Sector Scatterings}",
    eprint = "1806.04689",
    archivePrefix = "arXiv",
    primaryClass = "hep-ph",
    doi = "10.1140/epjc/s10052-020-8052-1",
    journal = "Eur. Phys. J. C",
    volume = "80",
    number = "6",
    pages = "498",
    year = "2020"
}

@article{Borah:2019epq,
    author = "Borah, Debasish and Dasgupta, Arnab and Kang, Sin Kyu",
    title = "{Two-component dark matter with cogenesis of the baryon asymmetry of the Universe}",
    eprint = "1903.10516",
    archivePrefix = "arXiv",
    primaryClass = "hep-ph",
    doi = "10.1103/PhysRevD.100.103502",
    journal = "Phys. Rev. D",
    volume = "100",
    number = "10",
    pages = "103502",
    year = "2019"
}

@article{Dasgupta:2019lha,
    author = "Dasgupta, Arnab and Dev, P. S. Bhupal and Kang, Sin Kyu and Zhang, Yongchao",
    title = "{New mechanism for matter-antimatter asymmetry and connection with dark matter}",
    eprint = "1911.03013",
    archivePrefix = "arXiv",
    primaryClass = "hep-ph",
    doi = "10.1103/PhysRevD.102.055009",
    journal = "Phys. Rev. D",
    volume = "102",
    number = "5",
    pages = "055009",
    year = "2020"
}

@article{Boucenna:2013wba,
    author = "Boucenna, S. M. and Morisi, S.",
    title = "{Theories relating baryon asymmetry and dark matter: A mini review}",
    eprint = "1310.1904",
    archivePrefix = "arXiv",
    primaryClass = "hep-ph",
    doi = "10.3389/fphy.2013.00033",
    journal = "Front. in Phys.",
    volume = "1",
    pages = "33",
    year = "2014"
}

@article{Barman:2021ost,
    author = "Barman, Basabendu and Borah, Debasish and Das, Suruj Jyoti and Roshan, Rishav",
    title = "{Non-thermal origin of asymmetric dark matter from inflaton and primordial black holes}",
    eprint = "2111.08034",
    archivePrefix = "arXiv",
    primaryClass = "hep-ph",
    reportNumber = "PI/UAN-2021-705FT",
    doi = "10.1088/1475-7516/2022/03/031",
    journal = "JCAP",
    volume = "03",
    number = "03",
    pages = "031",
    year = "2022"
}

@article{Borah:2020ivi,
    author = "Borah, Debasish and Dasgupta, Arnab and Mahanta, Devabrat",
    title = "{Dark sector assisted low scale leptogenesis from three body decay}",
    eprint = "2008.10627",
    archivePrefix = "arXiv",
    primaryClass = "hep-ph",
    doi = "10.1103/PhysRevD.105.015015",
    journal = "Phys. Rev. D",
    volume = "105",
    number = "1",
    pages = "015015",
    year = "2022"
}

@article{Dasgupta:2022isg,
    author = "Dasgupta, Arnab and Dev, P. S. Bhupal and Ghoshal, Anish and Mazumdar, Anupam",
    title = "{Gravitational wave pathway to testable leptogenesis}",
    eprint = "2206.07032",
    archivePrefix = "arXiv",
    primaryClass = "hep-ph",
    doi = "10.1103/PhysRevD.106.075027",
    journal = "Phys. Rev. D",
    volume = "106",
    number = "7",
    pages = "075027",
    year = "2022"
}

@article{Buchmuller:2002rq,
    author = "Buchmuller, W. and Di Bari, P. and Plumacher, M.",
    title = "{Cosmic microwave background, matter - antimatter asymmetry and neutrino masses}",
    eprint = "hep-ph/0205349",
    archivePrefix = "arXiv",
    reportNumber = "DESY-02-058, OUTP-02-23-P",
    doi = "10.1016/S0550-3213(02)00737-X",
    journal = "Nucl. Phys. B",
    volume = "643",
    pages = "367--390",
    year = "2002",
    note = "[Erratum: Nucl.Phys.B 793, 362 (2008)]"
}

@article{Das:2021esm,
    author = "Das, Arindam and Dev, P. S. Bhupal and Hosotani, Yutaka and Mandal, Sanjoy",
    title = "{Probing the minimal U(1)X model at future electron-positron colliders via fermion pair-production channels}",
    eprint = "2104.10902",
    archivePrefix = "arXiv",
    primaryClass = "hep-ph",
    reportNumber = "OU-HET-1085",
    doi = "10.1103/PhysRevD.105.115030",
    journal = "Phys. Rev. D",
    volume = "105",
    number = "11",
    pages = "115030",
    year = "2022"
}

@article{Kersten:2007vk,
    author = {Kersten, J{\"o}rn and Smirnov, Alexei Yu.},
    title = "{Right-Handed Neutrinos at CERN LHC and the Mechanism of Neutrino Mass Generation}",
    eprint = "0705.3221",
    archivePrefix = "arXiv",
    primaryClass = "hep-ph",
    doi = "10.1103/PhysRevD.76.073005",
    journal = "Phys. Rev. D",
    volume = "76",
    pages = "073005",
    year = "2007"
}

@article{Pilaftsis:1991ug,
    author = "Pilaftsis, Apostolos",
    title = "{Radiatively induced neutrino masses and large Higgs neutrino couplings in the standard model with Majorana fields}",
    eprint = "hep-ph/9901206",
    archivePrefix = "arXiv",
    reportNumber = "MZ-TH-91-32",
    doi = "10.1007/BF01482590",
    journal = "Z. Phys. C",
    volume = "55",
    pages = "275--282",
    year = "1992"
}

@article{Huang:2022vkf,
    author = "Huang, Peisi and Xie, Ke-Pan",
    title = "{Leptogenesis triggered by a first-order phase transition}",
    eprint = "2206.04691",
    archivePrefix = "arXiv",
    primaryClass = "hep-ph",
    doi = "10.1007/JHEP09(2022)052",
    journal = "JHEP",
    volume = "09",
    pages = "052",
    year = "2022"
}

@article{Borah:2024wos,
    author = "Borah, Debasish and Mahapatra, Satyabrata and Paul, Partha Kumar and Sahu, Narendra and Shukla, Prashant",
    title = "{Asymmetric self-interacting dark matter with a canonical seesaw model}",
    eprint = "2404.14912",
    archivePrefix = "arXiv",
    primaryClass = "hep-ph",
    doi = "10.1103/PhysRevD.110.035033",
    journal = "Phys. Rev. D",
    volume = "110",
    number = "3",
    pages = "035033",
    year = "2024"
}

@article{Borah:2023god,
    author = "Borah, Debasish and Dasgupta, Arnab and Knauss, Matthew and Saha, Indrajit",
    title = "{Baryon asymmetry from dark matter decay in the vicinity of a phase transition}",
    eprint = "2306.05459",
    archivePrefix = "arXiv",
    primaryClass = "hep-ph",
    doi = "10.1103/PhysRevD.108.L091701",
    journal = "Phys. Rev. D",
    volume = "108",
    number = "9",
    pages = "L091701",
    year = "2023"
}

@article{Chu:2021qwk,
    author = "Chu, Xiaoyong and Cui, Yanou and Pradler, Josef and Shamma, Michael",
    title = "{Dark freeze-out cogenesis}",
    eprint = "2112.10784",
    archivePrefix = "arXiv",
    primaryClass = "hep-ph",
    doi = "10.1007/JHEP03(2022)031",
    journal = "JHEP",
    volume = "03",
    pages = "031",
    year = "2022"
}

@article{Mahanta:2022gsi,
    author = "Mahanta, Devabrat and Borah, Debasish",
    title = "{WIMPy leptogenesis in non-standard cosmologies}",
    eprint = "2208.11295",
    archivePrefix = "arXiv",
    primaryClass = "hep-ph",
    doi = "10.1088/1475-7516/2023/03/049",
    journal = "JCAP",
    volume = "03",
    pages = "049",
    year = "2023"
}

@article{Roszkowski:2006kw,
    author = "Roszkowski, Leszek and Seto, Osamu",
    title = "{Axino dark matter from Q-balls in Affleck-Dine baryogenesis and the Omega(b) - Omega(DM) coincidence problem}",
    eprint = "hep-ph/0608013",
    archivePrefix = "arXiv",
    doi = "10.1103/PhysRevLett.98.161304",
    journal = "Phys. Rev. Lett.",
    volume = "98",
    pages = "161304",
    year = "2007"
}

@article{Seto:2007ym,
    author = "Seto, Osamu and Yamaguchi, Masahide",
    title = "{Axino warm dark matter and Omega(b) - Omega(DM) coincidence}",
    eprint = "0704.0510",
    archivePrefix = "arXiv",
    primaryClass = "hep-ph",
    reportNumber = "IFT-UAM-CSIC-07-15",
    doi = "10.1103/PhysRevD.75.123506",
    journal = "Phys. Rev. D",
    volume = "75",
    pages = "123506",
    year = "2007"
}

@article{Cheung:2011if,
    author = "Cheung, Clifford and Zurek, Kathryn M.",
    title = "{Affleck-Dine Cogenesis}",
    eprint = "1105.4612",
    archivePrefix = "arXiv",
    primaryClass = "hep-ph",
    doi = "10.1103/PhysRevD.84.035007",
    journal = "Phys. Rev. D",
    volume = "84",
    pages = "035007",
    year = "2011"
}

@article{vonHarling:2012yn,
    author = "von Harling, Benedict and Petraki, Kalliopi and Volkas, Raymond R.",
    title = "{Affleck-Dine dynamics and the dark sector of pangenesis}",
    eprint = "1201.2200",
    archivePrefix = "arXiv",
    primaryClass = "hep-ph",
    doi = "10.1088/1475-7516/2012/05/021",
    journal = "JCAP",
    volume = "05",
    pages = "021",
    year = "2012"
}

@article{Borah:2022qln,
    author = "Borah, Debasish and Jyoti Das, Suruj and Okada, Nobuchika",
    title = "{Affleck-Dine cogenesis of baryon and dark matter}",
    eprint = "2212.04516",
    archivePrefix = "arXiv",
    primaryClass = "hep-ph",
    doi = "10.1007/JHEP05(2023)004",
    journal = "JHEP",
    volume = "05",
    pages = "004",
    year = "2023"
}

@article{Borah:2023saq,
    author = "Borah, Debasish and Dasgupta, Arnab and Saha, Indrajit",
    title = "{LIGO-Virgo constraints on dark matter and leptogenesis triggered by a first order phase transition at high scale}",
    eprint = "2304.08888",
    archivePrefix = "arXiv",
    primaryClass = "hep-ph",
    doi = "10.1103/PhysRevD.109.095034",
    journal = "Phys. Rev. D",
    volume = "109",
    number = "9",
    pages = "095034",
    year = "2024"
}

@article{Cui:2020dly,
    author = "Cui, Yanou and Shamma, Michael",
    title = "{WIMP Cogenesis for Asymmetric Dark Matter and the Baryon Asymmetry}",
    eprint = "2002.05170",
    archivePrefix = "arXiv",
    primaryClass = "hep-ph",
    doi = "10.1007/JHEP12(2020)046",
    journal = "JHEP",
    volume = "12",
    pages = "046",
    year = "2020"
}

@article{Borah:2023qag,
    author = "Borah, Debasish and Jyoti Das, Suruj and Roshan, Rishav",
    title = "{Baryon asymmetry from dark matter decay}",
    eprint = "2305.13367",
    archivePrefix = "arXiv",
    primaryClass = "hep-ph",
    doi = "10.1103/PhysRevD.108.075025",
    journal = "Phys. Rev. D",
    volume = "108",
    number = "7",
    pages = "075025",
    year = "2023"
}

@article{Chun:2023ezg,
    author = "Chun, Eung Jin and Dutka, Tomasz P. and Jung, Tae Hyun and Nagels, Xander and Vanvlasselaer, Miguel",
    title = "{Bubble-assisted leptogenesis}",
    eprint = "2305.10759",
    archivePrefix = "arXiv",
    primaryClass = "hep-ph",
    reportNumber = "CTPU-PTC-23-17",
    doi = "10.1007/JHEP09(2023)164",
    journal = "JHEP",
    volume = "09",
    pages = "164",
    year = "2023"
}

@article{Arakawa:2024bkv,
    author = "Arakawa, Jason and Lu, Philip and Takhistov, Volodymyr",
    title = "{Phase Separation Baryogenesis}",
    eprint = "2409.12228",
    archivePrefix = "arXiv",
    primaryClass = "hep-ph",
    reportNumber = "KEK-QUP-2024-0022, KEK-TH-2655, KEK-Cosmo-0359",
    month = "9",
    year = "2024"
}

@article{Heisig:2024mwr,
    author = "Heisig, Jan",
    title = "{Conversion-Driven Leptogenesis: A Testable Theory of Dark Matter and Baryogenesis at the Electroweak Scale}",
    eprint = "2404.12428",
    archivePrefix = "arXiv",
    primaryClass = "hep-ph",
    reportNumber = "TTK-24-12",
    doi = "10.1103/PhysRevLett.133.191803",
    journal = "Phys. Rev. Lett.",
    volume = "133",
    number = "19",
    pages = "191803",
    year = "2024"
}

@article{Steigman:2012nb,
    author = "Steigman, Gary and Dasgupta, Basudeb and Beacom, John F.",
    title = "{Precise Relic WIMP Abundance and its Impact on Searches for Dark Matter Annihilation}",
    eprint = "1204.3622",
    archivePrefix = "arXiv",
    primaryClass = "hep-ph",
    doi = "10.1103/PhysRevD.86.023506",
    journal = "Phys. Rev. D",
    volume = "86",
    pages = "023506",
    year = "2012"
}

@article{Viana:2019ucn,
    author = "Viana, Aion and Schoorlemmer, Harm and Albert, Andrea and de Souza, Vitor and Harding, J. Patrick and Hinton, Jim",
    title = "{Searching for Dark Matter in the Galactic Halo with a Wide Field of View TeV Gamma-ray Observatory in the Southern Hemisphere}",
    eprint = "1906.03353",
    archivePrefix = "arXiv",
    primaryClass = "astro-ph.HE",
    doi = "10.1088/1475-7516/2019/12/061",
    journal = "JCAP",
    volume = "12",
    pages = "061",
    year = "2019"
}

\end{document}